\documentclass[aps,prb,10pt,twocolumn,longbibliography,superscriptaddress]{revtex4-2}
\usepackage{multirow}
\usepackage{graphicx}
\usepackage{dcolumn}
\usepackage{bm}
\usepackage{hyperref}
\hypersetup{
    colorlinks=true,
    urlcolor= blue,
    citecolor=blue,
linkcolor= blue}
\usepackage{amsmath,amssymb,amsfonts}%
\usepackage{amsthm}%
\usepackage{mathrsfs}%
\usepackage{appendix}
\usepackage{soul}
\usepackage{xr}
\begin{document}

\preprint{APS/123-QED}

\newcommand{\rmnote}[1]{\textcolor{blue}{\bf RM: #1}}
\newcommand{\SRgnote}[1]{\textcolor{green}{\bf SR: #1}}
\newcommand{\mknote}[1]{\textcolor{red}{\bf MK: #1}}

\title{Nonequilibrium Andreev resonances in ballistic graphene Andreev interferometers}

\author{Asmaul Smitha Rashid}
\affiliation{Department of Electrical Engineering, The Pennsylvania State University, University Park, Pennsylvania 16802, USA}
\affiliation{Materials Research Institute, The Pennsylvania State University, University Park, Pennsylvania 16802, USA}

\author{Le Yi}
\affiliation{Department of Physics, The Pennsylvania State University, University Park, Pennsylvania 16802, USA}
\affiliation{Materials Research Institute, The Pennsylvania State University, University Park, Pennsylvania 16802, USA}

\author{Takashi Taniguchi}
\affiliation{Research Center for Materials Nanoarchitectonics, National Institute for Materials Science,  1-1 Namiki, Tsukuba 305-0044, Japan}

\author{Kenji Watanabe}
\affiliation{Research Center for Electronic and Optical Materials, National Institute for Materials Science, 1-1 Namiki, Tsukuba 305-0044, Japan}

\author{Nitin Samarth}
\affiliation{Department of Physics, The Pennsylvania State University, University Park, Pennsylvania 16802, USA}
\affiliation{Department of Materials Science and Engineering, The Pennsylvania
State University, University Park, Pennsylvania 16802, USA}
\affiliation{Materials Research Institute, The Pennsylvania State University, University Park, Pennsylvania 16802, USA}

\author{R\'egis M\'elin}

\affiliation{Université Grenoble-Alpes, CNRS, Grenoble INP, Institut
  NEEL, Grenoble, France}

\author{Morteza Kayyalha}
\thanks{Corresponding author: mzk463@psu.edu}
\affiliation{Department of Electrical Engineering, The Pennsylvania State University, University Park, Pennsylvania 16802, USA}
\affiliation{Materials Research Institute, The Pennsylvania State University, University Park, Pennsylvania 16802, USA}

\date{\today}

\begin{abstract}
We investigate the nonequilibrium populations of Andreev bound states (ABSs) in ballistic graphene three-terminal Josephson junctions (JJs). Our measurements reveal periodic resistance oscillations as a function of the  superconducting phase, which is modulated by the applied magnetic flux. Additionally, we observe several voltage-induced crossovers, where resistance oscillations alternate between minima and maxima at integer flux quanta, corresponding to $0$ and $\pi$ phase shifts, respectively. We utilize a microscopic model for long ballistic Andreev interferometers, accounting for both perturbative and non-perturbative scattering processes. This model attributes the $0-\pi$ crossovers to phase-sensitive Andreev reflections (ARs), which convert normal current into ABSs with nonequilibrium populations. Our combined theoretical and experimental approach suggests that unconventional energy-phase relations, such as $\pi$-shifted ABSs, can be created in multiterminal JJs without relying on complex Cooper quartet or Floquet physics.

\end{abstract}

\maketitle

In superconductor-normal-superconductor (SNS) Josephson junctions (JJs), Andreev reflection at N-S interfaces leads to the formation of Andreev bound states (ABSs) {\cite{Furusaki1991, Kulik1970, Ishii1970, Bagwell1992}}. At equilibrium, these ABSs occupy the lowest energy states and are fundamental to the emergent properties of hybrid superconducting systems, such as spinful Andreev levels {\cite{Zazunov2003, Chtchelkatchev2003, Padurariu2010, HaysAndreevspin2021, PIta2023, Gingrich2016}}  and topological phases {\cite{Quitopologicalreview}}. Introducing additional superconducting leads in multiterminal JJs adds extra phase degrees of freedom to the ABS spectrum {\cite{Sauls2018, Nazarov2019}}, thereby enabling the exploration of higher-dimensional topological phases {\cite{Riwar2016,Meyer2017, Nazarov2019, Peralta2019, Chandrasekhar2022,  Peralta2023}}, Andreev modes hybridization {\cite{Chtchelkatchev2003, Coraiola2023, Matsuophase2023, MatsuoScience2023, Pillet2019}}, and Cooper quartets {\cite{Melin2023, Melininterferometer2023, gupta2024evidence}}.

When JJs are driven out of equilibrium by an external perturbation, such as a voltage bias {\cite{Melinvoltagebias2019, Baselmans1999}} or microwave irradiation {\cite{vanWoerkom2017, Rudner2020, Melinfloquet2017, Park2022, Haxell2023}}, the lowest-energy ABSs are excited to higher energy levels. In this scenario, the external perturbation often introduces a time-dependent superconducting phase, triggering a nonadiabatic transition that significantly impacts the population and dynamics of the ABSs. These nonequilibrium ABSs offer a unique platform to explore a range of phenomena, including non-trivial current-phase relations \cite{Golikovafluid2014, Chtchelkatchev2003} and the formation of Floquet-Andreev states \cite{Oka2019, Liu2019, Park2022, Carrad2022}.

The nonequilibrium dynamics of ABSs have been theoretically investigated in various systems, including diffusive Andreev interferometers {\cite{Zaikindiffusive2019}}, one-dimensional long junctions {\cite{vanWees1991}}, quantum dots {\cite{Melinquantumdot2020}}, and  two and multi-terminal JJs (MTJJs) {\cite{Melinquartet2020, Chandrasekhar2022}. We note that the last category of devices, MTJJs, have been extensively explored with a different focus {\cite{Pankratova2020,Graziano2020, Draelos2019, Arnault2021, MatutequantumCircuit2023}}, uncovering a variety of phenomena, including topological phases {\cite{Strambini2016, Meyer2017, Nazarov2019, Peralta2019, Chandrasekhar2022, Riwar2016, Peralta2023}}, Cooper quartets {\cite{Pfeffer2014, Arnaultdynamic2022, Grazianoselective2022, Zhang2023, cohen2018nonlocal,gupta2024evidence}}, hybridization of Andreev modes {\cite{Pillet2019, Coraiola2023, Matsuophase2023, MatsuoScience2023, Coraiola2023spin}}, and nonreciprocal superconducting transport {\cite{gupta2023superconducting, Coraioladiode, Chilesnonreciprocal2023, Zhang2024nonreciprocal, Matsuodiode2023}}. Experimental studies of nonequilibrium ABSs have largely centered on diffusive JJs, where voltage-induced shifts in the chemical potential broaden the quasiparticle distribution function, leading to the observation of anomalous supercurrents {\cite{Baselmans1999,Crosser2008, Dolgirev2019, Margineda2023}}. However, nonequilibrium processes in diffusive JJs are often short-lived and are thus challenging to study in detail {\cite{Dynes1978, Jaworskilifetime176}}.} In contrast, ballistic JJs are expected to exhibit ABSs with enhanced lifetime \cite{Melin3TJJ}, offering a more robust platform for studying nonequilibrium processes in quantum transport. Recent experiments have explored the inverse AC Josephson effect  \cite{Arnault2021} and Floquet-induced Cooper quartets \cite{huang2022evidence} in ballistic graphene multiterminal JJs. Despite these recent advances, the underlying dynamics and population of the ABSs in ballistic junctions remain largely unexplored.

In this work, we gain insights into nonequilibrium Andreev processes by investigating the interplay between the Fermi surface, superconducting phases, and voltage biases in ballistic graphene three-terminal Andreev interferometers. We investigate two distinct configurations: an all superconducting interferometer (SSS) and a phased-biased two-terminal JJ with an additional normal contact (SSN). In both cases, we observe periodic oscillations of resistance ($R$) with respect to the superconducting phase difference $\varphi$. Notably, in the all superconducting interferometer, we find \textit{multiple} voltage-induced crossovers where the oscillations alternate between minima ($0$ phase) and maxima ($\pi$ phase) at integer multiples of the magnetic flux quantum. 

To interpret our results, we employ a model, accounting for both ABSs and the nonequilibrium Fermi surface of graphene \cite{Melin3TJJ}. 
Unlike a recent report attributing the observed $0-\pi$ crossovers to nonequilibrium Cooper quartets \cite{huang2022evidence}, this model instead ascribes these crossovers to a phase-sensitive Andreev reflection process (phase-AR). In this framework, phase-ARs generate ABSs with nonequilibrium populations due to the finite chemical potential in graphene. 
In contrast to prior studies in diffusive JJs with normal contacts {\cite{Baselmans1999,Crosser2008, Dolgirev2019, Margineda2023}}, in ballistic JJs, we theoretically find that ABSs have longer lifetimes and smaller linewidth broadening  {\cite{Dynes1978, Jaworskilifetime176, Melin3TJJ}}. This enhanced  lifetime is consistent with the observation of multiple $0-\pi$ crossovers in our experiments. Our combined theoretical and experimental approach shows that conductance spectroscopy in SSS interferometers directly probes the nonequilibrium population of Andreev levels. Furthermore, our findings reveal that the crossovers are a unique feature of voltage-biased multiterminal JJs without the need for relying on more exotic phenomena, such as Cooper quartets and Floquet ABSs. 

We fabricate all superconducting (SSS) Andreev interferometers on hBN/graphene/hBN van der Waals heterostructures, which are edge-contacted by Ti (10 nm)/Al (80 nm) electrodes {\cite{wang2013one}}. The graphene channel is typically around 1.2 $\mu$m long and 0.5 $\mu$m wide. Figure~\ref{Fig1.jpg}(a) shows a scanning electron microscope (SEM) image of a representative three-terminal Andreev interferometer. The left, right, and top terminals are denoted as $S_L, S_R$, and $S_T$, respectively. Furthermore, terminals $S_R$ and $S_T$ in this geometry are connected by a superconducting loop. This loop allows us to control the superconducting phase $\varphi$ between $S_R$ and $S_T$ via an external magnetic flux as $\varphi = 2\pi\Phi/\Phi_0$, where $\Phi$ is the magnetic flux and $\Phi_0 = h/2e$ is the flux quantum. We apply current  $I_{dc}$  to $S_L$ and measure voltage $V_{LT}$ between $S_L$ and $S_T$. As a result, the three-terminal Andreev interferometer ($S_L, S_R$, $S_T$) is at ($V_{LT}$, 0, 0) bias voltage configuration. We note that the Al tunneling probe in this device exhibits resistance on the order of tens of M$\Omega$ and hence does not play any role in the transport properties of our Andreev interferometer; see results for another device without any tunneling probe in 
the supporting information (SI; see also Refs. \cite{Jessen2019, Vinay2024, Caroli1971, Caroli1972, Cuevas,Klapwijk2012} therein). To reduce Joule heating, we perform all the measurements at 300 mK if not otherwise specified {\cite{TomiJoule2021}}. 

We select graphene as a material of choice for our experiments because it offers ballistic transport, supports transparent superconducting contacts, and allows gate-tunability of the superconducting coupling {\cite{Borzenets2016, Knoch2012, DanneauCPS20221}}. Signatures of ballistic transport including Fabry-P\'erot oscillations of the critical current in a two-terminal JJ, and gate dependence of the characteristic energy $\date{E}$ in a SSS device are summarized in section {V} of the SI.  Finally, we design T-shaped devices such that all terminals are equidistantly connected to the graphene channel \cite{gupta2023superconducting} with the distance between each pair of terminals from the graphene center around 0.6 $\mu$m as depicted in Fig~\ref{Fig1.jpg}(a).

We first explore the critical current modulation as a function of the superconducting phase in SSS interferometers. Figures~\ref{Fig1.jpg}(b) and (c) show the color plot of the differential resistance $R$ as a function of the magnetic field $B$ and, respectively, the bias voltage $V_{LT}$ and the bias current $I_{dc}$ at 
gate voltage $V_g$ = 0 $V$. The dark blue region corresponds to supercurrent with $R$ = 0. We observe modulations of the supercurrent as a function of the magnetic flux. The period of the modulations is approximately $\Delta$B = 0.8 G which is consistent with $\Delta$B = 0.827 G, corresponding to one flux quantum for the enclosed superconducting loop area of 25 $\mu$m$^2$. We further observe a series of dark-bright regions (indicated by arrows) appearing outside of the superconducting region. To better illustrate these regions, we plot line cuts of $R$ versus $\Phi/\Phi_0$ (corresponding to the red arrow in Figs.~\ref{Fig1.jpg}(b) and (c)) at various $V_{LT}$’s in Fig.~\ref{Fig1.jpg}(d). For $|V_{LT}| < 35 \mu$V, we observe minima in the resistance oscillations at multiples of the flux quantum with a periodicity of $\Phi_0$. As we increase the bias voltage ($|V_{LT}|$), we observe frequency doubling at $V_{LT}\approx$ -35 $\mu$V, beyond which $\pi$-phase shifted oscillations with maxima at multiples of flux quantum are dominant. Thus, changing the bias voltage produces $0-\pi$ crossovers in $R$ versus $\varphi$. In total, we observe four crossovers for bias voltages $V_{LT}<\Delta = 180~\mu$eV, where $\Delta$ is the superconducting gap of aluminum; see section {VII} of the SI for line cuts corresponding to purple, yellow, and black arrows in Figs. \ref{Fig1.jpg}(b) and (c).

{To understand the role of the superconducting terminal $S_L$, we fabricate an SSN Andreev interferometer, where a normal terminal $N_L$, fabricated with Cr (10nm)/Au (80nm), replaces $S_L$. Here, we also apply current $I_N$ to $N_L$ and measure voltage $V_{LT}$ between $N_L$ and $S_T$ as shown in Fig.~\ref{Fig2SSN.jpg}(a). We observe similar periodic oscillations of the resistance with the superconducting phase marked by a black arrow in Fig.~\ref{Fig2SSN.jpg}(b). These maps are generated by sweeping $I_{dc}$ at different $B$ fields with step sizes of 1 nA and 0.05 G, respectively. Because of the large step size of $B$, the cuts from Fig.~\ref{Fig2SSN.jpg}(a) have a poor resolution (see section XII of the SI). To improve the quality of the data, we separately measure the differential resistance $R$ by sweeping the $B$ field with a step size of 0.01 G at different values of $I_{dc}$ near the crossover as shown in Fig.~\ref{Fig2SSN.jpg}(b). Figure~\ref{FigS2SSN.jpg}(c) shows cuts at constant $I_{dc}$’s near the crossover extracted directly from Fig.~\ref{Fig2SSN.jpg}(a). We note that due to flux trapping in between measurements, the curves show a horizontal shift ($\approx$ 0.5 G) in Fig.~\ref{Fig2SSN.jpg}(b). However, unlike the SSS junction, which shows multiple $0-\pi$ crossovers, we only observe a single crossover in the SSN device. We attribute this finding to the broadening of the distribution function and elastic scattering at the normal metal–graphene interface {\cite{ DanneauCPS20221, Baselmans1999}}, leading to shorter Andreev resonance lifetimes. The shorter lifetime is consistent with quasiparticle poisoning arising from parity-nonconserving scattering processes occurring in the SSN device; see sections {II} and {III} of the SI for more details.

\begin{figure}[h]
\centering
\includegraphics[width=7.5cm]{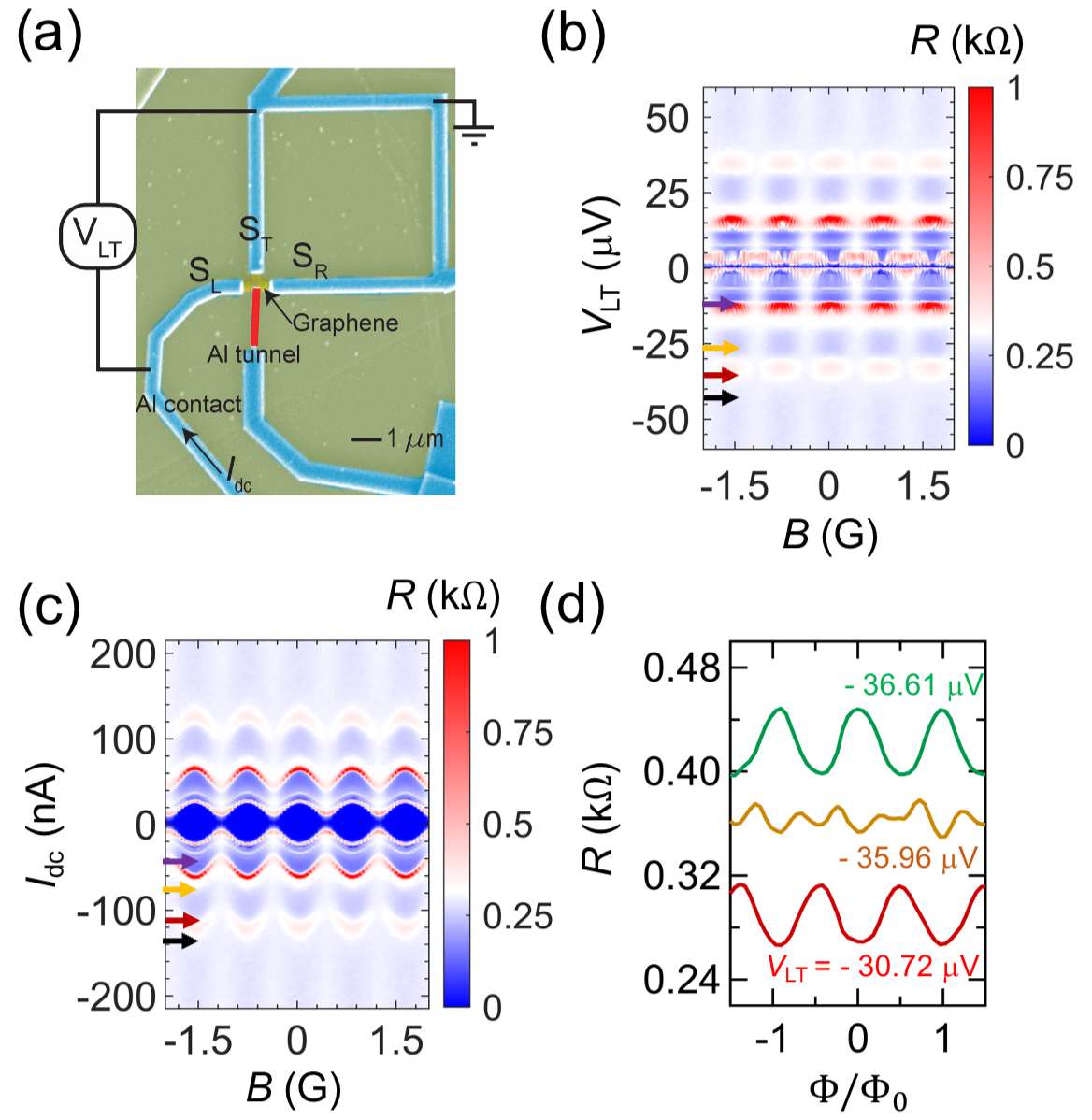}
 \caption{\label{Fig1.jpg} (a) Scanning electron microscope (SEM) image of a representative three-terminal Andreev interferometer along with the measurement configuration. Current $I_{dc}$ is applied to terminal $S_L$ and voltage $V_{LT}$ is measured between terminals $S_L$ and $S_T$. The superconducting loop is grounded. (b, c) Differential resistance $R$ as a function of the magnetic field $B$ and $V_{LT}$ (b) and bias current $I_{dc}$ (c). Arrows mark $0-\pi$ crossovers. (d) Differential resistance $R$ as a function of the magnetic flux $\Phi$ for the crossovers marked by the red arrows in (b) and (c). The oscillations show an evolution from minima to maxima with increasing $|V_{LT}|$. Curves are shifted vertically by 0.03 $k\Omega$ for clarity. Average value of $V_{LT}$ is used in (d). Color maps are saturated to better highlight the $0-\pi$ crossovers. These maps are generated by sweeping $B$ and $I_{dc}$ with step sizes of 0.05 G and 1 nA, respectively.}
\end{figure}

\begin{figure}[h]
\centering
\includegraphics[width=8.8cm]{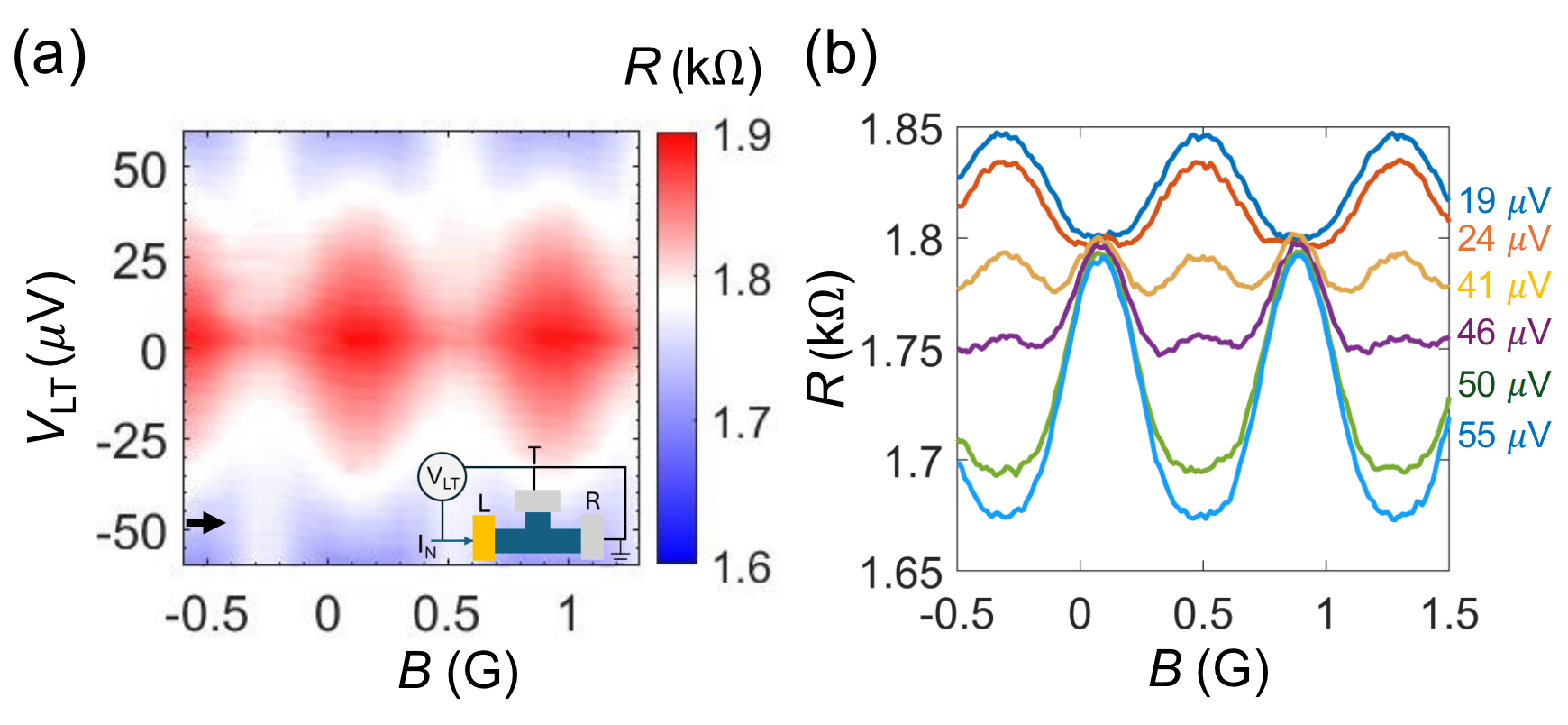}
 \caption{\label{Fig2SSN.jpg}{(a) Differential resistance $R$ as a function of the magnetic field $B$ and $V_{LT}$ for a three-terminal SSN device at a back gate voltage $V_g = $ 0 $V$ and temperature $ T$ = 12 mK. The Dirac point is at $V_{CNP}$ $\approx$ -4.5 $V$. The inset shows the measurement setup. The superconducting terminals are denoted as $S_T$ and $S_R $, and the normal terminal is denoted as $N_L$. Current $I_N$ is applied to $N_L$ and voltage $V_{LT}$ is measurable between terminals $N_L$ and $S_T$. (b)  Differential resistance $R$ versus $B$ at different $V_{LT}$'s measured separately near the $0-\pi$ crossover.  $R$ is generated by sweeping $B$ field with a step size of 0.01 G at different $V_{LT}$'s. The average value of $V_{LT}$ is used in panel (b). The colormap is saturated to better highlight $0-\pi$ crossovers. Colormap is generated by sweeping $B$ and $I_{dc}$ with step sizes of 0.1 G and 1 nA, respectively.}}
 \end{figure}

We explore several potential mechanisms for $0-\pi$ crossovers in our experiment, including multiple Andreev reflections (MARs), normal current, self-induced Shapiro steps, and the phase-AR process. We note that the voltage range, where we observe the resonances, are very small ($\leq$ 50 $\mu$V), which if coming from MARs, would correspond to $n \geq$ 6 (and even to $n \approx$ 27), where $n$ = 2$\Delta/eV_{LT}$ and $\Delta \approx$ 180 $\mu $eV. Additionally, we do not find any resonance peak for $n$ = 1, 2, 3, 4, and 5, further supporting our conclusion that MARs are likely not responsible for $0-\pi$ crossovers; see section {VIII} of the SI for more details. 

To understand the impact of the normal current injection in graphene JJs, we fabricate two-terminal JJs with two additional normal contacts following Ref. {\cite{Baselmans1999}}. Unlike Ref. {\cite{Baselmans1999}}, we observe that the critical current monotonically decreases with increasing normal current without any $0-\pi$ transition or supercurrent sign reversal; see section {IX} of the SI for more details.

Finally, we investigate the potential contribution of self-induced Shapiro steps \cite{Klapwijk1977,Fiske1965} to the observed resonances by fabricating a two-terminal graphene JJ adjacent to the SSS interferometer. In the two-terminal JJ, we observe two self-induced Shapiro resonances for bias voltages within $\pm 25 \mu$V ($|V_{bias} | \leq 25 \mu$V). However, additional resonance peaks emerge in the SSS interferometer for $|V_{LT} |\ge 25 \mu$; see section {X} of the SI for more details. Furthermore, in our SSN interferometer, where terminals $S_R$ and $S_T$ are grounded, Shapiro steps are entirely suppressed, consistent with our conclusion that self-induced Shapiro steps are likely not the source of the crossovers.

In the following, we describe the phase-AR process, which provides the most straightforward explanation for the observed resonance peaks and $0-\pi$ crossovers in both SSS and SSN interferometers. Figure~\ref{Fig2theory}(a) shows the schematics of the phase-AR process in a voltage-biased three-terminal Andreev interferometer. In two-terminal JJs, Cooper pair transfers are facilitated by low-energy ABSs with equilibrium populations, as both superconductors are at the same chemical potential. Likewise, in the SSS interferometer, when $S_L$ is at zero bias voltage ($V_{LT}$= 0 V), Cooper pairs transmit between $S_L$ and ($S_R$, $S_T$) via equilibrium ABSs. In contrast, a bias voltage applied to $S_L$ changes the chemical potential of graphene. However, phase-dependent Cooper pair transfers still occur between terminals $S_R$ and $S_T$, which remain grounded. This transfer process requires non-standard phase-sensitive ARs, giving rise to ABSs at higher energy levels. We have schematically represented this phase-AR process in  Fig.~\ref{Fig2theory}(a), where the wavy line terminated by ($\rho$) denotes nonlocal coupling via the Keldysh Green's function in our model; see sections {II} and {III} of the SI for more details.

Our theoretical model considers the device structure of Fig.~\ref{Fig1.jpg}(a) with superconducting phase variables $\varphi_R \simeq \varphi_0 + \Phi/2$ and $\varphi_T \simeq \varphi_0 - \Phi/2$, where $\varphi_0$ is the independent phase variable and $\Phi$ is the external magnetic flux threading the loop. In this case, the phase variable $\varphi_R - \varphi_T \simeq \Phi$ is the only static variable in our calculations, and other phase variables, e.g., $\varphi_{R,T} - \varphi_L$, have AC components when $V_{LT} \neq $ 0.  We note that our three-terminal Andreev interferometer operates in the AR regime for $V_{LT} \neq $ 0, and the transport of Cooper pairs is primarily through AC Josephson oscillations of the current from $S_L$ to $S_R$ and $S_T$. The dissipative component of this current results in a change of the chemical potential by $\delta \mu_N = e V_{LT}$, which in turn leads to voltage-sensitive jumps in the nonequilibrium distribution function of graphene {\cite{Melin3TJJ}}.

\begin{figure}[h]
 \centering
\includegraphics[width=8.5cm]{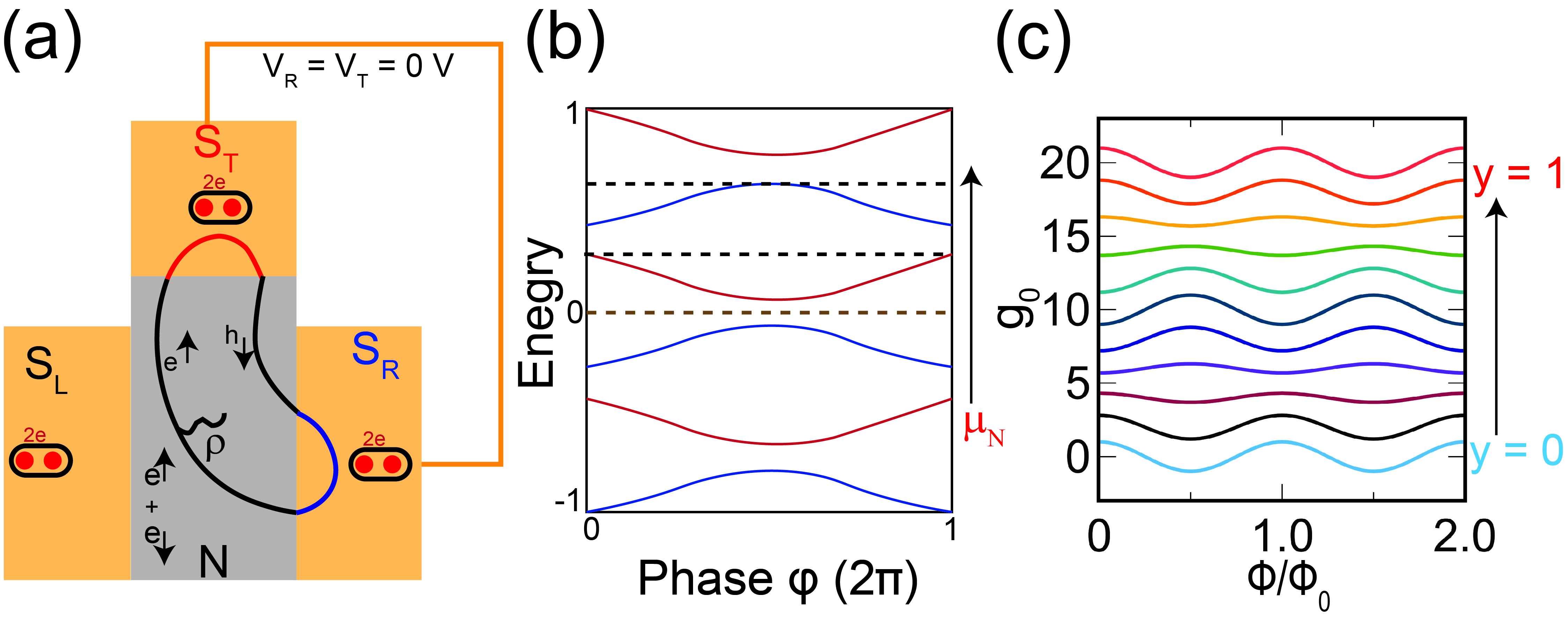}
\caption{\label{Fig2theory} (a) Schematics of the phase-Andreev refection (AR) process in a three-terminal Andreev interferometer where the wavy line represents coupling to the density of state in the normal region,
via the nonlocal Keldysh Green’s function. (b) The schematic diagram of energy versus phase relation of Andreev bound states (ABSs) with nonequilibrium populations. The dashed lines mark the chemical potential $\mu_N$ which changes with $V_{LT}$. (c) Theoretically calculated dimensionless conductance $g_0$ as a function of $\Phi/\Phi_0$ at various effective Fermi energies ($y$'s).}

\end{figure}

Figure~\ref{Fig2theory}(b) schematically shows the spectrum of the nonequilibrium ABSs arising from the voltage-sensitive jumps in the chemical potential. These ABSs resemble those expected in the long-junction limit {\cite{Bagwell1992, Kulik1970, Ishii1970}}.
Our model predicts that the chemical potential $\mu_N$ (dash horizontal lines) increases with increasing $V_{LT}$ and eventually crosses the spectrum at higher energy ABSs when $E(\varphi_T - \varphi_R) = \mu_N$. The high-energy ABSs exhibit $\pi$-shifted energy-phase relation, thereby resulting in the observation of $\pi$-shifted resonances in $R$ versus $\varphi$. Alternating between different spectra of ABSs via further increase of $\mu_N$ will yield crossovers between resonances with maxima (minima) at multiples of flux quantum as shown by red (blue) curves. Experimentally, we observe that DC voltages ($V_n$’s) corresponding to peaks in $R$ are independent of $V_g$; see section {V} of the SI for more details. In long JJs, the energy level spacing of Andreev modes is proportional to the Fermi velocity $v_F$ {\cite{Kulik1970, Ishii1970, Bagwell1992}}. Therefore, $v_F$ and, consequently, $V_n$’s remain approximately constant with $V_g$ due to the linear Dirac dispersion relation in graphene.

To calculate the differential conductance of the graphene Andreev interferometer, we assume an infinite planar geometry. We further assume the graphene interferometer is spatially uniform without any geometrical resonance. We calculate the current $I_{LT}$ defined by Keldysh Green’s functions and obtain the differential conductance $dI_{LT}/dV_{LT}$  within the lowest-order perturbation theory as a function of bias voltage $V_{LT}$ and current $I_{LT}$. We obtain a dimensionless differential conductance ($g_0 \propto  dI_{LT}/dV_{LT}$ ) by taking the derivative of $I_{LT}$ with respect to $V_{LT}$, where $g_0(\Phi/2\pi,V_{LT} R/\pi \hbar v_F)$ = $\cos{\Phi} \cos{(2eV_{LT}R/\hbar v_F)}$. 
We note that unlike Ref. \cite{huang2022evidence} where the quartet critical current ($I_q \propto \sin{\varphi_q}$) is responsible for crossovers, in our system, the conductance is proportional to $\cos{\Phi}$ instead of the expected $\sin{\Phi}$ for the equilibrium
DC Josephson current between $S_T$ and $S_R$. This $\cos{\Phi}$ term arises from phase-ARs, which are responsible for the conversion of normal current into supercurrent.

In ballistic devices, this model predicts that Andreev resonances are undamped and follow the $\cos{(2eV_{LT}R/\hbar v_F)}$, where $E=eV_{LT}$ is the bias voltage energy, $\tau = R/v_F$ is the ballistic propagation delay between $S_R$ and $S_T$, and $v_F$ is the Fermi velocity {\cite{Melin3TJJ}}. These undamped conductance oscillations are distinct from the damping predicted in the diffusive limit \cite{Zaikindiffusive2019} and exhibit multiple sign changes and $0-\pi$ crossovers, consistent with our experiments; see section {IV} of the SI for more details. Figure~\ref{Fig2theory}(c) plots the calculated $g_0$ as a function of $\Phi/\Phi_0$ and effective Fermi energy $y$ = $V_{LT}$ $R/\pi\hbar v_F$. We observe multiple $0-\pi$ crossovers with increasing bias voltage $V_{LT}$ in this theoretical plot, corroborating the experimental findings in Figs.~\ref{Fig1.jpg} and ~\ref{Fig2SSN.jpg}. Our model predicts equally spaced resonances as shown in Fig.~\ref{Fig2theory}(c), which result from specific assumptions about linearized dispersion relations within a perfect waveguide geometry. However, replacing the rectangular graphene section with a chaotic cavity is expected to yield unequally spaced Andreev resonances {\cite{Chaoticcavity}}, more closely resembling the experimental observations in Figs.~\ref{Fig1.jpg} and ~\ref{Fig2SSN.jpg}. Finally, we find theoretically that the line width broadening of ABSs is reduced by a factor of $\eta_s/\Delta$ in SSS devices compared to SSN interferometers, where $\eta_s$ is Dynes parameter. The emergence of a long timescale and small linewidth broadening points to the protection of the ABSs against the short relaxation time of the quasiparticle continua.

In conclusion, we used conductance spectroscopy to probe the nonequilibrium occupation of ABSs in voltage-biased three-terminal graphene JJs. We investigated three-terminal Andreev interferometers with superconducting contacts (SSS) and two-terminal JJs with normal contacts (SSN).
In both cases, we observed periodic resonances in $R$ versus $\varphi$, which we attributed to the energy-phase relation of nonequilibrium ABSs. In the SSS interferometer, we further observed several $0-\pi$ crossovers, where the resistance oscillations exhibit alternating $0$ and $\pi$ phase shifts with minima and maxima at integer flux quanta, respectively. Conversely, in the SSN interferometer, the crossovers are suppressed and only one crossover is observed.
To understand the microscopic mechanisms of these crossovers, we employed a model based on  the  phase-AR mechanism that converts normal current to supercurrent. This model predicts that Andreev resonances are undamped in ballistic JJs, resulting in multiple $0-\pi$ crossovers in the conductance, consistent with our experimental observations. We considered several possible mechanisms for the observed crossovers, including MARs and self-induced Shapiro steps, and concluded that the phase-AR process provides the simplest explanation for both SSN and SSS configurations. However, the comparison between our theoretical predictions and experimental data highlights the need for future investigations into inelastic scattering mechanisms and quasiparticle poisoning, employing more realistic modeling techniques. One promising direction for such future theoretical work is the adoption of density matrix calculations to provide a more accurate description. Future research may focus on direct measurements of the local density of states, e.g., through conductance and tunneling spectroscopy \cite{Bretheau2017, Bretheau2018}, to elucidate the connection between phase-ARs and the theoretically predicted Floquet-Andreev states in multiterminal JJs.

The data supporting the conclusions of this letter is
available on Zenodo {\cite{Zenodo}}.
\begin{acknowledgements}
We acknowledge funding from the Pennsylvania State University Materials Research Science and Engineering Center supported by the US National Science Foundation (DMR 2011839). R.M. acknowledges support from the French CNRS in Grenoble and the German KIT in Karlsruhe, which is funded by the International Research Project SUPRADEVMAT. K.W. and T.T. acknowledge support from the JSPS KAKENHI (Grant Numbers 21H05233 and 23H02052) and World Premier International Research Center Initiative (WPI), MEXT, Japan.

\end{acknowledgements}

\bibliography{References}

\begin{thebibliography}{82}%
\makeatletter
\providecommand \@ifxundefined [1]{%
 \@ifx{#1\undefined}
}%
\providecommand \@ifnum [1]{%
 \ifnum #1\expandafter \@firstoftwo
 \else \expandafter \@secondoftwo
 \fi
}%
\providecommand \@ifx [1]{%
 \ifx #1\expandafter \@firstoftwo
 \else \expandafter \@secondoftwo
 \fi
}%
\providecommand \natexlab [1]{#1}%
\providecommand \enquote  [1]{``#1''}%
\providecommand \bibnamefont  [1]{#1}%
\providecommand \bibfnamefont [1]{#1}%
\providecommand \citenamefont [1]{#1}%
\providecommand \href@noop [0]{\@secondoftwo}%
\providecommand \href [0]{\begingroup \@sanitize@url \@href}%
\providecommand \@href[1]{\@@startlink{#1}\@@href}%
\providecommand \@@href[1]{\endgroup#1\@@endlink}%
\providecommand \@sanitize@url [0]{\catcode `\\12\catcode `\$12\catcode `\&12\catcode `\#12\catcode `\^12\catcode `\_12\catcode `\%12\relax}%
\providecommand \@@startlink[1]{}%
\providecommand \@@endlink[0]{}%
\providecommand \url  [0]{\begingroup\@sanitize@url \@url }%
\providecommand \@url [1]{\endgroup\@href {#1}{\urlprefix }}%
\providecommand \urlprefix  [0]{URL }%
\providecommand \Eprint [0]{\href }%
\providecommand \doibase [0]{https://doi.org/}%
\providecommand \selectlanguage [0]{\@gobble}%
\providecommand \bibinfo  [0]{\@secondoftwo}%
\providecommand \bibfield  [0]{\@secondoftwo}%
\providecommand \translation [1]{[#1]}%
\providecommand \BibitemOpen [0]{}%
\providecommand \bibitemStop [0]{}%
\providecommand \bibitemNoStop [0]{.\EOS\space}%
\providecommand \EOS [0]{\spacefactor3000\relax}%
\providecommand \BibitemShut  [1]{\csname bibitem#1\endcsname}%
\let\auto@bib@innerbib\@empty
\bibitem [{\citenamefont {Furusaki}\ and\ \citenamefont {Tsukada}(1991)}]{Furusaki1991}%
  \BibitemOpen
  \bibfield  {author} {\bibinfo {author} {\bibfnamefont {A.}~\bibnamefont {Furusaki}}\ and\ \bibinfo {author} {\bibfnamefont {M.}~\bibnamefont {Tsukada}},\ }\bibfield  {title} {\bibinfo {title} {Current-carrying states in josephson junctions},\ }\href {https://doi.org/10.1103/PhysRevB.43.10164} {\bibfield  {journal} {\bibinfo  {journal} {Phys. Rev. B}\ }\textbf {\bibinfo {volume} {43}},\ \bibinfo {pages} {10164} (\bibinfo {year} {1991})}\BibitemShut {NoStop}%
\bibitem [{\citenamefont {{Kulik}}(1969)}]{Kulik1970}%
  \BibitemOpen
  \bibfield  {author} {\bibinfo {author} {\bibfnamefont {I.~O.}\ \bibnamefont {{Kulik}}},\ }\bibfield  {title} {\bibinfo {title} {Macroscopic quantization and the proximity effect in s-n-s junctions},\ }\href@noop {} {\bibfield  {journal} {\bibinfo  {journal} {Soviet Journal of Experimental and Theoretical Physics}\ }\textbf {\bibinfo {volume} {30}},\ \bibinfo {pages} {944} (\bibinfo {year} {1969})}\BibitemShut {NoStop}%
\bibitem [{\citenamefont {Ishii}(1970)}]{Ishii1970}%
  \BibitemOpen
  \bibfield  {author} {\bibinfo {author} {\bibfnamefont {C.}~\bibnamefont {Ishii}},\ }\bibfield  {title} {\bibinfo {title} {Josephson currents through junctions with normal metal barriers},\ }\href {https://doi.org/10.1143/PTP.44.1525} {\bibfield  {journal} {\bibinfo  {journal} {Progress of Theoretical Physics}\ }\textbf {\bibinfo {volume} {44}},\ \bibinfo {pages} {1525} (\bibinfo {year} {1970})}\BibitemShut {NoStop}%
\bibitem [{\citenamefont {Bagwell}(1992)}]{Bagwell1992}%
  \BibitemOpen
  \bibfield  {author} {\bibinfo {author} {\bibfnamefont {P.~F.}\ \bibnamefont {Bagwell}},\ }\bibfield  {title} {\bibinfo {title} {Suppression of the josephson current through a narrow, mesoscopic, semiconductor channel by a single impurity},\ }\href {https://doi.org/10.1103/PhysRevB.46.12573} {\bibfield  {journal} {\bibinfo  {journal} {Phys. Rev. B}\ }\textbf {\bibinfo {volume} {46}},\ \bibinfo {pages} {12573} (\bibinfo {year} {1992})}\BibitemShut {NoStop}%
\bibitem [{\citenamefont {Zazunov}\ \emph {et~al.}(2003)\citenamefont {Zazunov}, \citenamefont {Shumeiko}, \citenamefont {Bratus'}, \citenamefont {Lantz},\ and\ \citenamefont {Wendin}}]{Zazunov2003}%
  \BibitemOpen
  \bibfield  {author} {\bibinfo {author} {\bibfnamefont {A.}~\bibnamefont {Zazunov}}, \bibinfo {author} {\bibfnamefont {V.~S.}\ \bibnamefont {Shumeiko}}, \bibinfo {author} {\bibfnamefont {E.~N.}\ \bibnamefont {Bratus'}}, \bibinfo {author} {\bibfnamefont {J.}~\bibnamefont {Lantz}},\ and\ \bibinfo {author} {\bibfnamefont {G.}~\bibnamefont {Wendin}},\ }\bibfield  {title} {\bibinfo {title} {Andreev level qubit},\ }\href {https://doi.org/10.1103/PhysRevLett.90.087003} {\bibfield  {journal} {\bibinfo  {journal} {Phys. Rev. Lett.}\ }\textbf {\bibinfo {volume} {90}},\ \bibinfo {pages} {087003} (\bibinfo {year} {2003})}\BibitemShut {NoStop}%
\bibitem [{\citenamefont {Chtchelkatchev}\ and\ \citenamefont {Nazarov}(2003)}]{Chtchelkatchev2003}%
  \BibitemOpen
  \bibfield  {author} {\bibinfo {author} {\bibfnamefont {N.~M.}\ \bibnamefont {Chtchelkatchev}}\ and\ \bibinfo {author} {\bibfnamefont {Y.~V.}\ \bibnamefont {Nazarov}},\ }\bibfield  {title} {\bibinfo {title} {Andreev quantum dots for spin manipulation},\ }\href {https://doi.org/10.1103/PhysRevLett.90.226806} {\bibfield  {journal} {\bibinfo  {journal} {Phys. Rev. Lett.}\ }\textbf {\bibinfo {volume} {90}},\ \bibinfo {pages} {226806} (\bibinfo {year} {2003})}\BibitemShut {NoStop}%
\bibitem [{\citenamefont {Padurariu}\ and\ \citenamefont {Nazarov}(2010)}]{Padurariu2010}%
  \BibitemOpen
  \bibfield  {author} {\bibinfo {author} {\bibfnamefont {C.}~\bibnamefont {Padurariu}}\ and\ \bibinfo {author} {\bibfnamefont {Y.~V.}\ \bibnamefont {Nazarov}},\ }\bibfield  {title} {\bibinfo {title} {Theoretical proposal for superconducting spin qubits},\ }\href {https://doi.org/10.1103/PhysRevB.81.144519} {\bibfield  {journal} {\bibinfo  {journal} {Phys. Rev. B}\ }\textbf {\bibinfo {volume} {81}},\ \bibinfo {pages} {144519} (\bibinfo {year} {2010})}\BibitemShut {NoStop}%
\bibitem [{\citenamefont {Hays}\ \emph {et~al.}(2021)\citenamefont {Hays}, \citenamefont {Fatemi}, \citenamefont {Bouman}, \citenamefont {Cerrillo}, \citenamefont {Diamond}, \citenamefont {Serniak}, \citenamefont {Connolly}, \citenamefont {Krogstrup}, \citenamefont {Nygård}, \citenamefont {Yeyati}, \citenamefont {Geresdi},\ and\ \citenamefont {Devoret}}]{HaysAndreevspin2021}%
  \BibitemOpen
  \bibfield  {author} {\bibinfo {author} {\bibfnamefont {M.}~\bibnamefont {Hays}}, \bibinfo {author} {\bibfnamefont {V.}~\bibnamefont {Fatemi}}, \bibinfo {author} {\bibfnamefont {D.}~\bibnamefont {Bouman}}, \bibinfo {author} {\bibfnamefont {J.}~\bibnamefont {Cerrillo}}, \bibinfo {author} {\bibfnamefont {S.}~\bibnamefont {Diamond}}, \bibinfo {author} {\bibfnamefont {K.}~\bibnamefont {Serniak}}, \bibinfo {author} {\bibfnamefont {T.}~\bibnamefont {Connolly}}, \bibinfo {author} {\bibfnamefont {P.}~\bibnamefont {Krogstrup}}, \bibinfo {author} {\bibfnamefont {J.}~\bibnamefont {Nygård}}, \bibinfo {author} {\bibfnamefont {A.~L.}\ \bibnamefont {Yeyati}}, \bibinfo {author} {\bibfnamefont {A.}~\bibnamefont {Geresdi}},\ and\ \bibinfo {author} {\bibfnamefont {M.~H.}\ \bibnamefont {Devoret}},\ }\bibfield  {title} {\bibinfo {title} {Coherent manipulation of an andreev spin qubit},\ }\href {https://doi.org/10.1126/science.abf0345} {\bibfield  {journal} {\bibinfo  {journal} {Science}\ }\textbf {\bibinfo {volume} {373}},\
  \bibinfo {pages} {430} (\bibinfo {year} {2021})}\BibitemShut {NoStop}%
\bibitem [{\citenamefont {Pita-Vidal}\ \emph {et~al.}(2023)\citenamefont {Pita-Vidal}, \citenamefont {Bargerbos}, \citenamefont {{\v{Z}}itko}, \citenamefont {Splitthoff}, \citenamefont {Gr{\"u}nhaupt}, \citenamefont {Wesdorp}, \citenamefont {Liu}, \citenamefont {Kouwenhoven}, \citenamefont {Aguado}, \citenamefont {van Heck}, \citenamefont {Kou},\ and\ \citenamefont {Andersen}}]{PIta2023}%
  \BibitemOpen
  \bibfield  {author} {\bibinfo {author} {\bibfnamefont {M.}~\bibnamefont {Pita-Vidal}}, \bibinfo {author} {\bibfnamefont {A.}~\bibnamefont {Bargerbos}}, \bibinfo {author} {\bibfnamefont {R.}~\bibnamefont {{\v{Z}}itko}}, \bibinfo {author} {\bibfnamefont {L.~J.}\ \bibnamefont {Splitthoff}}, \bibinfo {author} {\bibfnamefont {L.}~\bibnamefont {Gr{\"u}nhaupt}}, \bibinfo {author} {\bibfnamefont {J.~J.}\ \bibnamefont {Wesdorp}}, \bibinfo {author} {\bibfnamefont {Y.}~\bibnamefont {Liu}}, \bibinfo {author} {\bibfnamefont {L.~P.}\ \bibnamefont {Kouwenhoven}}, \bibinfo {author} {\bibfnamefont {R.}~\bibnamefont {Aguado}}, \bibinfo {author} {\bibfnamefont {B.}~\bibnamefont {van Heck}}, \bibinfo {author} {\bibfnamefont {A.}~\bibnamefont {Kou}},\ and\ \bibinfo {author} {\bibfnamefont {C.~K.}\ \bibnamefont {Andersen}},\ }\bibfield  {title} {\bibinfo {title} {Direct manipulation of a superconducting spin qubit strongly coupled to a transmon qubit},\ }\href {https://doi.org/10.1038/s41567-023-02071-x} {\bibfield  {journal}
  {\bibinfo  {journal} {Nature Physics}\ }\textbf {\bibinfo {volume} {19}},\ \bibinfo {pages} {1110} (\bibinfo {year} {2023})}\BibitemShut {NoStop}%
\bibitem [{\citenamefont {Gingrich}\ \emph {et~al.}(2016)\citenamefont {Gingrich}, \citenamefont {Niedzielski}, \citenamefont {Glick}, \citenamefont {Wang}, \citenamefont {Miller}, \citenamefont {Loloee}, \citenamefont {Pratt~Jr},\ and\ \citenamefont {Birge}}]{Gingrich2016}%
  \BibitemOpen
  \bibfield  {author} {\bibinfo {author} {\bibfnamefont {E.~C.}\ \bibnamefont {Gingrich}}, \bibinfo {author} {\bibfnamefont {B.~M.}\ \bibnamefont {Niedzielski}}, \bibinfo {author} {\bibfnamefont {J.~A.}\ \bibnamefont {Glick}}, \bibinfo {author} {\bibfnamefont {Y.}~\bibnamefont {Wang}}, \bibinfo {author} {\bibfnamefont {D.~L.}\ \bibnamefont {Miller}}, \bibinfo {author} {\bibfnamefont {R.}~\bibnamefont {Loloee}}, \bibinfo {author} {\bibfnamefont {W.~P.}\ \bibnamefont {Pratt~Jr}},\ and\ \bibinfo {author} {\bibfnamefont {N.~O.}\ \bibnamefont {Birge}},\ }\bibfield  {title} {\bibinfo {title} {Controllable 0--$\pi$ josephson junctions containing a ferromagnetic spin valve},\ }\href {https://doi.org/10.1038/nphys3681} {\bibfield  {journal} {\bibinfo  {journal} {Nature Physics}\ }\textbf {\bibinfo {volume} {12}},\ \bibinfo {pages} {564} (\bibinfo {year} {2016})}\BibitemShut {NoStop}%
\bibitem [{\citenamefont {Qi}\ and\ \citenamefont {Zhang}(2011)}]{Quitopologicalreview}%
  \BibitemOpen
  \bibfield  {author} {\bibinfo {author} {\bibfnamefont {X.-L.}\ \bibnamefont {Qi}}\ and\ \bibinfo {author} {\bibfnamefont {S.-C.}\ \bibnamefont {Zhang}},\ }\bibfield  {title} {\bibinfo {title} {Topological insulators and superconductors},\ }\href {https://doi.org/10.1103/RevModPhys.83.1057} {\bibfield  {journal} {\bibinfo  {journal} {Rev. Mod. Phys.}\ }\textbf {\bibinfo {volume} {83}},\ \bibinfo {pages} {1057} (\bibinfo {year} {2011})}\BibitemShut {NoStop}%
\bibitem [{\citenamefont {Sauls}(2018)}]{Sauls2018}%
  \BibitemOpen
  \bibfield  {author} {\bibinfo {author} {\bibfnamefont {J.~A.}\ \bibnamefont {Sauls}},\ }\bibfield  {title} {\bibinfo {title} {Andreev bound states and their signatures},\ }\href {https://doi.org/10.1098/rsta.2018.0140} {\bibfield  {journal} {\bibinfo  {journal} {Philosophical Transactions of the Royal Society A: Mathematical, Physical and Engineering Sciences}\ }\textbf {\bibinfo {volume} {376}},\ \bibinfo {pages} {20180140} (\bibinfo {year} {2018})}\BibitemShut {NoStop}%
\bibitem [{\citenamefont {Repin}\ \emph {et~al.}(2019)\citenamefont {Repin}, \citenamefont {Chen},\ and\ \citenamefont {Nazarov}}]{Nazarov2019}%
  \BibitemOpen
  \bibfield  {author} {\bibinfo {author} {\bibfnamefont {E.~V.}\ \bibnamefont {Repin}}, \bibinfo {author} {\bibfnamefont {Y.}~\bibnamefont {Chen}},\ and\ \bibinfo {author} {\bibfnamefont {Y.~V.}\ \bibnamefont {Nazarov}},\ }\bibfield  {title} {\bibinfo {title} {Topological properties of multiterminal superconducting nanostructures: Effect of a continuous spectrum},\ }\href {https://doi.org/10.1103/PhysRevB.99.165414} {\bibfield  {journal} {\bibinfo  {journal} {Phys. Rev. B}\ }\textbf {\bibinfo {volume} {99}},\ \bibinfo {pages} {165414} (\bibinfo {year} {2019})}\BibitemShut {NoStop}%
\bibitem [{\citenamefont {Riwar}\ \emph {et~al.}(2016)\citenamefont {Riwar}, \citenamefont {Houzet}, \citenamefont {Meyer},\ and\ \citenamefont {Nazarov}}]{Riwar2016}%
  \BibitemOpen
  \bibfield  {author} {\bibinfo {author} {\bibfnamefont {R.-P.}\ \bibnamefont {Riwar}}, \bibinfo {author} {\bibfnamefont {M.}~\bibnamefont {Houzet}}, \bibinfo {author} {\bibfnamefont {J.~S.}\ \bibnamefont {Meyer}},\ and\ \bibinfo {author} {\bibfnamefont {Y.~V.}\ \bibnamefont {Nazarov}},\ }\bibfield  {title} {\bibinfo {title} {Multi-terminal josephson junctions as topological matter},\ }\href {https://doi.org/10.1038/ncomms11167} {\bibfield  {journal} {\bibinfo  {journal} {Nature Communications}\ }\textbf {\bibinfo {volume} {7}},\ \bibinfo {pages} {11167} (\bibinfo {year} {2016})}\BibitemShut {NoStop}%
\bibitem [{\citenamefont {Meyer}\ and\ \citenamefont {Houzet}(2017)}]{Meyer2017}%
  \BibitemOpen
  \bibfield  {author} {\bibinfo {author} {\bibfnamefont {J.~S.}\ \bibnamefont {Meyer}}\ and\ \bibinfo {author} {\bibfnamefont {M.}~\bibnamefont {Houzet}},\ }\bibfield  {title} {\bibinfo {title} {Nontrivial chern numbers in three-terminal josephson junctions},\ }\href {https://doi.org/10.1103/PhysRevLett.119.136807} {\bibfield  {journal} {\bibinfo  {journal} {Phys. Rev. Lett.}\ }\textbf {\bibinfo {volume} {119}},\ \bibinfo {pages} {136807} (\bibinfo {year} {2017})}\BibitemShut {NoStop}%
\bibitem [{\citenamefont {Peralta~Gavensky}\ \emph {et~al.}(2019)\citenamefont {Peralta~Gavensky}, \citenamefont {Usaj},\ and\ \citenamefont {Balseiro}}]{Peralta2019}%
  \BibitemOpen
  \bibfield  {author} {\bibinfo {author} {\bibfnamefont {L.}~\bibnamefont {Peralta~Gavensky}}, \bibinfo {author} {\bibfnamefont {G.}~\bibnamefont {Usaj}},\ and\ \bibinfo {author} {\bibfnamefont {C.~A.}\ \bibnamefont {Balseiro}},\ }\bibfield  {title} {\bibinfo {title} {Topological phase diagram of a three-terminal josephson junction: From the conventional to the majorana regime},\ }\href {https://doi.org/10.1103/PhysRevB.100.014514} {\bibfield  {journal} {\bibinfo  {journal} {Phys. Rev. B}\ }\textbf {\bibinfo {volume} {100}},\ \bibinfo {pages} {014514} (\bibinfo {year} {2019})}\BibitemShut {NoStop}%
\bibitem [{\citenamefont {Chandrasekhar}(2022)}]{Chandrasekhar2022}%
  \BibitemOpen
  \bibfield  {author} {\bibinfo {author} {\bibfnamefont {V.}~\bibnamefont {Chandrasekhar}},\ }\bibfield  {title} {\bibinfo {title} {{Probing the topological band structure of diffusive multiterminal Josephson junction devices with conductance measurements}},\ }\href {https://doi.org/10.1063/5.0125708} {\bibfield  {journal} {\bibinfo  {journal} {Applied Physics Letters}\ }\textbf {\bibinfo {volume} {121}},\ \bibinfo {pages} {222601} (\bibinfo {year} {2022})}\BibitemShut {NoStop}%
\bibitem [{\citenamefont {Gavensky}\ \emph {et~al.}(2023)\citenamefont {Gavensky}, \citenamefont {Usaj},\ and\ \citenamefont {Balseiro}}]{Peralta2023}%
  \BibitemOpen
  \bibfield  {author} {\bibinfo {author} {\bibfnamefont {L.~P.}\ \bibnamefont {Gavensky}}, \bibinfo {author} {\bibfnamefont {G.}~\bibnamefont {Usaj}},\ and\ \bibinfo {author} {\bibfnamefont {C.~A.}\ \bibnamefont {Balseiro}},\ }\bibfield  {title} {\bibinfo {title} {Multi-terminal josephson junctions: A road to topological flux networks},\ }\href {https://doi.org/10.1209/0295-5075/acb2f6} {\bibfield  {journal} {\bibinfo  {journal} {Europhysics Letters}\ }\textbf {\bibinfo {volume} {141}},\ \bibinfo {pages} {36001} (\bibinfo {year} {2023})}\BibitemShut {NoStop}%
\bibitem [{\citenamefont {Coraiola}\ \emph {et~al.}(2023{\natexlab{a}})\citenamefont {Coraiola}, \citenamefont {Haxell}, \citenamefont {Sabonis}, \citenamefont {Weisbrich}, \citenamefont {Svetogorov}, \citenamefont {Hinderling}, \citenamefont {ten Kate}, \citenamefont {Cheah}, \citenamefont {Krizek}, \citenamefont {Schott}, \citenamefont {Wegscheider}, \citenamefont {Cuevas}, \citenamefont {Belzig},\ and\ \citenamefont {Nichele}}]{Coraiola2023}%
  \BibitemOpen
  \bibfield  {author} {\bibinfo {author} {\bibfnamefont {M.}~\bibnamefont {Coraiola}}, \bibinfo {author} {\bibfnamefont {D.~Z.}\ \bibnamefont {Haxell}}, \bibinfo {author} {\bibfnamefont {D.}~\bibnamefont {Sabonis}}, \bibinfo {author} {\bibfnamefont {H.}~\bibnamefont {Weisbrich}}, \bibinfo {author} {\bibfnamefont {A.~E.}\ \bibnamefont {Svetogorov}}, \bibinfo {author} {\bibfnamefont {M.}~\bibnamefont {Hinderling}}, \bibinfo {author} {\bibfnamefont {S.~C.}\ \bibnamefont {ten Kate}}, \bibinfo {author} {\bibfnamefont {E.}~\bibnamefont {Cheah}}, \bibinfo {author} {\bibfnamefont {F.}~\bibnamefont {Krizek}}, \bibinfo {author} {\bibfnamefont {R.}~\bibnamefont {Schott}}, \bibinfo {author} {\bibfnamefont {W.}~\bibnamefont {Wegscheider}}, \bibinfo {author} {\bibfnamefont {J.~C.}\ \bibnamefont {Cuevas}}, \bibinfo {author} {\bibfnamefont {W.}~\bibnamefont {Belzig}},\ and\ \bibinfo {author} {\bibfnamefont {F.}~\bibnamefont {Nichele}},\ }\bibfield  {title} {\bibinfo {title} {Phase-engineering the andreev band structure of a
  three-terminal josephson junction},\ }\href {https://doi.org/10.1038/s41467-023-42356-6} {\bibfield  {journal} {\bibinfo  {journal} {Nature Communications}\ }\textbf {\bibinfo {volume} {14}},\ \bibinfo {pages} {6784} (\bibinfo {year} {2023}{\natexlab{a}})}\BibitemShut {NoStop}%
\bibitem [{\citenamefont {Matsuo}\ \emph {et~al.}(2023{\natexlab{a}})\citenamefont {Matsuo}, \citenamefont {Imoto}, \citenamefont {Yokoyama}, \citenamefont {Sato}, \citenamefont {Lindemann}, \citenamefont {Gronin}, \citenamefont {Gardner}, \citenamefont {Nakosai}, \citenamefont {Tanaka}, \citenamefont {Manfra},\ and\ \citenamefont {Tarucha}}]{Matsuophase2023}%
  \BibitemOpen
  \bibfield  {author} {\bibinfo {author} {\bibfnamefont {S.}~\bibnamefont {Matsuo}}, \bibinfo {author} {\bibfnamefont {T.}~\bibnamefont {Imoto}}, \bibinfo {author} {\bibfnamefont {T.}~\bibnamefont {Yokoyama}}, \bibinfo {author} {\bibfnamefont {Y.}~\bibnamefont {Sato}}, \bibinfo {author} {\bibfnamefont {T.}~\bibnamefont {Lindemann}}, \bibinfo {author} {\bibfnamefont {S.}~\bibnamefont {Gronin}}, \bibinfo {author} {\bibfnamefont {G.~C.}\ \bibnamefont {Gardner}}, \bibinfo {author} {\bibfnamefont {S.}~\bibnamefont {Nakosai}}, \bibinfo {author} {\bibfnamefont {Y.}~\bibnamefont {Tanaka}}, \bibinfo {author} {\bibfnamefont {M.~J.}\ \bibnamefont {Manfra}},\ and\ \bibinfo {author} {\bibfnamefont {S.}~\bibnamefont {Tarucha}},\ }\bibfield  {title} {\bibinfo {title} {Phase-dependent andreev molecules and superconducting gap closing in coherently-coupled josephson junctions},\ }\href {https://doi.org/10.1038/s41467-023-44111-3} {\bibfield  {journal} {\bibinfo  {journal} {Nature Communications}\ }\textbf {\bibinfo {volume}
  {14}},\ \bibinfo {pages} {8271} (\bibinfo {year} {2023}{\natexlab{a}})}\BibitemShut {NoStop}%
\bibitem [{\citenamefont {Matsuo}\ \emph {et~al.}(2023{\natexlab{b}})\citenamefont {Matsuo}, \citenamefont {Imoto}, \citenamefont {Yokoyama}, \citenamefont {Sato}, \citenamefont {Lindemann}, \citenamefont {Gronin}, \citenamefont {Gardner}, \citenamefont {Manfra},\ and\ \citenamefont {Tarucha}}]{MatsuoScience2023}%
  \BibitemOpen
  \bibfield  {author} {\bibinfo {author} {\bibfnamefont {S.}~\bibnamefont {Matsuo}}, \bibinfo {author} {\bibfnamefont {T.}~\bibnamefont {Imoto}}, \bibinfo {author} {\bibfnamefont {T.}~\bibnamefont {Yokoyama}}, \bibinfo {author} {\bibfnamefont {Y.}~\bibnamefont {Sato}}, \bibinfo {author} {\bibfnamefont {T.}~\bibnamefont {Lindemann}}, \bibinfo {author} {\bibfnamefont {S.}~\bibnamefont {Gronin}}, \bibinfo {author} {\bibfnamefont {G.~C.}\ \bibnamefont {Gardner}}, \bibinfo {author} {\bibfnamefont {M.~J.}\ \bibnamefont {Manfra}},\ and\ \bibinfo {author} {\bibfnamefont {S.}~\bibnamefont {Tarucha}},\ }\bibfield  {title} {\bibinfo {title} {Phase engineering of anomalous josephson effect derived from andreev molecules},\ }\href {https://doi.org/10.1126/sciadv.adj3698} {\bibfield  {journal} {\bibinfo  {journal} {Science Advances}\ }\textbf {\bibinfo {volume} {9}},\ \bibinfo {pages} {eadj3698} (\bibinfo {year} {2023}{\natexlab{b}})}\BibitemShut {NoStop}%
\bibitem [{\citenamefont {Pillet}\ \emph {et~al.}(2019)\citenamefont {Pillet}, \citenamefont {Benzoni}, \citenamefont {Griesmar}, \citenamefont {Smirr},\ and\ \citenamefont {Girit}}]{Pillet2019}%
  \BibitemOpen
  \bibfield  {author} {\bibinfo {author} {\bibfnamefont {J.}~\bibnamefont {Pillet}}, \bibinfo {author} {\bibfnamefont {V.}~\bibnamefont {Benzoni}}, \bibinfo {author} {\bibfnamefont {J.}~\bibnamefont {Griesmar}}, \bibinfo {author} {\bibfnamefont {J.-L.}\ \bibnamefont {Smirr}},\ and\ \bibinfo {author} {\bibfnamefont {C.~O.}\ \bibnamefont {Girit}},\ }\bibfield  {title} {\bibinfo {title} {Nonlocal josephson effect in andreev molecules},\ }\href {https://doi.org/10.1021/acs.nanolett.9b02686} {\bibfield  {journal} {\bibinfo  {journal} {Nano Letters}\ }\textbf {\bibinfo {volume} {19}},\ \bibinfo {pages} {7138} (\bibinfo {year} {2019})}\BibitemShut {NoStop}%
\bibitem [{\citenamefont {M\'elin}\ \emph {et~al.}(2023)\citenamefont {M\'elin}, \citenamefont {Danneau},\ and\ \citenamefont {Winkelmann}}]{Melin2023}%
  \BibitemOpen
  \bibfield  {author} {\bibinfo {author} {\bibfnamefont {R.}~\bibnamefont {M\'elin}}, \bibinfo {author} {\bibfnamefont {R.}~\bibnamefont {Danneau}},\ and\ \bibinfo {author} {\bibfnamefont {C.~B.}\ \bibnamefont {Winkelmann}},\ }\bibfield  {title} {\bibinfo {title} {Proposal for detecting the $\ensuremath{\pi}$-shifted cooper quartet supercurrent},\ }\href {https://doi.org/10.1103/PhysRevResearch.5.033124} {\bibfield  {journal} {\bibinfo  {journal} {Phys. Rev. Res.}\ }\textbf {\bibinfo {volume} {5}},\ \bibinfo {pages} {033124} (\bibinfo {year} {2023})}\BibitemShut {NoStop}%
\bibitem [{\citenamefont {M\'elin}\ and\ \citenamefont {Feinberg}(2023)}]{Melininterferometer2023}%
  \BibitemOpen
  \bibfield  {author} {\bibinfo {author} {\bibfnamefont {R.}~\bibnamefont {M\'elin}}\ and\ \bibinfo {author} {\bibfnamefont {D.}~\bibnamefont {Feinberg}},\ }\bibfield  {title} {\bibinfo {title} {Quantum interferometer for quartets in superconducting three-terminal josephson junctions},\ }\href {https://doi.org/10.1103/PhysRevB.107.L161405} {\bibfield  {journal} {\bibinfo  {journal} {Phys. Rev. B}\ }\textbf {\bibinfo {volume} {107}},\ \bibinfo {pages} {L161405} (\bibinfo {year} {2023})}\BibitemShut {NoStop}%
\bibitem [{\citenamefont {Gupta}\ \emph {et~al.}(2024)\citenamefont {Gupta}, \citenamefont {Khade}, \citenamefont {Riggert}, \citenamefont {Shani}, \citenamefont {Menning}, \citenamefont {Lueb}, \citenamefont {Jung}, \citenamefont {Mélin}, \citenamefont {Bakkers},\ and\ \citenamefont {Pribiag}}]{gupta2024evidence}%
  \BibitemOpen
  \bibfield  {author} {\bibinfo {author} {\bibfnamefont {M.}~\bibnamefont {Gupta}}, \bibinfo {author} {\bibfnamefont {V.}~\bibnamefont {Khade}}, \bibinfo {author} {\bibfnamefont {C.}~\bibnamefont {Riggert}}, \bibinfo {author} {\bibfnamefont {L.}~\bibnamefont {Shani}}, \bibinfo {author} {\bibfnamefont {G.}~\bibnamefont {Menning}}, \bibinfo {author} {\bibfnamefont {P.}~\bibnamefont {Lueb}}, \bibinfo {author} {\bibfnamefont {J.}~\bibnamefont {Jung}}, \bibinfo {author} {\bibfnamefont {R.}~\bibnamefont {Mélin}}, \bibinfo {author} {\bibfnamefont {E.~P. A.~M.}\ \bibnamefont {Bakkers}},\ and\ \bibinfo {author} {\bibfnamefont {V.~S.}\ \bibnamefont {Pribiag}},\ }\href@noop {} {\bibinfo {title} {Evidence for $\pi$-shifted cooper quartets and few-mode transport in pbte nanowire three-terminal josephson junctions}} (\bibinfo {year} {2024}),\ \Eprint {https://arxiv.org/abs/2312.17703} {arXiv:2312.17703 [cond-mat.mes-hall]} \BibitemShut {NoStop}%
\bibitem [{\citenamefont {M\'elin}\ \emph {et~al.}(2019)\citenamefont {M\'elin}, \citenamefont {Danneau}, \citenamefont {Yang}, \citenamefont {Caputo},\ and\ \citenamefont {Dou\ifmmode~\mbox{\c{c}}\else \c{c}\fi{}ot}}]{Melinvoltagebias2019}%
  \BibitemOpen
  \bibfield  {author} {\bibinfo {author} {\bibfnamefont {R.}~\bibnamefont {M\'elin}}, \bibinfo {author} {\bibfnamefont {R.}~\bibnamefont {Danneau}}, \bibinfo {author} {\bibfnamefont {K.}~\bibnamefont {Yang}}, \bibinfo {author} {\bibfnamefont {J.-G.}\ \bibnamefont {Caputo}},\ and\ \bibinfo {author} {\bibfnamefont {B.}~\bibnamefont {Dou\ifmmode~\mbox{\c{c}}\else \c{c}\fi{}ot}},\ }\bibfield  {title} {\bibinfo {title} {Engineering the floquet spectrum of superconducting multiterminal quantum dots},\ }\href {https://doi.org/10.1103/PhysRevB.100.035450} {\bibfield  {journal} {\bibinfo  {journal} {Phys. Rev. B}\ }\textbf {\bibinfo {volume} {100}},\ \bibinfo {pages} {035450} (\bibinfo {year} {2019})}\BibitemShut {NoStop}%
\bibitem [{\citenamefont {Baselmans}\ \emph {et~al.}(1999)\citenamefont {Baselmans}, \citenamefont {Morpurgo}, \citenamefont {van Wees},\ and\ \citenamefont {Klapwijk}}]{Baselmans1999}%
  \BibitemOpen
  \bibfield  {author} {\bibinfo {author} {\bibfnamefont {J.~J.~A.}\ \bibnamefont {Baselmans}}, \bibinfo {author} {\bibfnamefont {A.~F.}\ \bibnamefont {Morpurgo}}, \bibinfo {author} {\bibfnamefont {B.~J.}\ \bibnamefont {van Wees}},\ and\ \bibinfo {author} {\bibfnamefont {T.~M.}\ \bibnamefont {Klapwijk}},\ }\bibfield  {title} {\bibinfo {title} {Reversing the direction of the supercurrent in a controllable josephson junction},\ }\href {https://doi.org/10.1038/16204} {\bibfield  {journal} {\bibinfo  {journal} {Nature}\ }\textbf {\bibinfo {volume} {397}},\ \bibinfo {pages} {43} (\bibinfo {year} {1999})}\BibitemShut {NoStop}%
\bibitem [{\citenamefont {van Woerkom}\ \emph {et~al.}(2017)\citenamefont {van Woerkom}, \citenamefont {Proutski}, \citenamefont {van Heck}, \citenamefont {Bouman}, \citenamefont {V{\"a}yrynen}, \citenamefont {Glazman}, \citenamefont {Krogstrup}, \citenamefont {Nyg{\aa}rd}, \citenamefont {Kouwenhoven},\ and\ \citenamefont {Geresdi}}]{vanWoerkom2017}%
  \BibitemOpen
  \bibfield  {author} {\bibinfo {author} {\bibfnamefont {D.~J.}\ \bibnamefont {van Woerkom}}, \bibinfo {author} {\bibfnamefont {A.}~\bibnamefont {Proutski}}, \bibinfo {author} {\bibfnamefont {B.}~\bibnamefont {van Heck}}, \bibinfo {author} {\bibfnamefont {D.}~\bibnamefont {Bouman}}, \bibinfo {author} {\bibfnamefont {J.~I.}\ \bibnamefont {V{\"a}yrynen}}, \bibinfo {author} {\bibfnamefont {L.~I.}\ \bibnamefont {Glazman}}, \bibinfo {author} {\bibfnamefont {P.}~\bibnamefont {Krogstrup}}, \bibinfo {author} {\bibfnamefont {J.}~\bibnamefont {Nyg{\aa}rd}}, \bibinfo {author} {\bibfnamefont {L.~P.}\ \bibnamefont {Kouwenhoven}},\ and\ \bibinfo {author} {\bibfnamefont {A.}~\bibnamefont {Geresdi}},\ }\bibfield  {title} {\bibinfo {title} {Microwave spectroscopy of spinful andreev bound states in ballistic semiconductor josephson junctions},\ }\href {https://doi.org/10.1038/nphys4150} {\bibfield  {journal} {\bibinfo  {journal} {Nature Physics}\ }\textbf {\bibinfo {volume} {13}},\ \bibinfo {pages} {876} (\bibinfo {year}
  {2017})}\BibitemShut {NoStop}%
\bibitem [{\citenamefont {Rudner}\ and\ \citenamefont {Lindner}(2020)}]{Rudner2020}%
  \BibitemOpen
  \bibfield  {author} {\bibinfo {author} {\bibfnamefont {M.~S.}\ \bibnamefont {Rudner}}\ and\ \bibinfo {author} {\bibfnamefont {N.~H.}\ \bibnamefont {Lindner}},\ }\bibfield  {title} {\bibinfo {title} {Band structure engineering and non-equilibrium dynamics in floquet topological insulators},\ }\href {https://doi.org/10.1038/s42254-020-0170-z} {\bibfield  {journal} {\bibinfo  {journal} {Nature Reviews Physics}\ }\textbf {\bibinfo {volume} {2}},\ \bibinfo {pages} {229} (\bibinfo {year} {2020})}\BibitemShut {NoStop}%
\bibitem [{\citenamefont {M\'elin}\ \emph {et~al.}(2017)\citenamefont {M\'elin}, \citenamefont {Caputo}, \citenamefont {Yang},\ and\ \citenamefont {Dou\ifmmode~\mbox{\c{c}}\else \c{c}\fi{}ot}}]{Melinfloquet2017}%
  \BibitemOpen
  \bibfield  {author} {\bibinfo {author} {\bibfnamefont {R.}~\bibnamefont {M\'elin}}, \bibinfo {author} {\bibfnamefont {J.-G.}\ \bibnamefont {Caputo}}, \bibinfo {author} {\bibfnamefont {K.}~\bibnamefont {Yang}},\ and\ \bibinfo {author} {\bibfnamefont {B.}~\bibnamefont {Dou\ifmmode~\mbox{\c{c}}\else \c{c}\fi{}ot}},\ }\bibfield  {title} {\bibinfo {title} {Simple floquet-wannier-stark-andreev viewpoint and emergence of low-energy scales in a voltage-biased three-terminal josephson junction},\ }\href {https://doi.org/10.1103/PhysRevB.95.085415} {\bibfield  {journal} {\bibinfo  {journal} {Phys. Rev. B}\ }\textbf {\bibinfo {volume} {95}},\ \bibinfo {pages} {085415} (\bibinfo {year} {2017})}\BibitemShut {NoStop}%
\bibitem [{\citenamefont {Park}\ \emph {et~al.}(2022)\citenamefont {Park}, \citenamefont {Lee}, \citenamefont {Jang}, \citenamefont {Choi}, \citenamefont {Park}, \citenamefont {Jung}, \citenamefont {Watanabe}, \citenamefont {Taniguchi}, \citenamefont {Cho},\ and\ \citenamefont {Lee}}]{Park2022}%
  \BibitemOpen
  \bibfield  {author} {\bibinfo {author} {\bibfnamefont {S.}~\bibnamefont {Park}}, \bibinfo {author} {\bibfnamefont {W.}~\bibnamefont {Lee}}, \bibinfo {author} {\bibfnamefont {S.}~\bibnamefont {Jang}}, \bibinfo {author} {\bibfnamefont {Y.-B.}\ \bibnamefont {Choi}}, \bibinfo {author} {\bibfnamefont {J.}~\bibnamefont {Park}}, \bibinfo {author} {\bibfnamefont {W.}~\bibnamefont {Jung}}, \bibinfo {author} {\bibfnamefont {K.}~\bibnamefont {Watanabe}}, \bibinfo {author} {\bibfnamefont {T.}~\bibnamefont {Taniguchi}}, \bibinfo {author} {\bibfnamefont {G.~Y.}\ \bibnamefont {Cho}},\ and\ \bibinfo {author} {\bibfnamefont {G.-H.}\ \bibnamefont {Lee}},\ }\bibfield  {title} {\bibinfo {title} {Steady floquet--andreev states in graphene josephson junctions},\ }\href {https://doi.org/10.1038/s41586-021-04364-8} {\bibfield  {journal} {\bibinfo  {journal} {Nature}\ }\textbf {\bibinfo {volume} {603}},\ \bibinfo {pages} {421} (\bibinfo {year} {2022})}\BibitemShut {NoStop}%
\bibitem [{\citenamefont {Haxell}\ \emph {et~al.}(2023)\citenamefont {Haxell}, \citenamefont {Coraiola}, \citenamefont {Sabonis}, \citenamefont {Hinderling}, \citenamefont {ten Kate}, \citenamefont {Cheah}, \citenamefont {Krizek}, \citenamefont {Schott}, \citenamefont {Wegscheider}, \citenamefont {Belzig}, \citenamefont {Cuevas},\ and\ \citenamefont {Nichele}}]{Haxell2023}%
  \BibitemOpen
  \bibfield  {author} {\bibinfo {author} {\bibfnamefont {D.~Z.}\ \bibnamefont {Haxell}}, \bibinfo {author} {\bibfnamefont {M.}~\bibnamefont {Coraiola}}, \bibinfo {author} {\bibfnamefont {D.}~\bibnamefont {Sabonis}}, \bibinfo {author} {\bibfnamefont {M.}~\bibnamefont {Hinderling}}, \bibinfo {author} {\bibfnamefont {S.~C.}\ \bibnamefont {ten Kate}}, \bibinfo {author} {\bibfnamefont {E.}~\bibnamefont {Cheah}}, \bibinfo {author} {\bibfnamefont {F.}~\bibnamefont {Krizek}}, \bibinfo {author} {\bibfnamefont {R.}~\bibnamefont {Schott}}, \bibinfo {author} {\bibfnamefont {W.}~\bibnamefont {Wegscheider}}, \bibinfo {author} {\bibfnamefont {W.}~\bibnamefont {Belzig}}, \bibinfo {author} {\bibfnamefont {J.~C.}\ \bibnamefont {Cuevas}},\ and\ \bibinfo {author} {\bibfnamefont {F.}~\bibnamefont {Nichele}},\ }\bibfield  {title} {\bibinfo {title} {Microwave-induced conductance replicas in hybrid josephson junctions without floquet---andreev states},\ }\href {https://doi.org/10.1038/s41467-023-42357-5} {\bibfield  {journal}
  {\bibinfo  {journal} {Nature Communications}\ }\textbf {\bibinfo {volume} {14}},\ \bibinfo {pages} {6798} (\bibinfo {year} {2023})}\BibitemShut {NoStop}%
\bibitem [{\citenamefont {Golikova}\ \emph {et~al.}(2014)\citenamefont {Golikova}, \citenamefont {Wolf}, \citenamefont {Beckmann}, \citenamefont {Batov}, \citenamefont {Bobkova}, \citenamefont {Bobkov},\ and\ \citenamefont {Ryazanov}}]{Golikovafluid2014}%
  \BibitemOpen
  \bibfield  {author} {\bibinfo {author} {\bibfnamefont {T.~E.}\ \bibnamefont {Golikova}}, \bibinfo {author} {\bibfnamefont {M.~J.}\ \bibnamefont {Wolf}}, \bibinfo {author} {\bibfnamefont {D.}~\bibnamefont {Beckmann}}, \bibinfo {author} {\bibfnamefont {I.~E.}\ \bibnamefont {Batov}}, \bibinfo {author} {\bibfnamefont {I.~V.}\ \bibnamefont {Bobkova}}, \bibinfo {author} {\bibfnamefont {A.~M.}\ \bibnamefont {Bobkov}},\ and\ \bibinfo {author} {\bibfnamefont {V.~V.}\ \bibnamefont {Ryazanov}},\ }\bibfield  {title} {\bibinfo {title} {Nonlocal supercurrent in mesoscopic multiterminal sns josephson junction in the low-temperature limit},\ }\href {https://doi.org/10.1103/PhysRevB.89.104507} {\bibfield  {journal} {\bibinfo  {journal} {Phys. Rev. B}\ }\textbf {\bibinfo {volume} {89}},\ \bibinfo {pages} {104507} (\bibinfo {year} {2014})}\BibitemShut {NoStop}%
\bibitem [{\citenamefont {Oka}\ and\ \citenamefont {Kitamura}(2019)}]{Oka2019}%
  \BibitemOpen
  \bibfield  {author} {\bibinfo {author} {\bibfnamefont {T.}~\bibnamefont {Oka}}\ and\ \bibinfo {author} {\bibfnamefont {S.}~\bibnamefont {Kitamura}},\ }\bibfield  {title} {\bibinfo {title} {Floquet engineering of quantum materials},\ }\href {https://doi.org/https://doi.org/10.1146/annurev-conmatphys-031218-013423} {\bibfield  {journal} {\bibinfo  {journal} {Annual Review of Condensed Matter Physics}\ }\textbf {\bibinfo {volume} {10}},\ \bibinfo {pages} {387} (\bibinfo {year} {2019})}\BibitemShut {NoStop}%
\bibitem [{\citenamefont {Liu}\ \emph {et~al.}(2019)\citenamefont {Liu}, \citenamefont {Shabani},\ and\ \citenamefont {Mitra}}]{Liu2019}%
  \BibitemOpen
  \bibfield  {author} {\bibinfo {author} {\bibfnamefont {D.~T.}\ \bibnamefont {Liu}}, \bibinfo {author} {\bibfnamefont {J.}~\bibnamefont {Shabani}},\ and\ \bibinfo {author} {\bibfnamefont {A.}~\bibnamefont {Mitra}},\ }\bibfield  {title} {\bibinfo {title} {Floquet majorana zero and $\ensuremath{\pi}$ modes in planar josephson junctions},\ }\href {https://doi.org/10.1103/PhysRevB.99.094303} {\bibfield  {journal} {\bibinfo  {journal} {Phys. Rev. B}\ }\textbf {\bibinfo {volume} {99}},\ \bibinfo {pages} {094303} (\bibinfo {year} {2019})}\BibitemShut {NoStop}%
\bibitem [{\citenamefont {Carrad}\ \emph {et~al.}(2022)\citenamefont {Carrad}, \citenamefont {Stampfer}, \citenamefont {Olsteins}, \citenamefont {Petersen}, \citenamefont {Khan}, \citenamefont {Krogstrup},\ and\ \citenamefont {Jespersen}}]{Carrad2022}%
  \BibitemOpen
  \bibfield  {author} {\bibinfo {author} {\bibfnamefont {D.~J.}\ \bibnamefont {Carrad}}, \bibinfo {author} {\bibfnamefont {L.}~\bibnamefont {Stampfer}}, \bibinfo {author} {\bibfnamefont {D.}~\bibnamefont {Olsteins}}, \bibinfo {author} {\bibfnamefont {C.~E.~N.}\ \bibnamefont {Petersen}}, \bibinfo {author} {\bibfnamefont {S.~A.}\ \bibnamefont {Khan}}, \bibinfo {author} {\bibfnamefont {P.}~\bibnamefont {Krogstrup}},\ and\ \bibinfo {author} {\bibfnamefont {T.~S.}\ \bibnamefont {Jespersen}},\ }\bibfield  {title} {\bibinfo {title} {Photon-assisted tunneling of high-order multiple andreev reflections in epitaxial nanowire josephson junctions},\ }\href {https://doi.org/10.1021/acs.nanolett.2c01840} {\bibfield  {journal} {\bibinfo  {journal} {Nano Letters}\ }\textbf {\bibinfo {volume} {22}},\ \bibinfo {pages} {6262} (\bibinfo {year} {2022})},\ \bibinfo {note} {pMID: 35862144}\BibitemShut {NoStop}%
\bibitem [{\citenamefont {Dolgirev}\ \emph {et~al.}(2019{\natexlab{a}})\citenamefont {Dolgirev}, \citenamefont {Kalenkov}, \citenamefont {Tarkhov},\ and\ \citenamefont {Zaikin}}]{Zaikindiffusive2019}%
  \BibitemOpen
  \bibfield  {author} {\bibinfo {author} {\bibfnamefont {P.~E.}\ \bibnamefont {Dolgirev}}, \bibinfo {author} {\bibfnamefont {M.~S.}\ \bibnamefont {Kalenkov}}, \bibinfo {author} {\bibfnamefont {A.~E.}\ \bibnamefont {Tarkhov}},\ and\ \bibinfo {author} {\bibfnamefont {A.~D.}\ \bibnamefont {Zaikin}},\ }\bibfield  {title} {\bibinfo {title} {Phase-coherent electron transport in asymmetric crosslike andreev interferometers},\ }\href {https://doi.org/10.1103/PhysRevB.100.054511} {\bibfield  {journal} {\bibinfo  {journal} {Phys. Rev. B}\ }\textbf {\bibinfo {volume} {100}},\ \bibinfo {pages} {054511} (\bibinfo {year} {2019}{\natexlab{a}})}\BibitemShut {NoStop}%
\bibitem [{\citenamefont {van Wees}\ \emph {et~al.}(1991)\citenamefont {van Wees}, \citenamefont {Lenssen},\ and\ \citenamefont {Harmans}}]{vanWees1991}%
  \BibitemOpen
  \bibfield  {author} {\bibinfo {author} {\bibfnamefont {B.~J.}\ \bibnamefont {van Wees}}, \bibinfo {author} {\bibfnamefont {K.-M.~H.}\ \bibnamefont {Lenssen}},\ and\ \bibinfo {author} {\bibfnamefont {C.~J. P.~M.}\ \bibnamefont {Harmans}},\ }\bibfield  {title} {\bibinfo {title} {Transmission formalism for supercurrent flow in multiprobe superconductor-semiconductor-superconductor devices},\ }\href {https://doi.org/10.1103/PhysRevB.44.470} {\bibfield  {journal} {\bibinfo  {journal} {Phys. Rev. B}\ }\textbf {\bibinfo {volume} {44}},\ \bibinfo {pages} {470} (\bibinfo {year} {1991})}\BibitemShut {NoStop}%
\bibitem [{\citenamefont {M\'elin}\ and\ \citenamefont {Dou\ifmmode~\mbox{\c{c}}\else \c{c}\fi{}ot}(2020)}]{Melinquantumdot2020}%
  \BibitemOpen
  \bibfield  {author} {\bibinfo {author} {\bibfnamefont {R.}~\bibnamefont {M\'elin}}\ and\ \bibinfo {author} {\bibfnamefont {B.}~\bibnamefont {Dou\ifmmode~\mbox{\c{c}}\else \c{c}\fi{}ot}},\ }\bibfield  {title} {\bibinfo {title} {Inversion in a four-terminal superconducting device on the quartet line. ii. quantum dot and floquet theory},\ }\href {https://doi.org/10.1103/PhysRevB.102.245436} {\bibfield  {journal} {\bibinfo  {journal} {Phys. Rev. B}\ }\textbf {\bibinfo {volume} {102}},\ \bibinfo {pages} {245436} (\bibinfo {year} {2020})}\BibitemShut {NoStop}%
\bibitem [{\citenamefont {M\'elin}(2020)}]{Melinquartet2020}%
  \BibitemOpen
  \bibfield  {author} {\bibinfo {author} {\bibfnamefont {R.}~\bibnamefont {M\'elin}},\ }\bibfield  {title} {\bibinfo {title} {Inversion in a four-terminal superconducting device on the quartet line. i. two-dimensional metal and the quartet beam splitter},\ }\href {https://doi.org/10.1103/PhysRevB.102.245435} {\bibfield  {journal} {\bibinfo  {journal} {Phys. Rev. B}\ }\textbf {\bibinfo {volume} {102}},\ \bibinfo {pages} {245435} (\bibinfo {year} {2020})}\BibitemShut {NoStop}%
\bibitem [{\citenamefont {Pankratova}\ \emph {et~al.}(2020)\citenamefont {Pankratova}, \citenamefont {Lee}, \citenamefont {Kuzmin}, \citenamefont {Wickramasinghe}, \citenamefont {Mayer}, \citenamefont {Yuan}, \citenamefont {Vavilov}, \citenamefont {Shabani},\ and\ \citenamefont {Manucharyan}}]{Pankratova2020}%
  \BibitemOpen
  \bibfield  {author} {\bibinfo {author} {\bibfnamefont {N.}~\bibnamefont {Pankratova}}, \bibinfo {author} {\bibfnamefont {H.}~\bibnamefont {Lee}}, \bibinfo {author} {\bibfnamefont {R.}~\bibnamefont {Kuzmin}}, \bibinfo {author} {\bibfnamefont {K.}~\bibnamefont {Wickramasinghe}}, \bibinfo {author} {\bibfnamefont {W.}~\bibnamefont {Mayer}}, \bibinfo {author} {\bibfnamefont {J.}~\bibnamefont {Yuan}}, \bibinfo {author} {\bibfnamefont {M.~G.}\ \bibnamefont {Vavilov}}, \bibinfo {author} {\bibfnamefont {J.}~\bibnamefont {Shabani}},\ and\ \bibinfo {author} {\bibfnamefont {V.~E.}\ \bibnamefont {Manucharyan}},\ }\bibfield  {title} {\bibinfo {title} {Multiterminal josephson effect},\ }\href {https://doi.org/10.1103/PhysRevX.10.031051} {\bibfield  {journal} {\bibinfo  {journal} {Phys. Rev. X}\ }\textbf {\bibinfo {volume} {10}},\ \bibinfo {pages} {031051} (\bibinfo {year} {2020})}\BibitemShut {NoStop}%
\bibitem [{\citenamefont {Graziano}\ \emph {et~al.}(2020)\citenamefont {Graziano}, \citenamefont {Lee}, \citenamefont {Pendharkar}, \citenamefont {Palmstr},\ and\ \citenamefont {Pribiag}}]{Graziano2020}%
  \BibitemOpen
  \bibfield  {author} {\bibinfo {author} {\bibfnamefont {G.~V.}\ \bibnamefont {Graziano}}, \bibinfo {author} {\bibfnamefont {J.~S.}\ \bibnamefont {Lee}}, \bibinfo {author} {\bibfnamefont {M.}~\bibnamefont {Pendharkar}}, \bibinfo {author} {\bibfnamefont {C.~J.}\ \bibnamefont {Palmstr}},\ and\ \bibinfo {author} {\bibfnamefont {V.~S.}\ \bibnamefont {Pribiag}},\ }\bibfield  {title} {\bibinfo {title} {Transport studies in a gate-tunable three-terminal josephson junction},\ }\href {https://doi.org/10.1103/PhysRevB.101.054510} {\bibfield  {journal} {\bibinfo  {journal} {Phys. Rev. B}\ }\textbf {\bibinfo {volume} {101}},\ \bibinfo {pages} {054510} (\bibinfo {year} {2020})}\BibitemShut {NoStop}%
\bibitem [{\citenamefont {Draelos}\ \emph {et~al.}(2019)\citenamefont {Draelos}, \citenamefont {Wei}, \citenamefont {Seredinski}, \citenamefont {Li}, \citenamefont {Mehta}, \citenamefont {Watanabe}, \citenamefont {Taniguchi}, \citenamefont {Borzenets}, \citenamefont {Amet},\ and\ \citenamefont {Finkelstein}}]{Draelos2019}%
  \BibitemOpen
  \bibfield  {author} {\bibinfo {author} {\bibfnamefont {A.~W.}\ \bibnamefont {Draelos}}, \bibinfo {author} {\bibfnamefont {M.-T.}\ \bibnamefont {Wei}}, \bibinfo {author} {\bibfnamefont {A.}~\bibnamefont {Seredinski}}, \bibinfo {author} {\bibfnamefont {H.}~\bibnamefont {Li}}, \bibinfo {author} {\bibfnamefont {Y.}~\bibnamefont {Mehta}}, \bibinfo {author} {\bibfnamefont {K.}~\bibnamefont {Watanabe}}, \bibinfo {author} {\bibfnamefont {T.}~\bibnamefont {Taniguchi}}, \bibinfo {author} {\bibfnamefont {I.~V.}\ \bibnamefont {Borzenets}}, \bibinfo {author} {\bibfnamefont {F.}~\bibnamefont {Amet}},\ and\ \bibinfo {author} {\bibfnamefont {G.}~\bibnamefont {Finkelstein}},\ }\bibfield  {title} {\bibinfo {title} {Supercurrent flow in multiterminal graphene josephson junctions},\ }\href {https://doi.org/10.1021/acs.nanolett.8b04330} {\bibfield  {journal} {\bibinfo  {journal} {Nano Letters}\ }\textbf {\bibinfo {volume} {19}},\ \bibinfo {pages} {1039} (\bibinfo {year} {2019})}\BibitemShut {NoStop}%
\bibitem [{\citenamefont {Arnault}\ \emph {et~al.}(2021)\citenamefont {Arnault}, \citenamefont {Larson}, \citenamefont {Seredinski}, \citenamefont {Zhao}, \citenamefont {Idris}, \citenamefont {McConnell}, \citenamefont {Watanabe}, \citenamefont {Taniguchi}, \citenamefont {Borzenets}, \citenamefont {Amet},\ and\ \citenamefont {Finkelstein}}]{Arnault2021}%
  \BibitemOpen
  \bibfield  {author} {\bibinfo {author} {\bibfnamefont {E.~G.}\ \bibnamefont {Arnault}}, \bibinfo {author} {\bibfnamefont {T.~F.~Q.}\ \bibnamefont {Larson}}, \bibinfo {author} {\bibfnamefont {A.}~\bibnamefont {Seredinski}}, \bibinfo {author} {\bibfnamefont {L.}~\bibnamefont {Zhao}}, \bibinfo {author} {\bibfnamefont {S.}~\bibnamefont {Idris}}, \bibinfo {author} {\bibfnamefont {A.}~\bibnamefont {McConnell}}, \bibinfo {author} {\bibfnamefont {K.}~\bibnamefont {Watanabe}}, \bibinfo {author} {\bibfnamefont {T.}~\bibnamefont {Taniguchi}}, \bibinfo {author} {\bibfnamefont {I.}~\bibnamefont {Borzenets}}, \bibinfo {author} {\bibfnamefont {F.}~\bibnamefont {Amet}},\ and\ \bibinfo {author} {\bibfnamefont {G.}~\bibnamefont {Finkelstein}},\ }\bibfield  {title} {\bibinfo {title} {Multiterminal inverse ac josephson effect},\ }\href {https://doi.org/10.1021/acs.nanolett.1c03474} {\bibfield  {journal} {\bibinfo  {journal} {Nano Letters}\ }\textbf {\bibinfo {volume} {21}},\ \bibinfo {pages} {9668} (\bibinfo {year} {2021})},\
  \bibinfo {note} {pMID: 34779633}\BibitemShut {NoStop}%
\bibitem [{\citenamefont {Matute-Cañadas}\ \emph {et~al.}(2023)\citenamefont {Matute-Cañadas}, \citenamefont {Tosi},\ and\ \citenamefont {Yeyati}}]{MatutequantumCircuit2023}%
  \BibitemOpen
  \bibfield  {author} {\bibinfo {author} {\bibfnamefont {F.~J.}\ \bibnamefont {Matute-Cañadas}}, \bibinfo {author} {\bibfnamefont {L.}~\bibnamefont {Tosi}},\ and\ \bibinfo {author} {\bibfnamefont {A.~L.}\ \bibnamefont {Yeyati}},\ }\href@noop {} {\bibinfo {title} {Quantum circuits with multiterminal josephson-andreev junctions}} (\bibinfo {year} {2023}),\ \Eprint {https://arxiv.org/abs/2312.17305} {arXiv:2312.17305 [cond-mat.mes-hall]} \BibitemShut {NoStop}%
\bibitem [{\citenamefont {Strambini}\ \emph {et~al.}(2016)\citenamefont {Strambini}, \citenamefont {D'Ambrosio}, \citenamefont {Vischi}, \citenamefont {Bergeret}, \citenamefont {Nazarov},\ and\ \citenamefont {Giazotto}}]{Strambini2016}%
  \BibitemOpen
  \bibfield  {author} {\bibinfo {author} {\bibfnamefont {E.}~\bibnamefont {Strambini}}, \bibinfo {author} {\bibfnamefont {S.}~\bibnamefont {D'Ambrosio}}, \bibinfo {author} {\bibfnamefont {F.}~\bibnamefont {Vischi}}, \bibinfo {author} {\bibfnamefont {F.}~\bibnamefont {Bergeret}}, \bibinfo {author} {\bibfnamefont {Y.~V.}\ \bibnamefont {Nazarov}},\ and\ \bibinfo {author} {\bibfnamefont {F.}~\bibnamefont {Giazotto}},\ }\bibfield  {title} {\bibinfo {title} {The $\omega$-squipt as a tool to phase-engineer josephson topological materials},\ }\href {https://www.nature.com/articles/nnano.2016.157} {\bibfield  {journal} {\bibinfo  {journal} {Nature Nanotechnology}\ }\textbf {\bibinfo {volume} {11}},\ \bibinfo {pages} {1055} (\bibinfo {year} {2016})}\BibitemShut {NoStop}%
\bibitem [{\citenamefont {Pfeffer}\ \emph {et~al.}(2014{\natexlab{a}})\citenamefont {Pfeffer}, \citenamefont {Duvauchelle}, \citenamefont {Courtois}, \citenamefont {M\'elin}, \citenamefont {Feinberg},\ and\ \citenamefont {Lefloch}}]{Pfeffer2014}%
  \BibitemOpen
  \bibfield  {author} {\bibinfo {author} {\bibfnamefont {A.~H.}\ \bibnamefont {Pfeffer}}, \bibinfo {author} {\bibfnamefont {J.~E.}\ \bibnamefont {Duvauchelle}}, \bibinfo {author} {\bibfnamefont {H.}~\bibnamefont {Courtois}}, \bibinfo {author} {\bibfnamefont {R.}~\bibnamefont {M\'elin}}, \bibinfo {author} {\bibfnamefont {D.}~\bibnamefont {Feinberg}},\ and\ \bibinfo {author} {\bibfnamefont {F.}~\bibnamefont {Lefloch}},\ }\bibfield  {title} {\bibinfo {title} {Subgap structure in the conductance of a three-terminal josephson junction},\ }\href {https://doi.org/10.1103/PhysRevB.90.075401} {\bibfield  {journal} {\bibinfo  {journal} {Phys. Rev. B}\ }\textbf {\bibinfo {volume} {90}},\ \bibinfo {pages} {075401} (\bibinfo {year} {2014}{\natexlab{a}})}\BibitemShut {NoStop}%
\bibitem [{\citenamefont {Arnault}\ \emph {et~al.}(2022)\citenamefont {Arnault}, \citenamefont {Idris}, \citenamefont {McConnell}, \citenamefont {Zhao}, \citenamefont {Larson}, \citenamefont {Watanabe}, \citenamefont {Taniguchi}, \citenamefont {Finkelstein},\ and\ \citenamefont {Amet}}]{Arnaultdynamic2022}%
  \BibitemOpen
  \bibfield  {author} {\bibinfo {author} {\bibfnamefont {E.~G.}\ \bibnamefont {Arnault}}, \bibinfo {author} {\bibfnamefont {S.}~\bibnamefont {Idris}}, \bibinfo {author} {\bibfnamefont {A.}~\bibnamefont {McConnell}}, \bibinfo {author} {\bibfnamefont {L.}~\bibnamefont {Zhao}}, \bibinfo {author} {\bibfnamefont {T.~F.}\ \bibnamefont {Larson}}, \bibinfo {author} {\bibfnamefont {K.}~\bibnamefont {Watanabe}}, \bibinfo {author} {\bibfnamefont {T.}~\bibnamefont {Taniguchi}}, \bibinfo {author} {\bibfnamefont {G.}~\bibnamefont {Finkelstein}},\ and\ \bibinfo {author} {\bibfnamefont {F.}~\bibnamefont {Amet}},\ }\bibfield  {title} {\bibinfo {title} {Dynamical stabilization of multiplet supercurrents in multiterminal josephson junctions},\ }\href {https://doi.org/10.1021/acs.nanolett.2c01999} {\bibfield  {journal} {\bibinfo  {journal} {Nano Letters}\ }\textbf {\bibinfo {volume} {22}},\ \bibinfo {pages} {7073} (\bibinfo {year} {2022})},\ \bibinfo {note} {pMID: 35997531}\BibitemShut {NoStop}%
\bibitem [{\citenamefont {Graziano}\ \emph {et~al.}(2022)\citenamefont {Graziano}, \citenamefont {Gupta}, \citenamefont {Pendharkar}, \citenamefont {Dong}, \citenamefont {Dempsey}, \citenamefont {Palmstr{\o}m},\ and\ \citenamefont {Pribiag}}]{Grazianoselective2022}%
  \BibitemOpen
  \bibfield  {author} {\bibinfo {author} {\bibfnamefont {G.~V.}\ \bibnamefont {Graziano}}, \bibinfo {author} {\bibfnamefont {M.}~\bibnamefont {Gupta}}, \bibinfo {author} {\bibfnamefont {M.}~\bibnamefont {Pendharkar}}, \bibinfo {author} {\bibfnamefont {J.~T.}\ \bibnamefont {Dong}}, \bibinfo {author} {\bibfnamefont {C.~P.}\ \bibnamefont {Dempsey}}, \bibinfo {author} {\bibfnamefont {C.}~\bibnamefont {Palmstr{\o}m}},\ and\ \bibinfo {author} {\bibfnamefont {V.~S.}\ \bibnamefont {Pribiag}},\ }\bibfield  {title} {\bibinfo {title} {Selective control of conductance modes in multi-terminal josephson junctions},\ }\bibfield  {journal} {\bibinfo  {journal} {Nature Communications}\ }\textbf {\bibinfo {volume} {13}},\ \href {https://doi.org/10.1038/s41467-022-33682-2} {10.1038/s41467-022-33682-2} (\bibinfo {year} {2022})\BibitemShut {NoStop}%
\bibitem [{\citenamefont {Zhang}\ \emph {et~al.}(2023)\citenamefont {Zhang}, \citenamefont {Rashid}, \citenamefont {Ahari}, \citenamefont {Zhang}, \citenamefont {Ananthanarayanan}, \citenamefont {Xiao}, \citenamefont {de~Coster}, \citenamefont {Gilbert}, \citenamefont {Samarth},\ and\ \citenamefont {Kayyalha}}]{Zhang2023}%
  \BibitemOpen
  \bibfield  {author} {\bibinfo {author} {\bibfnamefont {F.}~\bibnamefont {Zhang}}, \bibinfo {author} {\bibfnamefont {A.~S.}\ \bibnamefont {Rashid}}, \bibinfo {author} {\bibfnamefont {M.~T.}\ \bibnamefont {Ahari}}, \bibinfo {author} {\bibfnamefont {W.}~\bibnamefont {Zhang}}, \bibinfo {author} {\bibfnamefont {K.~M.}\ \bibnamefont {Ananthanarayanan}}, \bibinfo {author} {\bibfnamefont {R.}~\bibnamefont {Xiao}}, \bibinfo {author} {\bibfnamefont {G.~J.}\ \bibnamefont {de~Coster}}, \bibinfo {author} {\bibfnamefont {M.~J.}\ \bibnamefont {Gilbert}}, \bibinfo {author} {\bibfnamefont {N.}~\bibnamefont {Samarth}},\ and\ \bibinfo {author} {\bibfnamefont {M.}~\bibnamefont {Kayyalha}},\ }\bibfield  {title} {\bibinfo {title} {Andreev processes in mesoscopic multiterminal graphene josephson junctions},\ }\href {https://doi.org/10.1103/PhysRevB.107.L140503} {\bibfield  {journal} {\bibinfo  {journal} {Phys. Rev. B}\ }\textbf {\bibinfo {volume} {107}},\ \bibinfo {pages} {L140503} (\bibinfo {year} {2023})}\BibitemShut {NoStop}%
\bibitem [{\citenamefont {Cohen}\ \emph {et~al.}(2018)\citenamefont {Cohen}, \citenamefont {Ronen}, \citenamefont {Kang}, \citenamefont {Heiblum}, \citenamefont {Feinberg}, \citenamefont {M{\'e}lin},\ and\ \citenamefont {Shtrikman}}]{cohen2018nonlocal}%
  \BibitemOpen
  \bibfield  {author} {\bibinfo {author} {\bibfnamefont {Y.}~\bibnamefont {Cohen}}, \bibinfo {author} {\bibfnamefont {Y.}~\bibnamefont {Ronen}}, \bibinfo {author} {\bibfnamefont {J.-H.}\ \bibnamefont {Kang}}, \bibinfo {author} {\bibfnamefont {M.}~\bibnamefont {Heiblum}}, \bibinfo {author} {\bibfnamefont {D.}~\bibnamefont {Feinberg}}, \bibinfo {author} {\bibfnamefont {R.}~\bibnamefont {M{\'e}lin}},\ and\ \bibinfo {author} {\bibfnamefont {H.}~\bibnamefont {Shtrikman}},\ }\bibfield  {title} {\bibinfo {title} {Nonlocal supercurrent of quartets in a three-terminal josephson junction},\ }\href {https://www.pnas.org/doi/10.1073/pnas.1800044115} {\bibfield  {journal} {\bibinfo  {journal} {Proceedings of the National Academy of Sciences}\ }\textbf {\bibinfo {volume} {115}},\ \bibinfo {pages} {6991} (\bibinfo {year} {2018})}\BibitemShut {NoStop}%
\bibitem [{\citenamefont {Coraiola}\ \emph {et~al.}(2023{\natexlab{b}})\citenamefont {Coraiola}, \citenamefont {Haxell}, \citenamefont {Sabonis}, \citenamefont {Hinderling}, \citenamefont {ten Kate}, \citenamefont {Cheah}, \citenamefont {Krizek}, \citenamefont {Schott}, \citenamefont {Wegscheider},\ and\ \citenamefont {Nichele}}]{Coraiola2023spin}%
  \BibitemOpen
  \bibfield  {author} {\bibinfo {author} {\bibfnamefont {M.}~\bibnamefont {Coraiola}}, \bibinfo {author} {\bibfnamefont {D.~Z.}\ \bibnamefont {Haxell}}, \bibinfo {author} {\bibfnamefont {D.}~\bibnamefont {Sabonis}}, \bibinfo {author} {\bibfnamefont {M.}~\bibnamefont {Hinderling}}, \bibinfo {author} {\bibfnamefont {S.~C.}\ \bibnamefont {ten Kate}}, \bibinfo {author} {\bibfnamefont {E.}~\bibnamefont {Cheah}}, \bibinfo {author} {\bibfnamefont {F.}~\bibnamefont {Krizek}}, \bibinfo {author} {\bibfnamefont {R.}~\bibnamefont {Schott}}, \bibinfo {author} {\bibfnamefont {W.}~\bibnamefont {Wegscheider}},\ and\ \bibinfo {author} {\bibfnamefont {F.}~\bibnamefont {Nichele}},\ }\href@noop {} {\bibinfo {title} {Spin-degeneracy breaking and parity transitions in three-terminal josephson junctions}} (\bibinfo {year} {2023}{\natexlab{b}}),\ \Eprint {https://arxiv.org/abs/2307.06715} {arXiv:2307.06715 [cond-mat.mes-hall]} \BibitemShut {NoStop}%
\bibitem [{\citenamefont {Gupta}\ \emph {et~al.}(2023)\citenamefont {Gupta}, \citenamefont {Graziano}, \citenamefont {Pendharkar}, \citenamefont {Dong}, \citenamefont {Dempsey}, \citenamefont {Palmstr{\o}m},\ and\ \citenamefont {Pribiag}}]{gupta2023superconducting}%
  \BibitemOpen
  \bibfield  {author} {\bibinfo {author} {\bibfnamefont {M.}~\bibnamefont {Gupta}}, \bibinfo {author} {\bibfnamefont {G.~V.}\ \bibnamefont {Graziano}}, \bibinfo {author} {\bibfnamefont {M.}~\bibnamefont {Pendharkar}}, \bibinfo {author} {\bibfnamefont {J.~T.}\ \bibnamefont {Dong}}, \bibinfo {author} {\bibfnamefont {C.~P.}\ \bibnamefont {Dempsey}}, \bibinfo {author} {\bibfnamefont {C.}~\bibnamefont {Palmstr{\o}m}},\ and\ \bibinfo {author} {\bibfnamefont {V.~S.}\ \bibnamefont {Pribiag}},\ }\bibfield  {title} {\bibinfo {title} {Gate-tunable superconducting diode effect in a three-terminal josephson device},\ }\href {https://doi.org/10.1038/s41467-023-38856-0} {\bibfield  {journal} {\bibinfo  {journal} {Nature Communications}\ }\textbf {\bibinfo {volume} {14}},\ \bibinfo {pages} {3078} (\bibinfo {year} {2023})}\BibitemShut {NoStop}%
\bibitem [{\citenamefont {Coraiola}\ \emph {et~al.}(2024)\citenamefont {Coraiola}, \citenamefont {Svetogorov}, \citenamefont {Haxell}, \citenamefont {Sabonis}, \citenamefont {Hinderling}, \citenamefont {ten Kate}, \citenamefont {Cheah}, \citenamefont {Krizek}, \citenamefont {Schott}, \citenamefont {Wegscheider}, \citenamefont {Cuevas}, \citenamefont {Belzig},\ and\ \citenamefont {Nichele}}]{Coraioladiode}%
  \BibitemOpen
  \bibfield  {author} {\bibinfo {author} {\bibfnamefont {M.}~\bibnamefont {Coraiola}}, \bibinfo {author} {\bibfnamefont {A.~E.}\ \bibnamefont {Svetogorov}}, \bibinfo {author} {\bibfnamefont {D.~Z.}\ \bibnamefont {Haxell}}, \bibinfo {author} {\bibfnamefont {D.}~\bibnamefont {Sabonis}}, \bibinfo {author} {\bibfnamefont {M.}~\bibnamefont {Hinderling}}, \bibinfo {author} {\bibfnamefont {S.~C.}\ \bibnamefont {ten Kate}}, \bibinfo {author} {\bibfnamefont {E.}~\bibnamefont {Cheah}}, \bibinfo {author} {\bibfnamefont {F.}~\bibnamefont {Krizek}}, \bibinfo {author} {\bibfnamefont {R.}~\bibnamefont {Schott}}, \bibinfo {author} {\bibfnamefont {W.}~\bibnamefont {Wegscheider}}, \bibinfo {author} {\bibfnamefont {J.~C.}\ \bibnamefont {Cuevas}}, \bibinfo {author} {\bibfnamefont {W.}~\bibnamefont {Belzig}},\ and\ \bibinfo {author} {\bibfnamefont {F.}~\bibnamefont {Nichele}},\ }\bibfield  {title} {\bibinfo {title} {Flux-tunable josephson diode effect in a hybrid four-terminal josephson junction},\ }\href
  {https://doi.org/10.1021/acsnano.4c01642} {\bibfield  {journal} {\bibinfo  {journal} {ACS Nano}\ }\textbf {\bibinfo {volume} {18}},\ \bibinfo {pages} {9221} (\bibinfo {year} {2024})},\ \bibinfo {note} {pMID: 38488287}\BibitemShut {NoStop}%
\bibitem [{\citenamefont {Chiles}\ \emph {et~al.}(2023)\citenamefont {Chiles}, \citenamefont {Arnault}, \citenamefont {Chen}, \citenamefont {Larson}, \citenamefont {Zhao}, \citenamefont {Watanabe}, \citenamefont {Taniguchi}, \citenamefont {Amet},\ and\ \citenamefont {Finkelstein}}]{Chilesnonreciprocal2023}%
  \BibitemOpen
  \bibfield  {author} {\bibinfo {author} {\bibfnamefont {J.}~\bibnamefont {Chiles}}, \bibinfo {author} {\bibfnamefont {E.~G.}\ \bibnamefont {Arnault}}, \bibinfo {author} {\bibfnamefont {C.-C.}\ \bibnamefont {Chen}}, \bibinfo {author} {\bibfnamefont {T.~F.~Q.}\ \bibnamefont {Larson}}, \bibinfo {author} {\bibfnamefont {L.}~\bibnamefont {Zhao}}, \bibinfo {author} {\bibfnamefont {K.}~\bibnamefont {Watanabe}}, \bibinfo {author} {\bibfnamefont {T.}~\bibnamefont {Taniguchi}}, \bibinfo {author} {\bibfnamefont {F.}~\bibnamefont {Amet}},\ and\ \bibinfo {author} {\bibfnamefont {G.}~\bibnamefont {Finkelstein}},\ }\bibfield  {title} {\bibinfo {title} {Nonreciprocal supercurrents in a field-free graphene josephson triode},\ }\href {https://doi.org/10.1021/acs.nanolett.3c01276} {\bibfield  {journal} {\bibinfo  {journal} {Nano Letters}\ }\textbf {\bibinfo {volume} {23}},\ \bibinfo {pages} {5257} (\bibinfo {year} {2023})},\ \bibinfo {note} {pMID: 37191404}\BibitemShut {NoStop}%
\bibitem [{\citenamefont {Zhang}\ \emph {et~al.}(2024)\citenamefont {Zhang}, \citenamefont {Rashid}, \citenamefont {Tanhayi~Ahari}, \citenamefont {de~Coster}, \citenamefont {Taniguchi}, \citenamefont {Watanabe}, \citenamefont {Gilbert}, \citenamefont {Samarth},\ and\ \citenamefont {Kayyalha}}]{Zhang2024nonreciprocal}%
  \BibitemOpen
  \bibfield  {author} {\bibinfo {author} {\bibfnamefont {F.}~\bibnamefont {Zhang}}, \bibinfo {author} {\bibfnamefont {A.~S.}\ \bibnamefont {Rashid}}, \bibinfo {author} {\bibfnamefont {M.}~\bibnamefont {Tanhayi~Ahari}}, \bibinfo {author} {\bibfnamefont {G.~J.}\ \bibnamefont {de~Coster}}, \bibinfo {author} {\bibfnamefont {T.}~\bibnamefont {Taniguchi}}, \bibinfo {author} {\bibfnamefont {K.}~\bibnamefont {Watanabe}}, \bibinfo {author} {\bibfnamefont {M.~J.}\ \bibnamefont {Gilbert}}, \bibinfo {author} {\bibfnamefont {N.}~\bibnamefont {Samarth}},\ and\ \bibinfo {author} {\bibfnamefont {M.}~\bibnamefont {Kayyalha}},\ }\bibfield  {title} {\bibinfo {title} {Magnetic-field-free nonreciprocal transport in graphene multiterminal josephson junctions},\ }\href {https://doi.org/10.1103/PhysRevApplied.21.034011} {\bibfield  {journal} {\bibinfo  {journal} {Phys. Rev. Appl.}\ }\textbf {\bibinfo {volume} {21}},\ \bibinfo {pages} {034011} (\bibinfo {year} {2024})}\BibitemShut {NoStop}%
\bibitem [{\citenamefont {Matsuo}\ \emph {et~al.}(2023{\natexlab{c}})\citenamefont {Matsuo}, \citenamefont {Imoto}, \citenamefont {Yokoyama}, \citenamefont {Sato}, \citenamefont {Lindemann}, \citenamefont {Gronin}, \citenamefont {Gardner}, \citenamefont {Manfra},\ and\ \citenamefont {Tarucha}}]{Matsuodiode2023}%
  \BibitemOpen
  \bibfield  {author} {\bibinfo {author} {\bibfnamefont {S.}~\bibnamefont {Matsuo}}, \bibinfo {author} {\bibfnamefont {T.}~\bibnamefont {Imoto}}, \bibinfo {author} {\bibfnamefont {T.}~\bibnamefont {Yokoyama}}, \bibinfo {author} {\bibfnamefont {Y.}~\bibnamefont {Sato}}, \bibinfo {author} {\bibfnamefont {T.}~\bibnamefont {Lindemann}}, \bibinfo {author} {\bibfnamefont {S.}~\bibnamefont {Gronin}}, \bibinfo {author} {\bibfnamefont {G.~C.}\ \bibnamefont {Gardner}}, \bibinfo {author} {\bibfnamefont {M.~J.}\ \bibnamefont {Manfra}},\ and\ \bibinfo {author} {\bibfnamefont {S.}~\bibnamefont {Tarucha}},\ }\bibfield  {title} {\bibinfo {title} {Josephson diode effect derived from short-range coherent coupling},\ }\href {https://doi.org/10.1038/s41567-023-02144-x} {\bibfield  {journal} {\bibinfo  {journal} {Nature Physics}\ }\textbf {\bibinfo {volume} {19}},\ \bibinfo {pages} {1636} (\bibinfo {year} {2023}{\natexlab{c}})}\BibitemShut {NoStop}%
\bibitem [{\citenamefont {Crosser}\ \emph {et~al.}(2008)\citenamefont {Crosser}, \citenamefont {Huang}, \citenamefont {Pierre}, \citenamefont {Virtanen}, \citenamefont {T.}, \citenamefont {Wilhelm},\ and\ \citenamefont {Birge}}]{Crosser2008}%
  \BibitemOpen
  \bibfield  {author} {\bibinfo {author} {\bibfnamefont {M.~S.}\ \bibnamefont {Crosser}}, \bibinfo {author} {\bibfnamefont {J.}~\bibnamefont {Huang}}, \bibinfo {author} {\bibfnamefont {F.}~\bibnamefont {Pierre}}, \bibinfo {author} {\bibfnamefont {P.}~\bibnamefont {Virtanen}}, \bibinfo {author} {\bibfnamefont {H.~T.}\ \bibnamefont {T.}}, \bibinfo {author} {\bibfnamefont {F.~K.}\ \bibnamefont {Wilhelm}},\ and\ \bibinfo {author} {\bibfnamefont {N.~O.}\ \bibnamefont {Birge}},\ }\bibfield  {title} {\bibinfo {title} {Nonequilibrium transport in mesoscopic multi-terminal sns josephson junctions},\ }\href {https://doi.org/10.1103/PhysRevB.77.014528} {\bibfield  {journal} {\bibinfo  {journal} {Phys. Rev. B}\ }\textbf {\bibinfo {volume} {77}},\ \bibinfo {pages} {014528} (\bibinfo {year} {2008})}\BibitemShut {NoStop}%
\bibitem [{\citenamefont {Dolgirev}\ \emph {et~al.}(2019{\natexlab{b}})\citenamefont {Dolgirev}, \citenamefont {Kalenkov},\ and\ \citenamefont {Zaikin}}]{Dolgirev2019}%
  \BibitemOpen
  \bibfield  {author} {\bibinfo {author} {\bibfnamefont {P.~E.}\ \bibnamefont {Dolgirev}}, \bibinfo {author} {\bibfnamefont {M.~S.}\ \bibnamefont {Kalenkov}},\ and\ \bibinfo {author} {\bibfnamefont {A.~D.}\ \bibnamefont {Zaikin}},\ }\bibfield  {title} {\bibinfo {title} {Interplay between josephson and aharonov-bohm effects in andreev interferometers},\ }\href {https://doi.org/10.1038/s41598-018-37653-w} {\bibfield  {journal} {\bibinfo  {journal} {Scientific Reports}\ }\textbf {\bibinfo {volume} {9}},\ \bibinfo {pages} {1301} (\bibinfo {year} {2019}{\natexlab{b}})}\BibitemShut {NoStop}%
\bibitem [{\citenamefont {Margineda}\ \emph {et~al.}(2023)\citenamefont {Margineda}, \citenamefont {Claydon}, \citenamefont {Qejvanaj},\ and\ \citenamefont {Checkley}}]{Margineda2023}%
  \BibitemOpen
  \bibfield  {author} {\bibinfo {author} {\bibfnamefont {D.}~\bibnamefont {Margineda}}, \bibinfo {author} {\bibfnamefont {J.~S.}\ \bibnamefont {Claydon}}, \bibinfo {author} {\bibfnamefont {F.}~\bibnamefont {Qejvanaj}},\ and\ \bibinfo {author} {\bibfnamefont {C.}~\bibnamefont {Checkley}},\ }\bibfield  {title} {\bibinfo {title} {Observation of anomalous josephson effect in nonequilibrium andreev interferometers},\ }\href {https://doi.org/10.1103/PhysRevB.107.L100502} {\bibfield  {journal} {\bibinfo  {journal} {Phys. Rev. B}\ }\textbf {\bibinfo {volume} {107}},\ \bibinfo {pages} {L100502} (\bibinfo {year} {2023})}\BibitemShut {NoStop}%
\bibitem [{\citenamefont {Dynes}\ \emph {et~al.}(1978)\citenamefont {Dynes}, \citenamefont {Narayanamurti},\ and\ \citenamefont {Garno}}]{Dynes1978}%
  \BibitemOpen
  \bibfield  {author} {\bibinfo {author} {\bibfnamefont {R.~C.}\ \bibnamefont {Dynes}}, \bibinfo {author} {\bibfnamefont {V.}~\bibnamefont {Narayanamurti}},\ and\ \bibinfo {author} {\bibfnamefont {J.~P.}\ \bibnamefont {Garno}},\ }\bibfield  {title} {\bibinfo {title} {Direct measurement of quasiparticle-lifetime broadening in a strong-coupled superconductor},\ }\href {https://doi.org/10.1103/PhysRevLett.41.1509} {\bibfield  {journal} {\bibinfo  {journal} {Phys. Rev. Lett.}\ }\textbf {\bibinfo {volume} {41}},\ \bibinfo {pages} {1509} (\bibinfo {year} {1978})}\BibitemShut {NoStop}%
\bibitem [{\citenamefont {Jaworski}\ \emph {et~al.}(1976)\citenamefont {Jaworski}, \citenamefont {Parker},\ and\ \citenamefont {Kaplan}}]{Jaworskilifetime176}%
  \BibitemOpen
  \bibfield  {author} {\bibinfo {author} {\bibfnamefont {F.}~\bibnamefont {Jaworski}}, \bibinfo {author} {\bibfnamefont {W.~H.}\ \bibnamefont {Parker}},\ and\ \bibinfo {author} {\bibfnamefont {S.~B.}\ \bibnamefont {Kaplan}},\ }\bibfield  {title} {\bibinfo {title} {Quasiparticle and phonon lifetimes in superconducting pb films},\ }\href {https://doi.org/10.1103/PhysRevB.14.4209} {\bibfield  {journal} {\bibinfo  {journal} {Phys. Rev. B}\ }\textbf {\bibinfo {volume} {14}},\ \bibinfo {pages} {4209} (\bibinfo {year} {1976})}\BibitemShut {NoStop}%
\bibitem [{\citenamefont {M\'elin}\ \emph {et~al.}(2024)\citenamefont {M\'elin}, \citenamefont {Rashid},\ and\ \citenamefont {Kayyalha}}]{Melin3TJJ}%
  \BibitemOpen
  \bibfield  {author} {\bibinfo {author} {\bibfnamefont {R.}~\bibnamefont {M\'elin}}, \bibinfo {author} {\bibfnamefont {A.~S.}\ \bibnamefont {Rashid}},\ and\ \bibinfo {author} {\bibfnamefont {M.}~\bibnamefont {Kayyalha}},\ }\bibfield  {title} {\bibinfo {title} {Ballistic andreev interferometers},\ }\href {https://doi.org/10.1103/PhysRevB.110.235419} {\bibfield  {journal} {\bibinfo  {journal} {Phys. Rev. B}\ }\textbf {\bibinfo {volume} {110}},\ \bibinfo {pages} {235419} (\bibinfo {year} {2024})}\BibitemShut {NoStop}%
\bibitem [{\citenamefont {Huang}\ \emph {et~al.}(2022)\citenamefont {Huang}, \citenamefont {Ronen}, \citenamefont {M{\'e}lin}, \citenamefont {Feinberg}, \citenamefont {Watanabe}, \citenamefont {Taniguchi},\ and\ \citenamefont {Kim}}]{huang2022evidence}%
  \BibitemOpen
  \bibfield  {author} {\bibinfo {author} {\bibfnamefont {K.-F.}\ \bibnamefont {Huang}}, \bibinfo {author} {\bibfnamefont {Y.}~\bibnamefont {Ronen}}, \bibinfo {author} {\bibfnamefont {R.}~\bibnamefont {M{\'e}lin}}, \bibinfo {author} {\bibfnamefont {D.}~\bibnamefont {Feinberg}}, \bibinfo {author} {\bibfnamefont {K.}~\bibnamefont {Watanabe}}, \bibinfo {author} {\bibfnamefont {T.}~\bibnamefont {Taniguchi}},\ and\ \bibinfo {author} {\bibfnamefont {P.}~\bibnamefont {Kim}},\ }\bibfield  {title} {\bibinfo {title} {Evidence for 4e charge of cooper quartets in a biased multi-terminal graphene-based josephson junction},\ }\href {https://www.nature.com/articles/s41467-022-30732-7} {\bibfield  {journal} {\bibinfo  {journal} {Nature Communications}\ }\textbf {\bibinfo {volume} {13}},\ \bibinfo {pages} {3032} (\bibinfo {year} {2022})}\BibitemShut {NoStop}%
\bibitem [{\citenamefont {Wang}\ \emph {et~al.}(2013)\citenamefont {Wang}, \citenamefont {Meric}, \citenamefont {Huang}, \citenamefont {Gao}, \citenamefont {Gao}, \citenamefont {Tran}, \citenamefont {Taniguchi}, \citenamefont {Watanabe}, \citenamefont {Campos}, \citenamefont {Muller} \emph {et~al.}}]{wang2013one}%
  \BibitemOpen
  \bibfield  {author} {\bibinfo {author} {\bibfnamefont {L.}~\bibnamefont {Wang}}, \bibinfo {author} {\bibfnamefont {I.}~\bibnamefont {Meric}}, \bibinfo {author} {\bibfnamefont {P.}~\bibnamefont {Huang}}, \bibinfo {author} {\bibfnamefont {Q.}~\bibnamefont {Gao}}, \bibinfo {author} {\bibfnamefont {Y.}~\bibnamefont {Gao}}, \bibinfo {author} {\bibfnamefont {H.}~\bibnamefont {Tran}}, \bibinfo {author} {\bibfnamefont {T.}~\bibnamefont {Taniguchi}}, \bibinfo {author} {\bibfnamefont {K.}~\bibnamefont {Watanabe}}, \bibinfo {author} {\bibfnamefont {L.}~\bibnamefont {Campos}}, \bibinfo {author} {\bibfnamefont {D.}~\bibnamefont {Muller}}, \emph {et~al.},\ }\bibfield  {title} {\bibinfo {title} {One-dimensional electrical contact to a two-dimensional material},\ }\href {https://www.science.org/doi/abs/10.1126/science.1244358?casa_token=bYaykP2VKecAAAAA:dFUHg9IyopT_IGJxTacpRtt1HLCY5tEtJPuVngo0BWug1YL0BJEW9kBz-bPh5tn_6bzdVg2ePZjiFYE} {\bibfield  {journal} {\bibinfo  {journal} {Science}\ }\textbf {\bibinfo {volume} {342}},\
  \bibinfo {pages} {614} (\bibinfo {year} {2013})}\BibitemShut {NoStop}%
\bibitem [{\citenamefont {Jessen}\ \emph {et~al.}(2019)\citenamefont {Jessen}, \citenamefont {Gammelgaard}, \citenamefont {Thomsen}, \citenamefont {Mackenzie}, \citenamefont {Thomsen}, \citenamefont {Caridad}, \citenamefont {Duegaard}, \citenamefont {Watanabe}, \citenamefont {Taniguchi}, \citenamefont {Booth}, \citenamefont {Pedersen}, \citenamefont {Jauho},\ and\ \citenamefont {B{\o}ggild}}]{Jessen2019}%
  \BibitemOpen
  \bibfield  {author} {\bibinfo {author} {\bibfnamefont {B.~S.}\ \bibnamefont {Jessen}}, \bibinfo {author} {\bibfnamefont {L.}~\bibnamefont {Gammelgaard}}, \bibinfo {author} {\bibfnamefont {M.~R.}\ \bibnamefont {Thomsen}}, \bibinfo {author} {\bibfnamefont {D.~M.~A.}\ \bibnamefont {Mackenzie}}, \bibinfo {author} {\bibfnamefont {J.~D.}\ \bibnamefont {Thomsen}}, \bibinfo {author} {\bibfnamefont {J.~M.}\ \bibnamefont {Caridad}}, \bibinfo {author} {\bibfnamefont {E.}~\bibnamefont {Duegaard}}, \bibinfo {author} {\bibfnamefont {K.}~\bibnamefont {Watanabe}}, \bibinfo {author} {\bibfnamefont {T.}~\bibnamefont {Taniguchi}}, \bibinfo {author} {\bibfnamefont {T.~J.}\ \bibnamefont {Booth}}, \bibinfo {author} {\bibfnamefont {T.~G.}\ \bibnamefont {Pedersen}}, \bibinfo {author} {\bibfnamefont {A.-P.}\ \bibnamefont {Jauho}},\ and\ \bibinfo {author} {\bibfnamefont {P.}~\bibnamefont {B{\o}ggild}},\ }\bibfield  {title} {\bibinfo {title} {Lithographic band structure engineering of graphene},\ }\href
  {https://doi.org/10.1038/s41565-019-0376-3} {\bibfield  {journal} {\bibinfo  {journal} {Nature Nanotechnology}\ }\textbf {\bibinfo {volume} {14}},\ \bibinfo {pages} {340} (\bibinfo {year} {2019})}\BibitemShut {NoStop}%
\bibitem [{\citenamefont {Kammarchedu}\ \emph {et~al.}(2024)\citenamefont {Kammarchedu}, \citenamefont {Butler}, \citenamefont {Rashid}, \citenamefont {Ebrahimi},\ and\ \citenamefont {Kayyalha}}]{Vinay2024}%
  \BibitemOpen
  \bibfield  {author} {\bibinfo {author} {\bibfnamefont {V.}~\bibnamefont {Kammarchedu}}, \bibinfo {author} {\bibfnamefont {D.}~\bibnamefont {Butler}}, \bibinfo {author} {\bibfnamefont {A.~S.}\ \bibnamefont {Rashid}}, \bibinfo {author} {\bibfnamefont {A.}~\bibnamefont {Ebrahimi}},\ and\ \bibinfo {author} {\bibfnamefont {M.}~\bibnamefont {Kayyalha}},\ }\bibfield  {title} {\bibinfo {title} {Understanding disorder in monolayer graphene devices with gate-defined superlattices},\ }\href {https://doi.org/10.1088/1361-6528/ad7853} {\bibfield  {journal} {\bibinfo  {journal} {Nanotechnology}\ }\textbf {\bibinfo {volume} {35}},\ \bibinfo {pages} {495701} (\bibinfo {year} {2024})}\BibitemShut {NoStop}%
\bibitem [{\citenamefont {Caroli}\ \emph {et~al.}(1971)\citenamefont {Caroli}, \citenamefont {Combescot}, \citenamefont {Nozieres},\ and\ \citenamefont {Saint-James}}]{Caroli1971}%
  \BibitemOpen
  \bibfield  {author} {\bibinfo {author} {\bibfnamefont {C.}~\bibnamefont {Caroli}}, \bibinfo {author} {\bibfnamefont {R.}~\bibnamefont {Combescot}}, \bibinfo {author} {\bibfnamefont {P.}~\bibnamefont {Nozieres}},\ and\ \bibinfo {author} {\bibfnamefont {D.}~\bibnamefont {Saint-James}},\ }\bibfield  {title} {\bibinfo {title} {Direct calculation of the tunneling current},\ }\href {https://doi.org/10.1088/0022-3719/4/8/018} {\bibfield  {journal} {\bibinfo  {journal} {Journal of Physics C: Solid State Physics}\ }\textbf {\bibinfo {volume} {4}},\ \bibinfo {pages} {916} (\bibinfo {year} {1971})}\BibitemShut {NoStop}%
\bibitem [{\citenamefont {Caroli}\ \emph {et~al.}(1972)\citenamefont {Caroli}, \citenamefont {Combescot}, \citenamefont {Nozieres},\ and\ \citenamefont {Saint-James}}]{Caroli1972}%
  \BibitemOpen
  \bibfield  {author} {\bibinfo {author} {\bibfnamefont {C.}~\bibnamefont {Caroli}}, \bibinfo {author} {\bibfnamefont {R.}~\bibnamefont {Combescot}}, \bibinfo {author} {\bibfnamefont {P.}~\bibnamefont {Nozieres}},\ and\ \bibinfo {author} {\bibfnamefont {D.}~\bibnamefont {Saint-James}},\ }\bibfield  {title} {\bibinfo {title} {A direct calculation of the tunnelling current: Iv. electron-phonon interaction effects},\ }\href {https://doi.org/10.1088/0022-3719/5/1/006} {\bibfield  {journal} {\bibinfo  {journal} {Journal of Physics C: Solid State Physics}\ }\textbf {\bibinfo {volume} {5}},\ \bibinfo {pages} {21} (\bibinfo {year} {1972})}\BibitemShut {NoStop}%
\bibitem [{\citenamefont {Cuevas}\ \emph {et~al.}(1996)\citenamefont {Cuevas}, \citenamefont {Mart\'{\i}n-Rodero},\ and\ \citenamefont {Yeyati}}]{Cuevas}%
  \BibitemOpen
  \bibfield  {author} {\bibinfo {author} {\bibfnamefont {J.~C.}\ \bibnamefont {Cuevas}}, \bibinfo {author} {\bibfnamefont {A.}~\bibnamefont {Mart\'{\i}n-Rodero}},\ and\ \bibinfo {author} {\bibfnamefont {A.~L.}\ \bibnamefont {Yeyati}},\ }\bibfield  {title} {\bibinfo {title} {Hamiltonian approach to the transport properties of superconducting quantum point contacts},\ }\href {https://doi.org/10.1103/PhysRevB.54.7366} {\bibfield  {journal} {\bibinfo  {journal} {Phys. Rev. B}\ }\textbf {\bibinfo {volume} {54}},\ \bibinfo {pages} {7366} (\bibinfo {year} {1996})}\BibitemShut {NoStop}%
\bibitem [{\citenamefont {Popinciuc}\ \emph {et~al.}(2012)\citenamefont {Popinciuc}, \citenamefont {Calado}, \citenamefont {Liu}, \citenamefont {Akhmerov}, \citenamefont {Klapwijk},\ and\ \citenamefont {Vandersypen}}]{Klapwijk2012}%
  \BibitemOpen
  \bibfield  {author} {\bibinfo {author} {\bibfnamefont {M.}~\bibnamefont {Popinciuc}}, \bibinfo {author} {\bibfnamefont {V.~E.}\ \bibnamefont {Calado}}, \bibinfo {author} {\bibfnamefont {X.~L.}\ \bibnamefont {Liu}}, \bibinfo {author} {\bibfnamefont {A.~R.}\ \bibnamefont {Akhmerov}}, \bibinfo {author} {\bibfnamefont {T.~M.}\ \bibnamefont {Klapwijk}},\ and\ \bibinfo {author} {\bibfnamefont {L.~M.~K.}\ \bibnamefont {Vandersypen}},\ }\bibfield  {title} {\bibinfo {title} {Zero-bias conductance peak and josephson effect in graphene-nbtin junctions},\ }\href {https://doi.org/10.1103/PhysRevB.85.205404} {\bibfield  {journal} {\bibinfo  {journal} {Phys. Rev. B}\ }\textbf {\bibinfo {volume} {85}},\ \bibinfo {pages} {205404} (\bibinfo {year} {2012})}\BibitemShut {NoStop}%
\bibitem [{\citenamefont {Tomi}\ \emph {et~al.}(2021)\citenamefont {Tomi}, \citenamefont {Samatov}, \citenamefont {Vasenko}, \citenamefont {Laitinen}, \citenamefont {Hakonen},\ and\ \citenamefont {Golubev}}]{TomiJoule2021}%
  \BibitemOpen
  \bibfield  {author} {\bibinfo {author} {\bibfnamefont {M.}~\bibnamefont {Tomi}}, \bibinfo {author} {\bibfnamefont {M.~R.}\ \bibnamefont {Samatov}}, \bibinfo {author} {\bibfnamefont {A.~S.}\ \bibnamefont {Vasenko}}, \bibinfo {author} {\bibfnamefont {A.}~\bibnamefont {Laitinen}}, \bibinfo {author} {\bibfnamefont {P.}~\bibnamefont {Hakonen}},\ and\ \bibinfo {author} {\bibfnamefont {D.~S.}\ \bibnamefont {Golubev}},\ }\bibfield  {title} {\bibinfo {title} {Joule heating effects in high-transparency josephson junctions},\ }\href {https://doi.org/10.1103/PhysRevB.104.134513} {\bibfield  {journal} {\bibinfo  {journal} {Phys. Rev. B}\ }\textbf {\bibinfo {volume} {104}},\ \bibinfo {pages} {134513} (\bibinfo {year} {2021})}\BibitemShut {NoStop}%
\bibitem [{\citenamefont {Borzenets}\ \emph {et~al.}(2016)\citenamefont {Borzenets}, \citenamefont {Amet}, \citenamefont {Ke}, \citenamefont {Draelos}, \citenamefont {Wei}, \citenamefont {Seredinski}, \citenamefont {Watanabe}, \citenamefont {Taniguchi}, \citenamefont {Bomze}, \citenamefont {Yamamoto}, \citenamefont {Tarucha},\ and\ \citenamefont {Finkelstein}}]{Borzenets2016}%
  \BibitemOpen
  \bibfield  {author} {\bibinfo {author} {\bibfnamefont {I.~V.}\ \bibnamefont {Borzenets}}, \bibinfo {author} {\bibfnamefont {F.}~\bibnamefont {Amet}}, \bibinfo {author} {\bibfnamefont {C.~T.}\ \bibnamefont {Ke}}, \bibinfo {author} {\bibfnamefont {A.~W.}\ \bibnamefont {Draelos}}, \bibinfo {author} {\bibfnamefont {M.~T.}\ \bibnamefont {Wei}}, \bibinfo {author} {\bibfnamefont {A.}~\bibnamefont {Seredinski}}, \bibinfo {author} {\bibfnamefont {K.}~\bibnamefont {Watanabe}}, \bibinfo {author} {\bibfnamefont {T.}~\bibnamefont {Taniguchi}}, \bibinfo {author} {\bibfnamefont {Y.}~\bibnamefont {Bomze}}, \bibinfo {author} {\bibfnamefont {M.}~\bibnamefont {Yamamoto}}, \bibinfo {author} {\bibfnamefont {S.}~\bibnamefont {Tarucha}},\ and\ \bibinfo {author} {\bibfnamefont {G.}~\bibnamefont {Finkelstein}},\ }\bibfield  {title} {\bibinfo {title} {Ballistic graphene josephson junctions from the short to the long junction regimes},\ }\href {https://doi.org/10.1103/PhysRevLett.117.237002} {\bibfield  {journal} {\bibinfo  {journal}
  {Phys. Rev. Lett.}\ }\textbf {\bibinfo {volume} {117}},\ \bibinfo {pages} {237002} (\bibinfo {year} {2016})}\BibitemShut {NoStop}%
\bibitem [{\citenamefont {Knoch}\ \emph {et~al.}(2012)\citenamefont {Knoch}, \citenamefont {Chen},\ and\ \citenamefont {Appenzeller}}]{Knoch2012}%
  \BibitemOpen
  \bibfield  {author} {\bibinfo {author} {\bibfnamefont {J.}~\bibnamefont {Knoch}}, \bibinfo {author} {\bibfnamefont {Z.}~\bibnamefont {Chen}},\ and\ \bibinfo {author} {\bibfnamefont {J.}~\bibnamefont {Appenzeller}},\ }\bibfield  {title} {\bibinfo {title} {Properties of metal–graphene contacts},\ }\href {https://doi.org/10.1109/TNANO.2011.2178611} {\bibfield  {journal} {\bibinfo  {journal} {IEEE Transactions on Nanotechnology}\ }\textbf {\bibinfo {volume} {11}},\ \bibinfo {pages} {513} (\bibinfo {year} {2012})}\BibitemShut {NoStop}%
\bibitem [{\citenamefont {Pandey}\ \emph {et~al.}(2021)\citenamefont {Pandey}, \citenamefont {Danneau},\ and\ \citenamefont {Beckmann}}]{DanneauCPS20221}%
  \BibitemOpen
  \bibfield  {author} {\bibinfo {author} {\bibfnamefont {P.}~\bibnamefont {Pandey}}, \bibinfo {author} {\bibfnamefont {R.}~\bibnamefont {Danneau}},\ and\ \bibinfo {author} {\bibfnamefont {D.}~\bibnamefont {Beckmann}},\ }\bibfield  {title} {\bibinfo {title} {Ballistic graphene cooper pair splitter},\ }\href {https://doi.org/10.1103/PhysRevLett.126.147701} {\bibfield  {journal} {\bibinfo  {journal} {Phys. Rev. Lett.}\ }\textbf {\bibinfo {volume} {126}},\ \bibinfo {pages} {147701} (\bibinfo {year} {2021})}\BibitemShut {NoStop}%
\bibitem [{\citenamefont {Klapwijk}\ and\ \citenamefont {Sepers}(1977)}]{Klapwijk1977}%
  \BibitemOpen
  \bibfield  {author} {\bibinfo {author} {\bibfnamefont {T.~M.}\ \bibnamefont {Klapwijk}}\ and\ \bibinfo {author} {\bibfnamefont {J.~E.}\ \bibnamefont {Sepers}, \bibfnamefont {M.and~Mooij}},\ }\bibfield  {title} {\bibinfo {title} {Regimes in the behavior of superconducting microbridges},\ }\href {https://doi.org/10.1007/BF00655709} {\bibfield  {journal} {\bibinfo  {journal} {Journal of Low Temperature Physics}\ }\textbf {\bibinfo {volume} {138}},\ \bibinfo {pages} {801} (\bibinfo {year} {1977})}\BibitemShut {NoStop}%
\bibitem [{\citenamefont {Coon}\ and\ \citenamefont {Fiske}(1965)}]{Fiske1965}%
  \BibitemOpen
  \bibfield  {author} {\bibinfo {author} {\bibfnamefont {D.~D.}\ \bibnamefont {Coon}}\ and\ \bibinfo {author} {\bibfnamefont {M.~D.}\ \bibnamefont {Fiske}},\ }\bibfield  {title} {\bibinfo {title} {Josephson ac and step structure in the supercurrent tunneling characteristic},\ }\href {https://doi.org/10.1103/PhysRev.138.A744} {\bibfield  {journal} {\bibinfo  {journal} {Phys. Rev.}\ }\textbf {\bibinfo {volume} {138}},\ \bibinfo {pages} {A744} (\bibinfo {year} {1965})}\BibitemShut {NoStop}%
\bibitem [{\citenamefont {Vanevi\ifmmode~\acute{c}\else \'{c}\fi{}}\ and\ \citenamefont {Belzig}(2005)}]{Chaoticcavity}%
  \BibitemOpen
  \bibfield  {author} {\bibinfo {author} {\bibfnamefont {M.}~\bibnamefont {Vanevi\ifmmode~\acute{c}\else \'{c}\fi{}}}\ and\ \bibinfo {author} {\bibfnamefont {W.}~\bibnamefont {Belzig}},\ }\bibfield  {title} {\bibinfo {title} {Full counting statistics of andreev scattering in an asymmetric chaotic cavity},\ }\href {https://doi.org/10.1103/PhysRevB.72.134522} {\bibfield  {journal} {\bibinfo  {journal} {Phys. Rev. B}\ }\textbf {\bibinfo {volume} {72}},\ \bibinfo {pages} {134522} (\bibinfo {year} {2005})}\BibitemShut {NoStop}%
\bibitem [{\citenamefont {Bretheau}\ \emph {et~al.}(2017)\citenamefont {Bretheau}, \citenamefont {Wang}, \citenamefont {Pisoni}, \citenamefont {Watanabe}, \citenamefont {Taniguchi},\ and\ \citenamefont {Jarillo-Herrero}}]{Bretheau2017}%
  \BibitemOpen
  \bibfield  {author} {\bibinfo {author} {\bibfnamefont {L.}~\bibnamefont {Bretheau}}, \bibinfo {author} {\bibfnamefont {J.~I.-J.}\ \bibnamefont {Wang}}, \bibinfo {author} {\bibfnamefont {R.}~\bibnamefont {Pisoni}}, \bibinfo {author} {\bibfnamefont {K.}~\bibnamefont {Watanabe}}, \bibinfo {author} {\bibfnamefont {T.}~\bibnamefont {Taniguchi}},\ and\ \bibinfo {author} {\bibfnamefont {P.}~\bibnamefont {Jarillo-Herrero}},\ }\bibfield  {title} {\bibinfo {title} {Tunnelling spectroscopy of andreev states in graphene},\ }\href {https://doi.org/10.1038/nphys4110} {\bibfield  {journal} {\bibinfo  {journal} {Nature Physics}\ }\textbf {\bibinfo {volume} {13}},\ \bibinfo {pages} {756} (\bibinfo {year} {2017})}\BibitemShut {NoStop}%
\bibitem [{\citenamefont {Wang}\ \emph {et~al.}(2018)\citenamefont {Wang}, \citenamefont {Bretheau}, \citenamefont {Rodan-Legrain}, \citenamefont {Pisoni}, \citenamefont {Watanabe}, \citenamefont {Taniguchi},\ and\ \citenamefont {Jarillo-Herrero}}]{Bretheau2018}%
  \BibitemOpen
  \bibfield  {author} {\bibinfo {author} {\bibfnamefont {J.~I.-J.}\ \bibnamefont {Wang}}, \bibinfo {author} {\bibfnamefont {L.}~\bibnamefont {Bretheau}}, \bibinfo {author} {\bibfnamefont {D.}~\bibnamefont {Rodan-Legrain}}, \bibinfo {author} {\bibfnamefont {R.}~\bibnamefont {Pisoni}}, \bibinfo {author} {\bibfnamefont {K.}~\bibnamefont {Watanabe}}, \bibinfo {author} {\bibfnamefont {T.}~\bibnamefont {Taniguchi}},\ and\ \bibinfo {author} {\bibfnamefont {P.}~\bibnamefont {Jarillo-Herrero}},\ }\bibfield  {title} {\bibinfo {title} {Tunneling spectroscopy of graphene nanodevices coupled to large-gap superconductors},\ }\href {https://doi.org/10.1103/PhysRevB.98.121411} {\bibfield  {journal} {\bibinfo  {journal} {Phys. Rev. B}\ }\textbf {\bibinfo {volume} {98}},\ \bibinfo {pages} {121411} (\bibinfo {year} {2018})}\BibitemShut {NoStop}%
\bibitem [{Zen()}]{Zenodo}%
  \BibitemOpen
  \href {https://doi.org/10.5281/zenodo.14749812} {\bibinfo {title} {{Nonequilibrium Andreev resonances in ballistic graphene Andreev interferometers [Data set]. Zenodo.}}}\BibitemShut {Stop}%
\bibitem [{\citenamefont {Pfeffer}\ \emph {et~al.}(2014{\natexlab{b}})\citenamefont {Pfeffer}, \citenamefont {Duvauchelle}, \citenamefont {Courtois}, \citenamefont {M\'elin}, \citenamefont {Feinberg},\ and\ \citenamefont {Lefloch}}]{Pfeffer2014subgap}%
  \BibitemOpen
  \bibfield  {author} {\bibinfo {author} {\bibfnamefont {A.~H.}\ \bibnamefont {Pfeffer}}, \bibinfo {author} {\bibfnamefont {J.~E.}\ \bibnamefont {Duvauchelle}}, \bibinfo {author} {\bibfnamefont {H.}~\bibnamefont {Courtois}}, \bibinfo {author} {\bibfnamefont {R.}~\bibnamefont {M\'elin}}, \bibinfo {author} {\bibfnamefont {D.}~\bibnamefont {Feinberg}},\ and\ \bibinfo {author} {\bibfnamefont {F.}~\bibnamefont {Lefloch}},\ }\bibfield  {title} {\bibinfo {title} {Subgap structure in the conductance of a three-terminal josephson junction},\ }\href {https://doi.org/10.1103/PhysRevB.90.075401} {\bibfield  {journal} {\bibinfo  {journal} {Phys. Rev. B}\ }\textbf {\bibinfo {volume} {90}},\ \bibinfo {pages} {075401} (\bibinfo {year} {2014}{\natexlab{b}})}\BibitemShut {NoStop}%
\end{thebibliography}%

\clearpage
\newpage
\onecolumngrid

\renewcommand\thefigure{S\arabic{figure}}    
\setcounter{figure}{0}

\centerline{\textbf{\large{Supporting Information}}}
\section{Andreev bound states in long Josephson junctions}

The spectrum of a long JJ for a normal metal $N$ of the dimension $n_0 a_0\times m_0 a_0$, where $a_0$ is the lattice constant, was calculated in Ref.~\onlinecite{Kulik1970} for highly transparent interfaces. Building on this model, we reduce the two-dimensional (2D)
Hamiltonian to a collection of $m_0$ one-dimensional (1D)
effective Hamiltonians with the conserved transverse quantum numbers $n_y$'s. In this case, the
energy levels evolve with
$\chi=\varphi_R-\varphi_L$, where $\chi$ is the superconducting phase difference between the terminals $S_R$ and
$S_L$. This analysis is consistent with our experimental observation in  Andreev interferometers, where we observe exponential decay of the critical current with temperature; see section {V} below for more details. This temperature dependence is a signature of long JJs {\cite{Borzenets2016}} and supports the use of the long-junction model  in our calculations. 
To investigate if the spectrum of graphene is continuous, we first calculate the disorder energy scale $\eta_G = \hbar v_F \sqrt{\pi  n_0} \approx$ 10 meV, where $n_0 \approx 10^{10}$ cm$^{-2}$ is the residual carrier concentration and $v_F = 10^6$ m/s {\cite{Jessen2019,Vinay2024}}. We also find the energy level spacing of the ABSs in long JJs as $s_{2D}\approx W_G (a_0/L)^2 \approx
5\times10^{-4}$ meV for $W_G$ = 0.5 eV and
$n_0=a_0/L\approx 10^{-3}$. Since, $\eta_G \gg s_{2D}$, we
conclude that the spectrum of graphene is continuous.

\section{The microscopic picture for the phase-AR process}

As we discussed in the main text, our three-terminal Andreev interferometers enable a nonstandard phase-AR process, resulting from the finite chemical potential of graphene. We model this microscopic phase-AR by applying a low-order perturbation theory to the tunneling amplitude. Specifically, to visualize the electron-electron and electron-hole scattering mechanisms, we employ closed-form Keldysh diagrams within the Nambu formalism. In this framework, the normal tunneling amplitude corresponds to the electron-electron scattering, while the anomalous tunneling amplitude represents the electron-hole scattering.
We first consider the Keldysh diagrams of orders 2 and 4 in an SSN ($N_L$, $S_T$, $S_R$) Andreev interferometer of Fig.~\ref{Fig2SSN.jpg}(a) of the main text. Figures ~\ref{FigSSN.pdf}(a) and (b) show scattering processes in which spin-up electrons undergo second and fourth order scattering at the graphene-$N_L$ interface, respectively. In contrast, Figs. ~\ref{FigSSN.pdf}(c) and (d) depict electron-hole scattering processes in which a spin-up electron and a spin-down hole undergo local Andreev reflections at the graphene-$S_L$ interface. Figure ~\ref{FigSSN.pdf}(e) further shows the lowest-order phase-AR process transmitting Cooper pairs into the grounded $S_R$ and $S_T$.

In the all superconducting SSS interferometer, the normal electron-electron processes at the graphene-$N_L$ interface are replaced by local Andreev reflections at the graphene-$S_L$ interface. To account for this change in the all superconducting ($S_L$, $S_T$, $S_R$) interferometer, we modify our Keldysh diagrams as shown in Fig. ~\ref{FigSSS.pdf}. Figures ~\ref{FigSSS.pdf}(a-c) show the lowest-order local Andreev reflections at graphene-$S_L$, graphene-$S_R$, and graphene-$S_T$ interfaces, respectively. Figure ~\ref{FigSSS.pdf}(d) shows the lowest order phase-AR process transmitting Cooper pairs into the grounded $S_T$ and $S_R$. Figures ~\ref{FigSSS.pdf}(e) and (f) show the lowest-order AC-Josephson oscillations under a voltage bias between ($S_L$ and $S_T$) and between ($S_L$ and $S_R$), respectively. In our approximation, we replace the normal tunneling differential conductance in SSN  with the Andreev differential conductance in SSS. We note that the AC Josephson processes can produce a DC signal as demonstrated in resistively coupled RSJ models {\cite{Zhang2023,gupta2023superconducting}. However, the connection between the phase coherent microscopic processes such as Andreev reflections and the RSJ model is not straightforward. While Landau-Zener processes may act as a potential bridge, in 2D metals such as graphene, these processes involve the low-energy Kulik-Ishii-Bagwell states {\cite{Kulik1970,Ishii1970,Bagwell1992}, rather than conventional ABSs. Consequently, the Landau-Zener transmitted quasiparticles are unable to directly escape into the quasiparticle continua. Therefore, we limit our theoretical analysis to the lowest-order perturbative scattering processes, recognizing that a comprehensive theory has yet to be developed. 

\begin{figure}
\centering
\includegraphics[width=13cm]{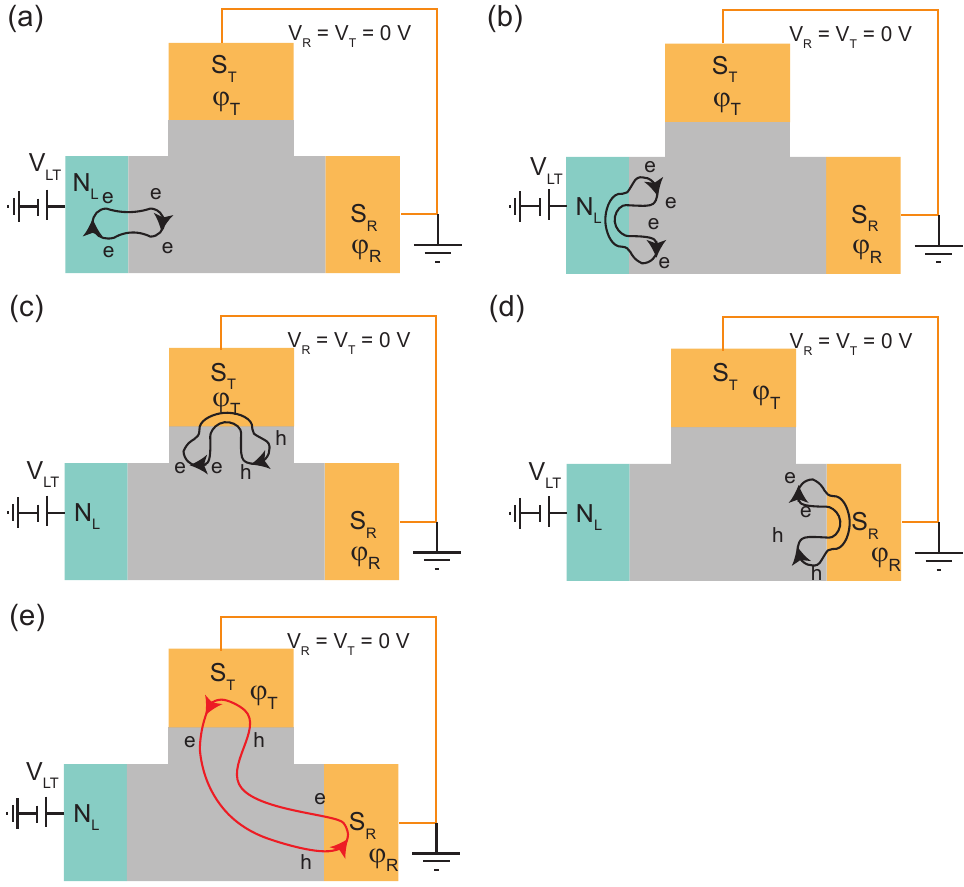}
\caption{\label{FigSSN.pdf} The low-order Keldysh diagrams for the ($N_L$, $S_T$, $S_R$) device. (a, b) Normal
electron scattering processes of orders 2 (a) and 4 (b). 
(c, d) Local Andreev reflection processes of order 4, enabling, at the lowest-order, the Cooper pair transfer between the
graphene-$N$ section and superconducting terminals $S_T$ (c) and $S_R$ (d). (e) The lowest-order phase-AR process, transferring
Cooper pairs from graphene into the superconducting terminals $S_T$ and $S_R$. This transfer process is sensitive to the flux through the loop and the distribution function in graphene.}
\end{figure}

\newpage
\section{Theoretical calculation of the phase-AR conductance}

Our calculation is based on a perturbative expansion of the current,
derived for the phase-AR diagrams shown in Figs. ~\ref{FigSSN.pdf}(e)
and ~\ref{FigSSS.pdf}(d). The superconducting leads
  $S_n$ are described by the standard BCS Hamiltonian with order
  parameter $\Delta$ and superconducting phase variable
  $\varphi_n$. The central normal metal is modeled by a tight-binding
  Hamiltonian, and the gate voltage allows the formation of a large
  metallic Fermi surface. The fully dressed advanced and retarded
  Nambu Green's functions $\hat{G}^A$ and $\hat{G}^R$ are obtained as
  the solution of the Dyson equations, that take the following form of
  a convolution over the time variables $t_1$, $t_2$ and $t_3$:
  \begin{eqnarray}
    \label{eq:convolution1}
    \hat{G}^A(t_1,t_2)&=&\hat{g}^A(t_1,t_2)
    + \int dt_3 \hat{g}^A(t_1,t_3) \hat{\Sigma}(t_3) \hat{G}^A(t_3,t_2)\\
     \hat{G}^R(t_1,t_2)&=&\hat{g}^R(t_1,t_2)
     + \int dt_3 \hat{g}^R(t_1,t_3) \hat{\Sigma}(t_3) \hat{G}^R(t_3,t_2)
     ,
     \label{eq:convolution2}
  \end{eqnarray}
where $\hat{g}^A$ and $\hat{g}^R$ are the bare advanced and retarded
Nambu Green's functions, i.e. the Green's functions with vanishingly
small hopping self-energy $\hat{\Sigma}$. Considering an SSN Andreev
interferometer, the Nambu self-energy $\Sigma(t_3)$ is independent on
its time-argument $t_3$ and the convolutions in
Eqs.~(\ref{eq:convolution1})-(\ref{eq:convolution2}) become simple
products after Fourier transforming to the frequency $\omega$:
\begin{eqnarray}
  \label{eq:Dyson1}
  \hat{G}^A(\omega)&=&\hat{g}^A(\omega)+\hat{g}^A(\omega)\hat{\Sigma}\hat{G}^A(\omega)\\
  \hat{G}^R(\omega)&=&\hat{g}^R(\omega)+\hat{g}^R(\omega)\hat{\Sigma}\hat{G}^R(\omega)
  .
  \label{eq:Dyson2}
\end{eqnarray}
Equations~(\ref{eq:Dyson1})-(\ref{eq:Dyson2}) are next expanded as a series in powers of the
hopping self-energy $\hat{\Sigma}$, for instance, the following is obtained at the order
$\Sigma^4$:
\begin{eqnarray}
  \hat{G}^{A/R}(\omega)&=&\hat{g}^{A/R}(\omega)\\
  &+& \hat{g}^{A/R}(\omega) \hat{\Sigma} \hat{g}^{A/R}(\omega)\\
  &+& \hat{g}^{A/R}(\omega) \hat{\Sigma} \hat{g}^{A/R}(\omega)\hat{\Sigma} \hat{g}^{A/R}(\omega)\\
  &+& \hat{g}^{A/R}(\omega) \hat{\Sigma} \hat{g}^{A/R}(\omega)\hat{\Sigma} \hat{g}^{A/R}(\omega)\hat{\Sigma} \hat{g}^{A/R}(\omega)\\
  &+& \hat{g}^{A/R}(\omega) \hat{\Sigma} \hat{g}^{A/R}(\omega)\hat{\Sigma} \hat{g}^{A/R}(\omega)\hat{\Sigma} \hat{g}^{A/R}(\omega)\hat{\Sigma} \hat{g}^{A/R}(\omega)\\
  &+& ...
\end{eqnarray}
This expansion can graphically be expanded as the diagrams similar to
Figs.~\ref{FigSSN.pdf} and~\ref{FigSSS.pdf}. The current per channel
through the $S_L$-$N$ interface is given by
\cite{Caroli1971,Caroli1972,Cuevas}
\begin{equation}
  \label{eq:I-a-alpha}
  I_{a,\alpha}(t)=\frac{ie}{\hbar} \Sigma_{a,\alpha} \sum_\sigma
  \left[\langle c_{\alpha,\sigma}^+(t) c_{a,\sigma}(t)\rangle
    - \langle c_{a,\sigma}^+(t) c_{\alpha,\sigma}(t)\rangle\right]
  ,
\end{equation}
where $\Sigma_{a,\alpha}$ is the hopping amplitude connecting the pair
of tight-binding sites $(a,\alpha)$ on the $S_L$ and $N$-sides of the
$S_L$-$N$ contact. The current in Eq.~(\ref{eq:I-a-alpha}) is calculated from the
Keldysh Green's function $\hat{G}^{+,-}$:
\begin{equation}
  \label{eq:G+-}
  \hat{G}^{+,-}_{a,\alpha}(t,t')=
  i\left(\begin{array}{cc}
    \langle c_{\alpha,\uparrow}^+(t') c_{a,\uparrow}(t)\rangle &
    \langle c_{\alpha,\downarrow}(t') c_{a,\uparrow}(t)\rangle\\
    \langle c_{\alpha,\uparrow}^+(t') c_{a,\downarrow}^+(t)\rangle &
    \langle c_{\alpha,\downarrow}(t') c_{a,\downarrow}^+(t)\rangle
  \end{array} \right)
  ,
\end{equation}
and Eqs.~(\ref{eq:I-a-alpha}) and~(\ref{eq:G+-}) are related by the following relation:
\begin{equation}
  I_{a,\alpha}(t)=\frac{e}{\hbar} \mbox{Nambu-trace}
  \left[ \hat{\Sigma}_{\alpha,a} \hat{G}^{+,-}_{a,\alpha}(t,t)
    - \hat{\Sigma}_{a,\alpha} \hat{G}^{+,-}_{\alpha,a}(t,t) \right]
  ,
\end{equation}
where the notation ``Nambu-trace'' stands for a summation over the
$1,1$ and the $2,2$ diagonal Nambu components. The Keldysh Green's
function in Eq.~(\ref{eq:G+-}) is calculated from the Dyson-Keldysh
equation that takes the following form in frequency $\omega$
\cite{Caroli1971,Caroli1972,Cuevas}:
\begin{equation}
  \label{eq:DK}
  \hat{G}^{+,-}(\omega)=
  \left(\hat{I}+\hat{G}^R(\omega)\hat{\Sigma}\right)
  \hat{g}^{+,-}(\omega)
  \left(\hat{I}+\hat{\Sigma} \hat{G}^A(\omega)\right)
  ,
\end{equation}
where $\hat{g}^{+,-}(\omega)=n_F(\omega)\left[\hat{g}^A(\omega)
  -\hat{g}^R(\omega)\right]$ is the bare Keldysh Green's function,
with $n_F(\omega)$ the Fermi-Dirac distribution function. The
Dyson-Keldysh equations given by Eq.~(\ref{eq:DK}) are also expanded
in powers of the tunneling self-energy $\hat{\Sigma}$, which can
graphically be represented as diagrams.

In the following, the current will be approximated as
  closed-loop diagrams, such as those appearing at the order
  $\Sigma_L^2 \Sigma_T^2$. In the absence of electron-electron
  interactions and electron-phonon coupling, the distribution function
  of the two-dimensional metal results in a voltage-sensitive step at
  the energy $\delta\mu_N$=$eV_{LT}$. We calculate in Ref.~\onlinecite{Melin3TJJ} the differential
  conductance by taking the partial derivative of the current with
  respect to $\delta\mu_N$:

\begin{equation}\label{g0}
    g_0(\Phi/2\pi,V_{LT} R/\pi \hbar v_F) = \cos{\Phi} \cos{(2eV_{LT}R/\hbar v_F)}.
\end{equation}
 
Following Ref.~\onlinecite{Melin3TJJ}, we next combine the perturbative and nonperturbative
  approaches to generate numerical results for the conductance.
Figure~\ref{FigS3eta} shows the resulting
differential conductance as a function of normalized voltage
2$eV_{LT}$/$\varepsilon_1$ and magnetic flux $\Phi/\Phi_0$ for various
values of the normalized linewidth broadening $\eta/\varepsilon_1$,
where $\varepsilon_1$ = $\hbar$$v_F/L$ is the 1D gap
{\cite{Kulik1970}}, $v_F$ is the fermi velocity, and $L$ is the
separation between $S_R$ and $S_T$. For the SSS interferometer with
small $\eta/\varepsilon_1$, e.g., $\eta/\varepsilon_1 = 0.1$ in
(Fig. ~\ref{FigS3eta}(a)), we observe narrow Andreev resonances, which
evolve into a regular phase-sensitive conductance oscillations for the
SSN device when $\eta/\varepsilon_1$ is larger than unity but persists
even up to $\eta/\varepsilon_1 =100$ (Fig. ~\ref{FigS3eta}(d)).  We
note that, in our calculations, the effect of finite linewidth broadening is incorporated through the inclusion of an imaginary component in the energy. This crude modeling is likely to be an oversimplification of the microscopic process of inelastic scattering in SSN. Furthermore, the Keldysh diagrams for the SSS interferometer
(Fig.~\ref{FigSSS.pdf}) represent \textcolor{blue}{the} parity-conserving transitions $m
\rightarrow m \pm 2$, where $m$ is the number of fermions. Conversely,
the quasiparticle processes in the SSN interferometer result in
parity non-conserving transitions $m \rightarrow m \pm 1$
(Fig. ~\ref{FigSSN.pdf}). Therefore, even though oscillating patterns
persist up to large $\eta/\varepsilon_1$ = 100 in the SSN device, the
parity-nonconserving processes likely lead to quasiparticle poisoning,
which in turn suppresses the coherent Andreev oscillations. The
mechanisms of quasiparticle poisoning and their effects on the
non-equilibrium populations of the Andreev modes will be the focus of
future investigations.

\begin{figure}[h]
\centering
\includegraphics[width=13cm]{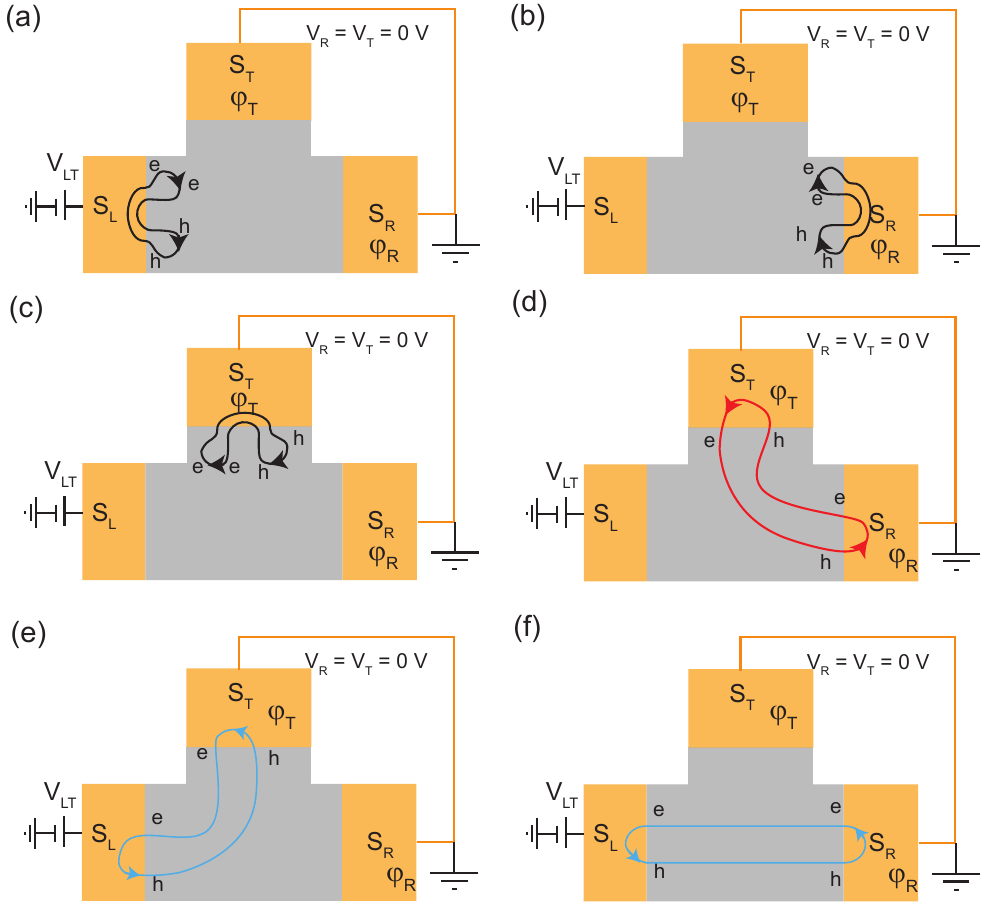}
\caption{\label{FigSSS.pdf} The low-order Keldysh diagrams for the ($S_L$, $S_T$, $S_R$) Andreev interferometer. (a-c) The lowest-order Andreev reflection Keldysh diagrams for graphene-$S_L$ (a), graphene-$S_R$ (b), and graphene-$S_T$ (c) interfaces. (d) The lowest-order phase-AR Keldysh diagram. (e, f) The lowest-order AC-Josephson Keldysh diagrams, enabling AC Josephson oscillations between $S_L$ and $S_R$ (e) and $S_L$ and $S_T$ (f). }
\end{figure}

\begin{figure}[h]
\centering
\includegraphics[width=10cm]{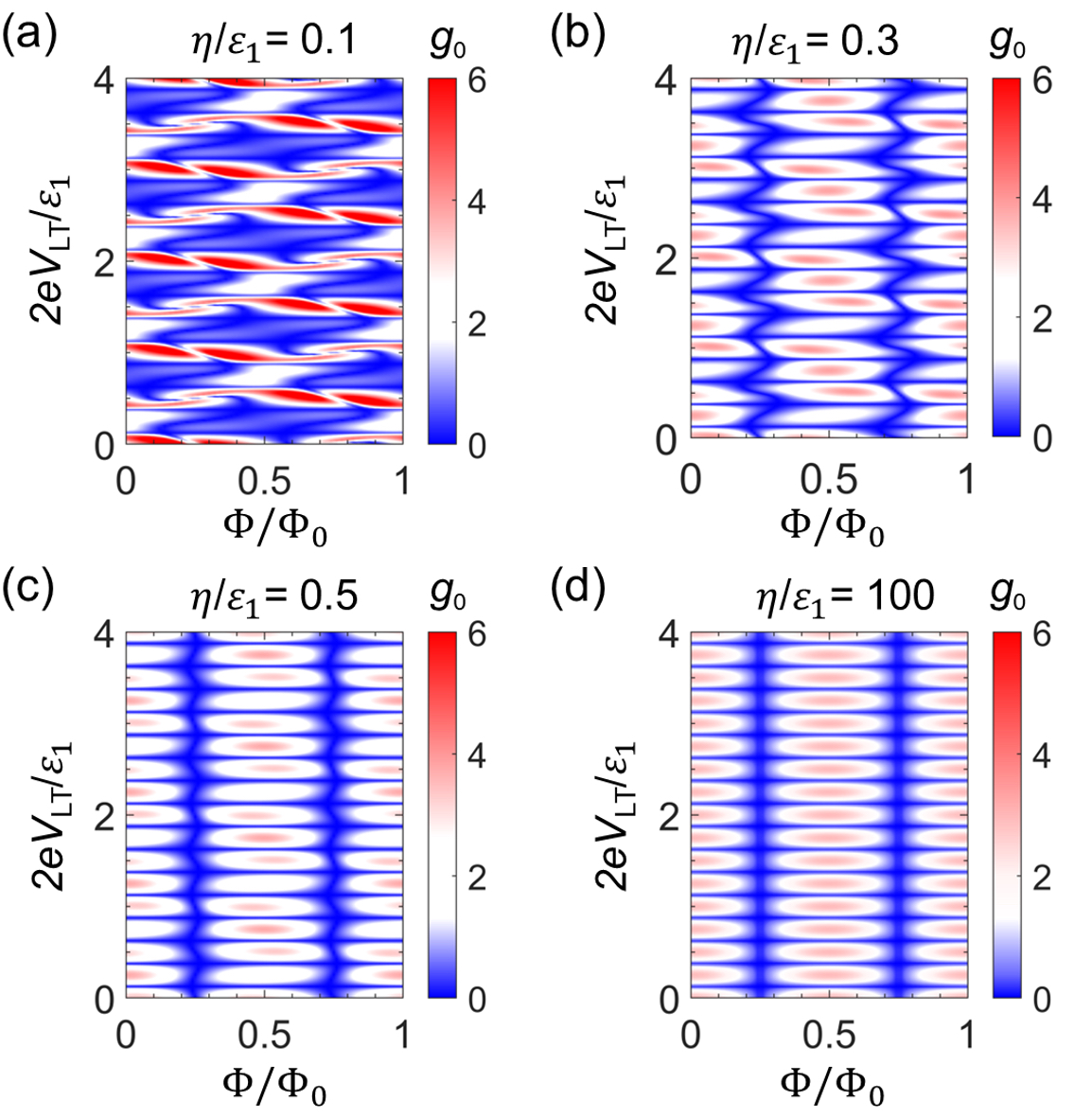}
\caption{\label{FigS3eta}(a-d) The calculated differential conductance $g_0$ as a function of normalized voltage $2eV_{LT} / \varepsilon_1$ and magnetic flux $\Phi/\Phi_0$ for various values of normalized linewidth broadening $\eta/\epsilon_1$ = 0.1 (a), 0.3 (b), 0.5 (c), and 100 (d).
  \label{fig:G}
}
\end{figure}

\newpage
\section{Ballistic or diffusive normal metal
  in the short-junction limit}

We consider a simple two-terminal $S_RINIS_b$ JJ in the short-junction limit, where $I$ is an insulating layer. The spectral current has two
poles at the energies $\omega=\pm \Delta$ in this JJ. We denote the pairs of the tight-binding sites making the
contacts at the $S_RIN$ and $NIS_L$ interfaces by $a$-$\alpha$
and $b$-$\beta$, respectively. At equilibrium, the
current through the $a$-$\alpha$ weak link takes the following
expression:
\begin{eqnarray}
  I_S&=&\frac{e}{\hbar}
  \int d\omega \mbox{Nambu-trace}\left[\hat{\Sigma}_{a,\alpha}
  \hat{G}^{+,-}_{\alpha,a} - \hat{\Sigma}_{\alpha,a}
  \hat{G}^{+,-}_{a,\alpha}\right]\\
  &=& \frac{e}{\hbar}
  \int d\omega n_F(\omega) \mbox{Nambu-trace}\left[
  \hat{\Sigma}_{a,\alpha}\left(\hat{G}^A_{\alpha,a}-
  \hat{G}^R_{\alpha,a}\right)
  -
  \hat{\Sigma}_{\alpha,a}\left(\hat{G}^A_{a,\alpha}-
  \hat{G}^R_{a,\alpha}\right)\right]
  ,
  \label{eq:I2}
\end{eqnarray}
where both $S_R$ and $S_L$ are grounded, and
$n_F(\omega)$ is the Fermi-Dirac distribution function.
Combining Eq.~(\ref{eq:I2}) with the Dyson equation leads to:
\begin{equation}
  I_S=\Sigma_a^2 \Sigma_b^2 \frac{e}{\hbar}
  \int d\omega n_F(\omega) \left\{
  g^{A,2,1}_{a,a}(\omega) g^{A,1,2}_{b,b}(\omega)
  g^{A,1,1}_{\alpha,\beta}(\omega) g^{A,2,2}_{\beta,\alpha}(\omega) -
  g^{A,1,2}_{a,a}(\omega) g^{A,2,1}_{b,b}(\omega)
  g^{A,2,2}_{\alpha,\beta}(\omega) g^{A,1,1}_{\beta,\alpha}(\omega)
  \right\}
  .
\end{equation}
Using the following expressions for the local Nambu Green's functions:
\begin{eqnarray}
  g_{a,a}^{A,1,2}(\omega)&=&\frac{\exp(i\varphi_a)}{BW\sqrt{\Delta^2-(\omega-i\eta)^2}}\\
  g_{a,a}^{A,2,1}(\omega)&=&\frac{\exp(-i\varphi_a)}{BW\sqrt{\Delta^2-(\omega-i\eta)^2}}\\
  g_{b,b}^{A,1,2}(\omega)&=&\frac{\exp(i\varphi_b)}{BW\sqrt{\Delta^2-(\omega-i\eta)^2}}\\
  g_{b,b}^{A,2,1}(\omega)&=&\frac{\exp(-i\varphi_b)}{BW\sqrt{\Delta^2-(\omega-i\eta)^2}}
,
\end{eqnarray}
we obtain the  supercurrent:
\begin{equation}
  \label{eq:IS-result}
  I_S=4\pi\frac{e\Delta}{\hbar}
  \frac{\Sigma_a^2 \Sigma_b^2}{BW^2} \sin(\varphi_a-\varphi_b)
  \mbox{Re}\left[ g^{A,1,1}_{\alpha,\beta}(-\Delta)
    g^{A,2,2}_{\beta,\alpha}(-\Delta) \right]
  ,
\end{equation}
where $BW$ is the bandwidth.

The Green's functions of a ballistic 2D metal are:
\begin{eqnarray}
  g^{A,1,1}&\simeq& \frac{i}{BW\sqrt{k_F L_{\alpha,\beta}}} \cos \left[
    \left(k_F+\frac{\omega}{v_F}\right) L_{\alpha,\beta} \right]\\
 g^{A,2,2}&\simeq& \frac{i}{BW\sqrt{k_F L_{\alpha,\beta}}} \cos \left[
    \left(k_F-\frac{\omega}{v_F}\right) L_{\alpha,\beta} \right]
  .
\end{eqnarray}
By averaging over the oscillations at the smallest scale of the Fermi wavelength, we find the following expression for the supercurrent :
\begin{equation}
  \label{eq:IS-ball-2D}
  I_S^{(ball, 2D)}\simeq 2\pi\frac{e\Delta}{\hbar}
  \frac{\Sigma_a^2
    \Sigma_b^2}{BW^4(k_F L_{\alpha,\beta})}
  \sin\left(\varphi_R-\varphi_L\right) \cos\left(\frac{2\Delta
    L}{v_F}\right).
\end{equation}
Eq.~(\ref{eq:IS-ball-2D}) features undamped oscillations as a
function of $L$ in the short-junction limit. 

The Green's functions of a ballistic 3D metal are:
\begin{eqnarray}
  g^{A,1,1}\simeq \frac{1}{BW(k_F L_{\alpha,\beta})} \exp\left[i
    \left(k_F+\frac{\omega}{v_F}\right) L_{\alpha,\beta}\right]
  \exp\left(\frac{i\omega L_{\alpha,\beta}}{v_F}\right)\\
 g^{A,2,2}\simeq \frac{1}{BW(k_F L_{\alpha,\beta})} \exp\left[i
    \left(k_F-\frac{\omega}{v_F}\right) L_{\alpha,\beta}\right]
 \exp\left(\frac{i\omega L_{\alpha,\beta}}{v_F}\right)
  .
\end{eqnarray}
The expression of the supercurrent is then similar to
Eq.~(\ref{eq:IS-ball-2D}), with $1/k_F L_{\alpha,\beta}$ replaced by
$1/(k_F L_{\alpha,\beta})^2$ in the prefactor.

Finally, considering a weakly disordered 3D metal, we evaluate the
diffusion probability $P_d(L)=\langle \langle g^{A,1,1}(L)
g^{A,2,2}(L) \rangle \rangle$ averaged over the disorder, see for
instance Eq.~(A11) in Ref.~\onlinecite{Pfeffer2014subgap}.
\begin{equation}
  \label{eq:Pd}
  P_d(L_{\alpha,\beta})\sim\frac{1}{BWDR_{\alpha,\beta}}
  \exp\left(-\sqrt{\frac{\omega}{E_{Th}}}\right)
  \cos\left(\sqrt{\frac{\omega}{E_{Th}}}\right)
  ,
\end{equation}
where the Thouless energy is given by $E_{Th}=\hbar
D/L_{\alpha,\beta}^2$. We finally plug in Eq.~(\ref{eq:Pd}) into
Eq.~(\ref{eq:IS-result}) and deduce that, in the diffusive limit, the
supercurrent exponentially decays with the distance $L_{\alpha,\beta}$
between the interfaces.

From the above calculations, we conclude that in ballistic junctions the Andreev resonances are undamped compared to diffusive JJs where the supercurrent decays exponentially as a function of distance between $S_L$ and $S_R$. Even though our calculations are done in the short-junction limit, we note that similar arguments are valid for long JJs as they rely on the presence of an elastic scattering time in disordered systems.

\clearpage
\section{Ballistic transport in graphene JJs}
To demonstrate ballistic transport, we first study a two-terminal JJ in the short-junction limit as shown in Fig.~\ref{FigS1ballistic.pdf}(a) (channel length $\approx$ 0.3 $\mu$m). We observe Fabry-P\'erot oscillations of the critical current when graphene is hole-doped, consistent with previous experiments in ballistic JJs {\cite{Borzenets2016}}. Additionally, we study the critical current $I_c$ evolutions with gate voltage $V_g$ and temperature in a three-terminal graphene interferometer in Fig.~\ref{FigS1ballistic.pdf}(b). We find that $I_c$ exponentially decreases with increasing temperature as $I_c \propto \exp\left(-\frac{k_B T}{\Delta E}\right)$. From an exponential fit of $I_c$ vs $T$, we calculate $\Delta{E}$ and find that it is almost $V_g$-independent (Fig.~\ref{FigS1ballistic.pdf}(c)), indicating \textcolor{blue}{that} our SSS interferometer is in the ballistic limit {\cite{Borzenets2016}}. We further observe in this device that the critical voltage $V_c$, below which the junction is superconducting, does not change with $V_g$ (Fig.~\ref{FigS2ballistic.pdf} (b) and (c)). The value of $V_c$ is proportional to the $I_c$$R_N$ product, and its gate-independent behavior is related to ballistic transport as discussed in Ref.~\onlinecite{Borzenets2016}. Moreover, in this SSS interferometer, we observe that dc voltages ($V_n$’s) corresponding to resonance peaks in differential resistance $R$ is gate voltage independent as shown in Fig. ~\ref{FigS2ballistic.pdf}(c). We note that the energy level spacing of Andreev modes in long JJs is proportional to the Fermi velocity $v_F$ {\cite{Kulik1970, Ishii1970,Bagwell1992}}. Given the Dirac dispersion relation in graphene, we expect $v_F$ and, consequently, $V_n$’s to remain approximately constant with varying gate voltage.

\begin{figure}[h]
\centering
\includegraphics[width=17cm]{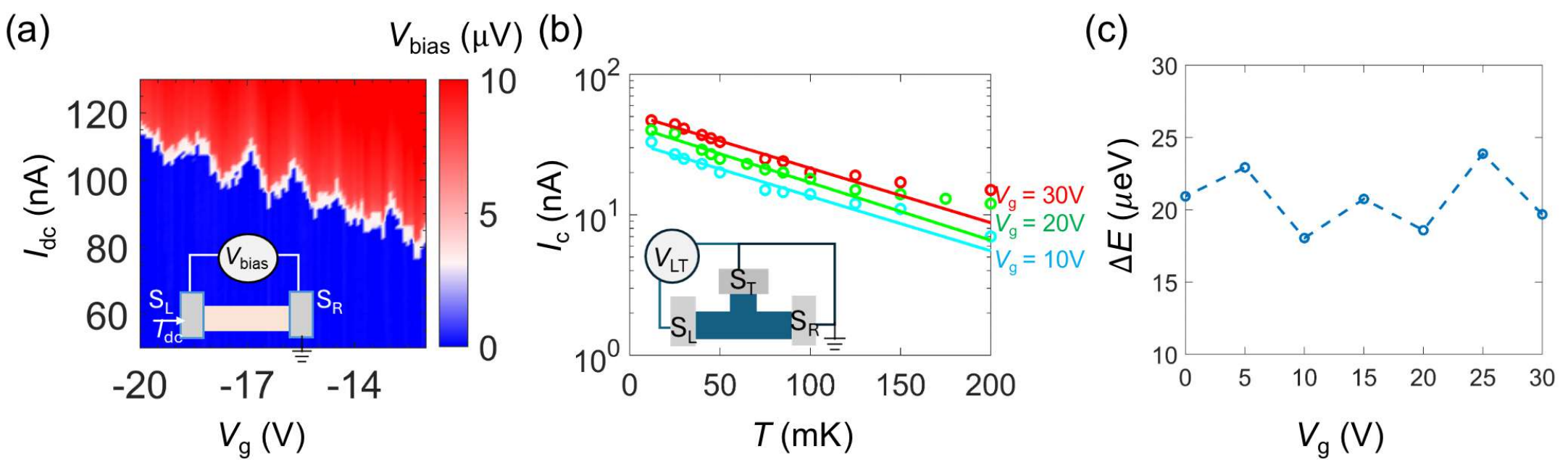}
\caption{\label{FigS1ballistic.pdf}(a) Bias voltage $V_{bias}$ as a function of dc current $I_{dc}$ and gate voltage $V_g$ for a two-terminal JJ (channel length $\approx$ 0.3 $\mu$m). (b) Critical current $I_c$ on a semi-log scale vs temperature at different gate voltages $V_g$'s in an three-terminal SSS interferometer. Insets in (a) and (b) show the device schematic and bias configuration. (c) Energy $\Delta{E}$, extracted from temperature dependence of $I_c$ in (b), as a function of $V_g$.}
\end{figure}

\begin{figure}[h]
\centering
\includegraphics[width=16cm]{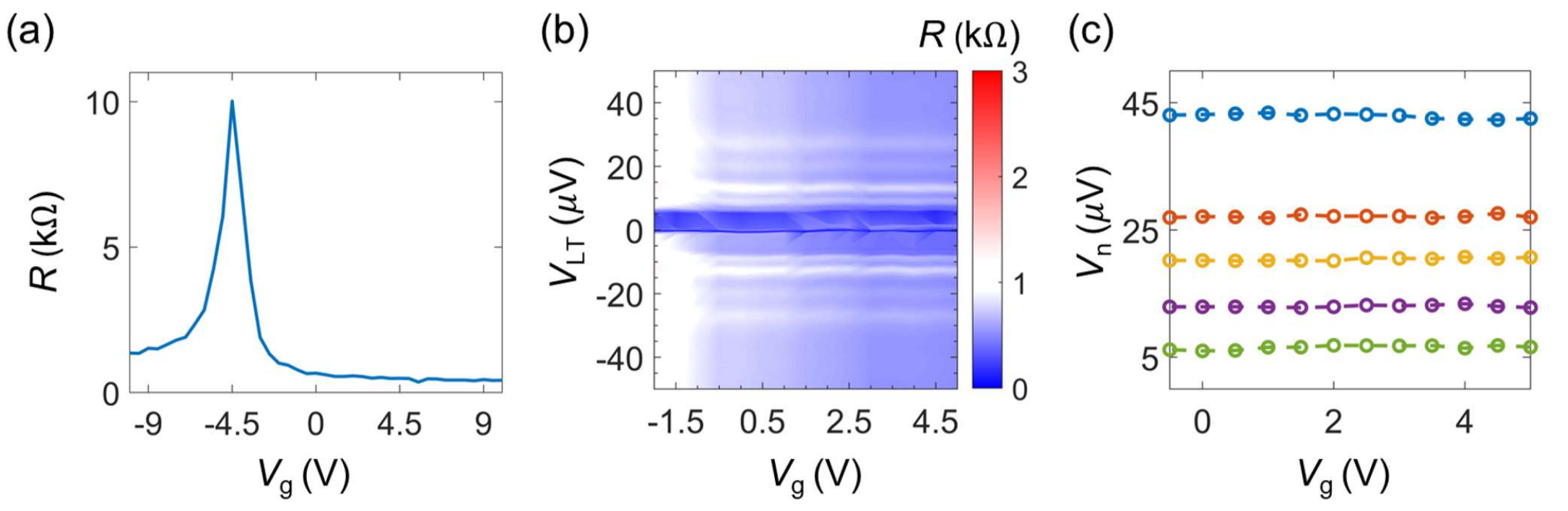}
\caption{\label{FigS2ballistic.pdf}(a) Differential resistance $R$ versus back gate $V_g$, indicating Dirac point at $V_{CNP}$ $\approx -4.5$ V. (b) Differential resistance $R$ as a function of bias voltage $V_{LT}$ and $V_g$ for the three-terminal SSS interferometer shown in Fig.~\ref{FigS1ballistic.pdf}(b). (c) Critical voltage $V_C$ and dc voltages $V_n$'s, corresponding to resonance peaks, as a function of $V_g$. Green symbols correspond to $V_c$, whereas blue, orange, yellow, and purple symbols correspond to $V_n$'s.}
\end{figure}

\clearpage
\section{Andreev resonances and crossovers in another SSS device}
We plot the differential resistance $R$ as functions of the magnetic field $B$ and bias voltage $V_{LT}$ for another three-terminal SSS device without any tunnel probe in Figs.~\ref{FigS1.jpg}(a-c). We see 0-$\pi$ crossovers marked by arrows in Figs.~\ref{FigS1.jpg}(b) and (c). The differential resistance $R$ is obtained by sweeping magnetic field $B$ at different values of $I_{dc}$ at 12 mK. Due to  stochastic switching and hysteresis caused by the $B$-field sweep, the colormap features appear slightly tilted. To mitigate this issue, our subsequent measurements [Fig.~\ref{Fig1.jpg}(b)] are performed at an elevated temperature $T$ = 300 mK, where the bias current $I_{dc}$ is swept at different $B$ fields to obtain differential resistance maps.
\begin{figure}[h]
\centering
\includegraphics[width=16cm]{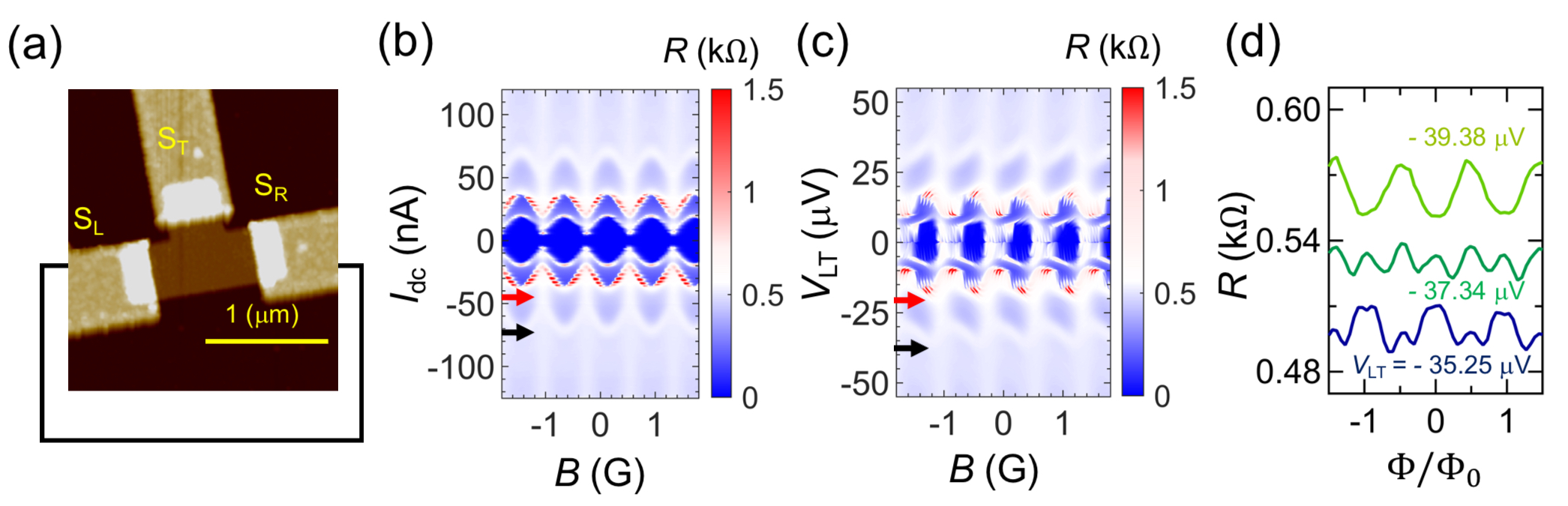}
\caption{\label{FigS1.jpg}  (a) Atomic force microscope (AFM) image of another three-terminal SSS interferometer. Unlike the device presented in the main text, this device does not have any tunneling probe. (b, c) Differential resistance $R$ as a function of the magnetic field $B$ and, respectively, bias current $I_{dc}$ (b) and bias voltage $V_{LT}$ (c) at $V_g$ = 20 V and $T$ = 12 mK. Arrows mark 0-$\pi$ crossovers. (d) Differential resistance $R$ as a function of the magnetic flux $\Phi$ for the crossovers marked by the black arrows in (b) and (c). The oscillations show an evolution from maxima to minima with decreasing $V_{LT}$. Curves are shifted vertically by 0.04 $k\Omega$ for clarity. Colormaps are saturated to better highlight crossovers. These maps are generated by sweeping $B$ and $I_{dc}$ with step sizes of 0.05 G and 1 nA, respectively.}
\end{figure}

\section{Additional line cuts of differential resistance maps}

\begin{figure}[h]
\centering
\includegraphics[width=13cm]{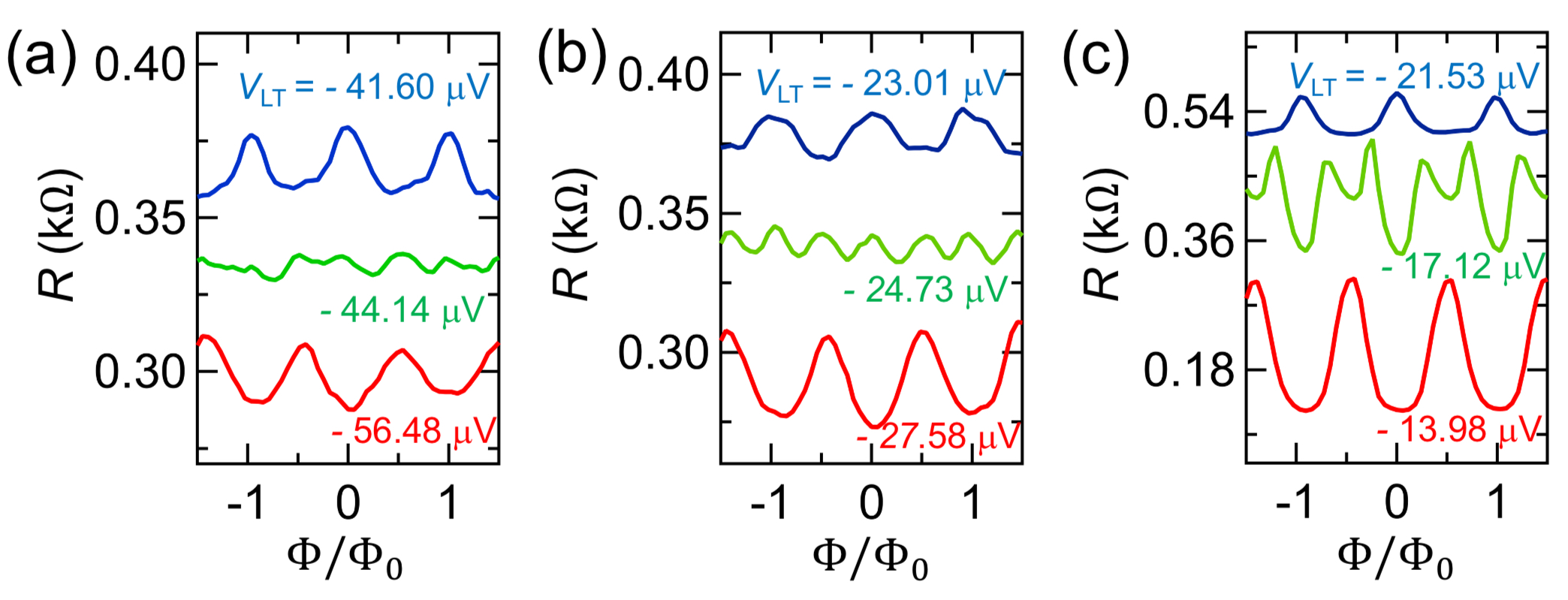}
\caption{\label{FigS2.jpg} (a-c) Differential resistance $R$ as a function of the magnetic flux $\Phi$ for the crossovers marked by the black (a), yellow (b), and purple (c) arrows in Figs. 1b and c of the main text. The oscillations show an evolution from maxima to minima with increasing $V_{LT}$. Curves are shifted vertically by 0.03 $k\Omega$ for clarity.}
\end{figure}
\newpage
\section{Multiple Andreev reflections (MARs)}
To investigate MARs, we measure \textcolor{blue}{the} differential conductance $G$ as a function of bias voltage $|V_{LT} |$. The differential conductance plots for $|V_{LT} |$ $\leq$ 400 $\mu$V do not show any sign of MAR peaks corresponding to $n$ = 1, 2, 3, 4, and 5, where $n$ = 2$\Delta/eV_{LT}$ and $\Delta \approx$ 180 $\mu$eV (Fig.~\ref{FigS5MARs.pdf}(a)). Therefore, we conclude that the resonances and their associated 0-$\pi$ crossovers are likely not related to MARs. Specifically, the voltage range where we observe the resonances are very small ($\leq$ 50 $\mu$V) as depicted in Fig.~\ref{FigS5MARs.pdf}(b), which if coming from MARs, would correspond to $n \geq$ 6 (and even to $n \approx$ 27 for $R_4$). Additionally, in Fig.~\ref{FigS5MARs.pdf}(c) we observe that the resonance peaks associated with the phase-AR mechanism show no temperature dependence. We attribute the lack of MARs to the long-junction nature of the SSS device (channel length $\approx$ 1.2 $\mu$m).
\begin{figure}[h]
\centering
\includegraphics[width=17.5cm]{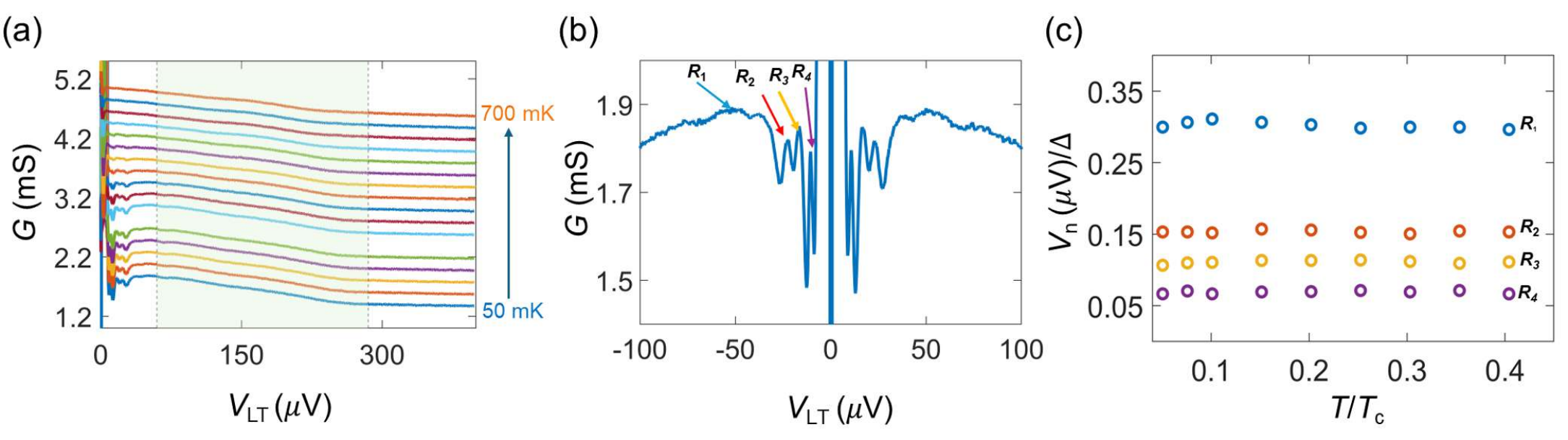}
 \caption{\label{FigS5MARs.pdf}(a) Differential conductance $G=1/R$ of a SSS interferometer as a function of bias voltage at various temperatures, ranging from 50 mK to 700 mK. (b) Differential conductance $G$ as a function of bias voltage at $T=50$ mK. Resonance peaks ($V_n$'s) are marked by arrows as $R_1$, $R_2$, $R_3$, and $R_4$. (c) Normalized $V_n/\Delta$ as a function of temperature for different resonance peaks.}
\end{figure}

\newpage
\section{Differential resistance map in a graphene four-terminal SNSN device}
To explore the impact of normal current injections in graphene JJs, we fabricate two-terminal JJs with additional normal contacts (Fig.~\ref{FigS6SNSN.pdf}). Previous experiments have reported supercurrent sign reversals and $\pi$-junctions in diffusive JJs with similar normal metal configurations {\cite{Baselmans1999}}. These observations have been attributed to modifications in the energy distribution of the current-carrying states induced by the normal current {\cite{Baselmans1999}}.  However, in our graphene SNSN device, we find that the critical current monotonically decreases with increasing normal current without any signature of a 0-$\pi$ transition or  supercurrent sign reversal. Notably, our results are consistent with a previous experiment on graphene JJs {\cite{DanneauCPS20221}}, and may be due to nonuniform doping density and potential barriers formed across the graphene channel {\cite{DanneauCPS20221}}.   
\begin{figure}[h]
\centering
\includegraphics[width=10cm]{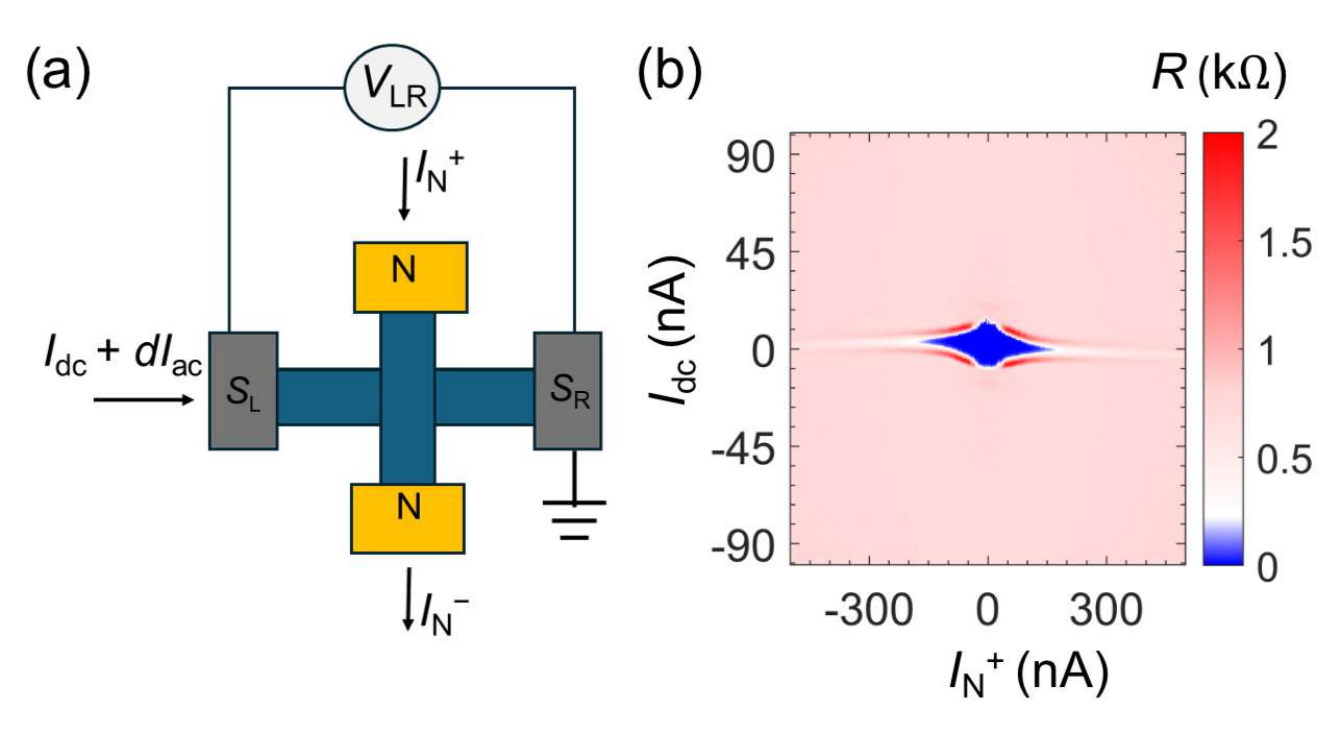}
 \caption{\label{FigS6SNSN.pdf}(a) Schematics of a four-terminal device with two normal and two superconducting terminals. (b) Colormap of differential resistance $R =  dV_{LR}/dI_{ac}$ as a function of normal current $I_{N+} = -I_{N-}$ and dc current $I_{dc}$ as  schematically shown in (a).}
\end{figure}
\clearpage
\section{Self-induced Shapiro Steps}
As discussed in the main text, in an Andreev interferometer, the voltage-biased terminal is analogous to an RF source. We note that coupling a two-terminal JJ to an electromagnetic (EM) field results in constant voltage steps in the current-voltage characteristics, known as Shapiro steps. However, even without any external EM field, the AC Josephson effect in a voltage-biased JJ, via coupling to the metallic enclosure of the cryostat, may generate an EM field. This EM field, in turn, may result in the formation of self-induced Shapiro-like steps {\cite{Klapwijk1977}} or Fiske resonances {\cite{Fiske1965}} in the $R$ versus $V$ curves. To distinguish self-induced Shapiro steps from the phase-ARs, we fabricate and characterize a two-terminal JJ, which is adjacent to our three-terminal Andreev interferometer. 

 Figures~\ref{Fig3.jpg}(a) and (b) show the differential resistance $R$ as a function of the superconducting flux and bias voltage in a two-terminal JJ and three-terminal SSS interferometer, respectively. The graphene channel for the two-terminal JJ is around 1.2 $\mu$m long and 0.5 $\mu$m wide. We observe two resonances above the critical current in the two-terminal JJ when the voltage bias is within $\pm 25 ~ \mu$V ($|V_{bias} | \leq 25 \mu$V). We attribute these resonances to self-induced Shapiro steps \cite{Klapwijk2012}. In contrast, in our SSS interferometer, we observe two additional dark-bright regions (red and blue arrows in Fig.~\ref{Fig3.jpg}(b), which fall outside of the bias voltage range for self-induced Shapiro steps. Moreover, we note that the observation of 0-$\pi$ crossovers in the SSN interferometer is inconsistent with the self-induced Shapiro steps, as both $S_R$ and $S_T$ are grounded in the SSN device. Therefore, we conclude that the crossovers are likely due to phase-AR processes which provide the most straightforward explanation for our observations in both SSN and SSS configurations. Figure~\ref{Fig3.jpg}(c) shows vertical cuts of $R$ versus $V_{bias}$ (red curve) and $V_{LT}$ (blue curve) at $\Phi$ = -0.15 G corresponding to vertical dashed lines in Figs.~\ref{Fig3.jpg}(a) and (b), respectively. Red and blue arrows mark the resonances corresponding to the same arrows in Fig.~\ref{Fig3.jpg}(b). In the three-terminal SSS interferometer, we still observe multiple crossovers for $|V_{LT} |\leq 25 \mu$V. However, it is challenging to identify whether these crossovers correspond to self-induced Shapiro steps or phase-ARs as both processes lead to similar signatures in the $R$ versus $\varphi$ data.
\begin{figure}[h]
 \centering
\includegraphics[width=8.7cm]{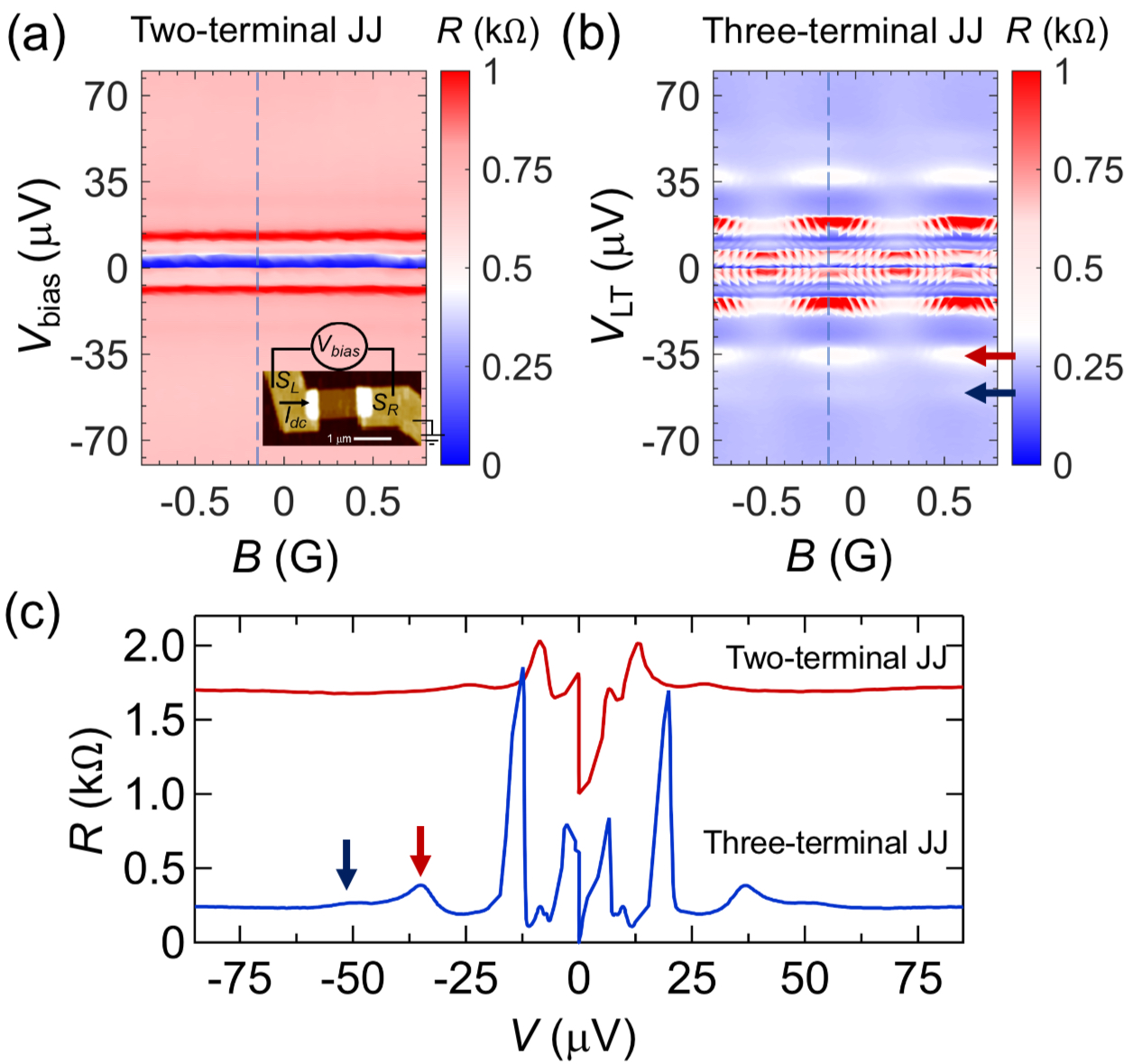}
\caption{\label{Fig3.jpg} (a-b) Differential resistance $R$ as a function of $B$ and the bias voltage $V_{bias}$ in a two-terminal JJ (a) and $V_{LT}$ in a three-terminal SSS interferometer (b). Inset in A shows the atomic force microscope (AFM) image of the two-terminal JJ alongside the measurement configuration. (c) Line cuts of $R$ versus $V$ for two-terminal JJ (red) and three-terminal  SSS interferometer (blue) at $B = -0.15$ G. Red and blue arrows mark the resonances corresponding to the same arrows in (b). Red curve is shifted vertically by 1 $k\Omega$ for clarity. Colormaps are saturated to better highlight 0-$\pi$ crossovers. These maps are generated by sweeping $B$ and $I_{dc}$ with step sizes of 0.05 G and 1 nA, respectively.}
\end{figure}
 
 \newpage
\section{Temperature dependence of the differential resistance maps}

\begin{figure}[h]
\centering
\includegraphics[width=18cm]{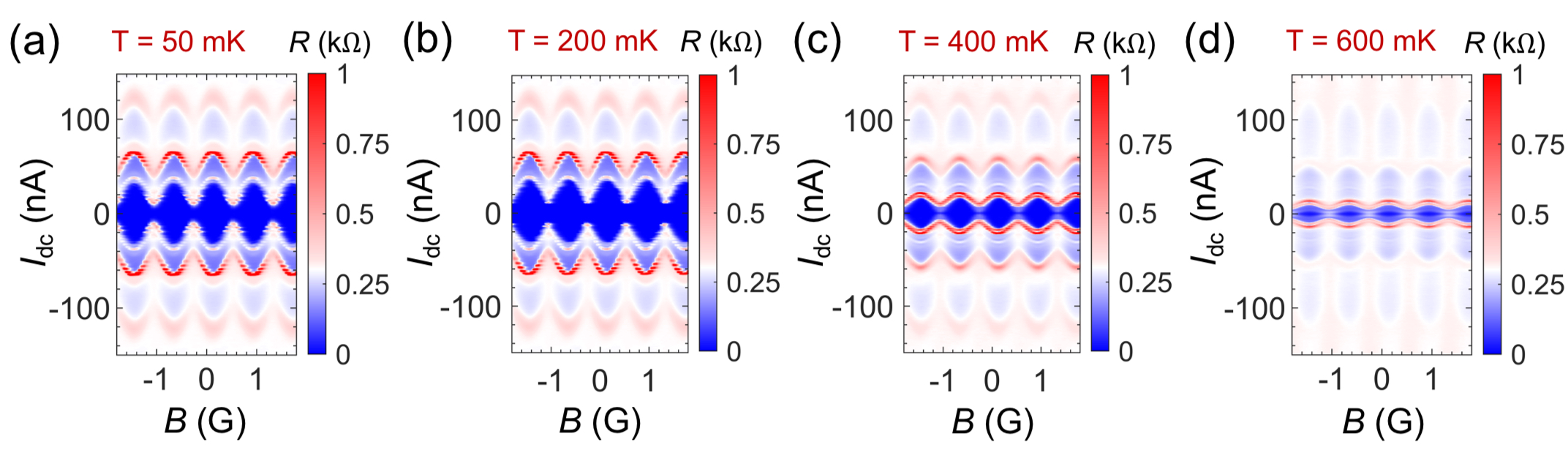}
 \caption{\label{FigS3.jpg} (a-d) Temperature dependence of differential resistance $R$ as a function of the magnetic field at 50 mK (a), 200 mK (b), 400 mK (c), and 600 mK (d). Colormaps are saturated to better highlight 0-$\pi$ crossovers.}
\end{figure}

\section{Line cuts in the SSN device}

\begin{figure}[h]
\centering
\includegraphics[width=10.5cm]{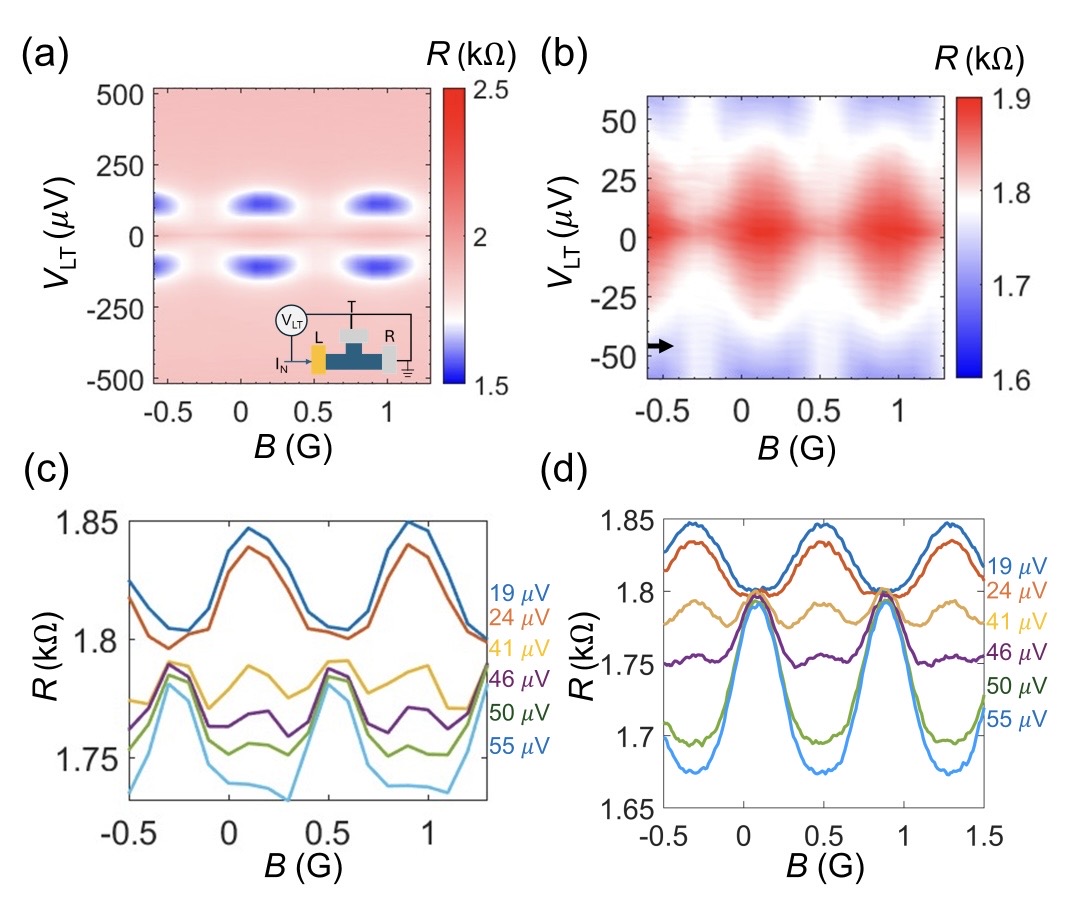}
 \caption{\label{FigS2SSN.jpg} (a) Differential resistance $R$ as a function of the magnetic field $B$ and $V_{LT}$ for a three-terminal SSN device at a back gate voltage $V_g = $ 0 $V$ and temperature $ T$ = 12 mK. The Dirac point is at $V_{CNP}$ $\approx$ -4.5 $V$. The inset shows the measurement setup. The superconducting terminals are denoted as $S_T$ and $S_R $, and the normal terminal is denoted as $N_L$. Current $I_N$ is applied to $N_L$ and voltage $V_{LT}$ is measurable between terminals $N_L$ and $S_T$. (b) Zoomed-in colormap of (a) showing differential resistance $R$ as a function of the magnetic field $B$ and $V_{LT}$. Black arrow indicates $0-\pi$ crossover.  (c) Line cuts of differential resistance $R$ versus $B$ at different $V_{LT}$ measured near $0-\pi$ crossover from (a). (d) Differential resistance $R$ versus $B$ at different $V_{LT}$'s measured separately near the $0-\pi$ crossover.  $R$ is measured by sweeping $B$ field with a step size of 0.01 G at different $V_{LT}$'s. Due to flux trapping in between measurements, the curves show a horizontal shift ($\sim$ 0.5 G) in Panel (d). The average value of $V_{LT}$ is used in panels (c) and (d). Color maps are saturated to better highlight $0-\pi$ crossovers. Colormaps are generated by sweeping $B$ and $I_{dc}$ with step sizes of 0.1 G and 1 nA, respectively.}
\end{figure}

\end{document}